\documentclass[twocolumn,prd,aps,floatfix,preprintnumbers,superscriptaddress,bibnotes]{revtex4-1}

\usepackage{epsfig}
\usepackage{amsmath}
\usepackage{latexsym}
\usepackage[psamsfonts]{amssymb}
\usepackage{graphicx}
\usepackage{longtable}
\usepackage{ulem}
\usepackage{longtable}
\usepackage{epstopdf}
\usepackage{bm}
\usepackage{color}
\usepackage{hyperref}

\newcommand{\be}{\begin{equation}}
\newcommand{\ee}{\end{equation}}
\newcommand{\ben}{\begin{eqnarray}}
\newcommand{\een}{\end{eqnarray}}
\newcommand{\bi}{\begin{itemize}}
\newcommand{\ei}{\end{itemize}}

\newcommand{\la}{\langle}
\newcommand{\ra}{\rangle}

\begin{document}

\preprint{UTHEP-721, UTCCS-P-113, HUPD-1806}


\title{Nucleon form factors on a large volume lattice\\
near the physical point in 2+1 flavor QCD}

\author{Ken-Ichi Ishikawa\:}
\affiliation{Graduate School of Science, Hiroshima University, Higashi-Hiroshima, 
Hiroshima 739-8526, Japan}
\affiliation{RIKEN Advanced Institute for Computational Science, Kobe, Hyogo 650-0047, Japan}
\altaffiliation{RIKEN Advanced Institute for Computational Science (AICS) was renamed to RIKEN Center for Computational Science (R-CCS) in April 2018.}
\author{Yoshinobu Kuramashi\:}
\email{kuramasi@het.ph.tsukuba.ac.jp}
\affiliation{Center for Computational Sciences, University of Tsukuba, Tsukuba, Ibaraki 305-8577, Japan}
\affiliation{RIKEN Advanced Institute for Computational Science, Kobe, Hyogo 650-0047, Japan}
\author{Shoichi Sasaki\:}
\email{ssasaki@nucl.phys.tohoku.ac.jp} 
\affiliation{Department of Physics, Tohoku University, Sendai 980-8578, Japan}
\affiliation{RIKEN Advanced Institute for Computational Science, Kobe, Hyogo 650-0047, Japan}
\author{Natsuki Tsukamoto\:}
\affiliation{Department of Physics, Tohoku University, Sendai 980-8578, Japan}
\author{Akira Ukawa\:}
\affiliation{Center for World Premier International Research Center Initiative (WPI), Japan Society for the Promotion of Science, Tokyo 102-0083, Japan}
\affiliation{RIKEN Advanced Institute for Computational Science, Kobe, Hyogo 650-0047, Japan}
\author{Takeshi Yamazaki\:}
\email{yamazaki@het.ph.tsukuba.ac.jp}
\affiliation{Faculty of Pure and Applied Sciences, University of Tsukuba, Tsukuba, Ibaraki, 305-8571, Japan}
\affiliation{Center for Computational Sciences, University of Tsukuba, Tsukuba, Ibaraki 305-8577, Japan}
\affiliation{RIKEN Advanced Institute for Computational Science, Kobe, Hyogo 650-0047, Japan}

\collaboration{PACS Collaboration}

\date{\today}
\begin{abstract}
We present results for the isovector nucleon form factors measured on a $96^4$ lattice 
at almost the physical pion mass with a lattice spacing of 0.085 fm in 2+1 flavor QCD. 
The configurations are generated with the stout-smeared $O(a)$-improved Wilson 
quark action and the Iwasaki gauge action at $\beta$=1.82. The pion mass at the simulation 
point is about 146 MeV. A large spatial volume of $(8.1~{\rm fm})^3$ allows us to investigate 
the form factors in the small momentum transfer region.
We determine the isovector electric radius and magnetic moment from nucleon electric ($G_E$) 
and magnetic ($G_M$) form factors as well as the axial-vector coupling $g_A$. 
We also report on the results of the axial-vector ($F_A$), induced pseudoscalar ($F_P$) 
and pseudoscalar ($G_P$) form factors in order to verify the axial Ward-Takahashi identity 
in terms of the nucleon matrix elements, which may be called as the generalized 
Goldberger-Treiman relation. 
\end{abstract}

\maketitle

 
\section{Introduction}
\label{sec:1}

The nucleon vector and axial elastic form factors are good probes to 
investigate the internal structure of the nucleon~\cite{Thomas:2001kw}.
Although great theoretical and experimental efforts have been 
devoted to improving our knowledge of the nucleon structure, 
there are several unsolved problems associated with fundamental 
properties of the proton and neutron.
The proton radius puzzle, where high-precision measurements
of the proton's electric charge radius from the muonic hydrogen 
Lamb shift~\cite{{Pohl:2010zza},{Pohl:2013yb}}
disagree with well established results of both electron-proton scattering 
and hydrogen spectroscopy~\cite{Tanabashi:2018oca},
is currently one of the most intriguing problems in this field~\cite{Sick:2018fzn}.
The neutron lifetime puzzle, where the
discrepancy between the results of beam experiments
and storage experiments remains unsolved, is another open question that deserves further investigation
in terms of the nucleon axial-vector coupling $g_A$~\cite{Czarnecki:2018okw}.

Much effort has been devoted to calculating the nucleon form factors with lattice QCD since 1980's. 
Unfortunately, satisfactory results, that are consistent with the experiments, have not yet been achieved 
for the nucleon structure in the previous lattice QCD simulations. The current situation is that we are still 
struggling to reproduce well-known experimental results, {\it e.g.}, the axial-vector coupling 
and the electric charge radius. This means that we have not yet achieved our final goal of properly generating
a single-nucleon state is properly generated in lattice QCD calculations. 
The discrepancy between existing lattice calculations 
and experimental values could be attributed to unresolved systematic errors in numerical simulations. 
An important source of systematic uncertainty should be due to the fact that the simulated quark mass used in 
simulations is heavier than the physical one. 
However, the particular quantities like the axial-vector coupling~\footnote{
After we complete this work, the new calculation method of the axial-vector coupling is developed 
based on the Feynman-Hellmann method in Ref.~\cite{Chang:2018uxx}. Their high-precision result is
quite consistent with the experimental value.} 
and the electric charge radius still show some discrepancies with respect to the experimental values 
in the previous lattice QCD simulations near the physical point~\cite{{Bhattacharya:2016zcn},{Alexandrou:2017hac},{Gupta:2018qil},{Green:2014xba},{Alexandrou:2017ypw}}.  

In this paper, we aim to calculate the nucleon form factors in a very large spatial volume 
in realistic lattice QCD, where the light quark masses are down to their physical values. 
For this purpose, we use the 2+1 flavor QCD gauge configurations generated 
on a $(8.1~ {\rm fm})^4$ lattice near the physical point (the pion mass $m_\pi$ 
at the simulation point is about 146 MeV) by the PACS Collaboration~\cite{Ishikawa:2015rho}.
There are three reasons for paying special attention to its large spatial volume of $(8.1~ {\rm fm})^3$.

First of all, the large finite volume effect, represented by the $m_\pi L$ scaling 
($L$ denotes the spatial extent of the lattice volume), on measurements of the axial-vector coupling $g_A$ 
was reported in Refs.~\cite{{Yamazaki:2008py},{Yamazaki:2009zq}}. 
According to their conclusion, one can 
estimate that spatial sizes of $7.7-9.4$ fm ($m_\pi L\approx 5.7-7.0$) 
for $m_\pi \approx 146$ MeV are necessary for keeping the finite volume effect 
on $g_A$ at or below 1\%~\footnote{There is some controversy about
significant finite-volume effect when $m_\pi L \ge 4$ as reported in recent unquenched
high-statistics calculations~\cite{{Chang:2018uxx},{Yoon:2016jzj},{Gupta:2018qil}}.
}. 

Second, thanks to the large spatial volume, we can access to the small momentum transfer 
region up to 152 MeV. The momentum squared ($q^2$) dependence of the nucleon form factors
can be carefully examined in the very large spatial volume. Indeed, the values of a given form factor 
at very low $q^2$ help to more accurately determine the slope of the form factor at $q^2=0$,
which is associated with the root-mean-square (RMS) radius $R$. In this context, uncertainties 
in the determination of $R$ are sensitive to the size of the spatial extent $L$ in physical units.
For the case of the electric form factor ($G_E$), $R$ corresponds to the charge radius.  

Finally we revisit the recent claim in the literature  
that the charge density of the proton, which has the shape of
an exponential in a classical argument, is widely distributed in space~\cite{Sick:2018fzn}.
The $q^2$ dependence of the Sachs form factors $G_E(q^2)$ and $G_M(q^2)$ 
are roughly described by the dipole shape: $G_D(q^2)=\Lambda^2/(q^2+\Lambda^2)^2$
with a single parameter $\Lambda$ called 
the dipole mass parameter around 0.84 GeV~\cite{Thomas:2001kw}.
The dipole form assumes that the spatial charge distribution is falling exponentially
at large $r$ as $\rho(r)\propto e^{-r\Lambda}$~\cite{Thomas:2001kw}. For the sake of simplicity,
we adopt the spatial charge distribution as $\rho(r)=\rho_0 e^{-r\Lambda}$,
where $\rho_0$ is determined by the normalization condition.
The RMS radius $R$ is defined in terms of the charge density as
%
%
\be
R^2\equiv 4\pi \int^\infty_0 \rho(r) r^4 dr,
\label{eq:RMS}
\ee
which gives a relation of $R=\sqrt{12}/\Lambda$.
We next introduce the truncated RMS radius $R(r_{\rm cut})$,
where the integral~(\ref{eq:RMS})
is stopped at $r=r_{\rm cut}$, and then calculate
the following ratio as a function of $r_{\rm cut}$~\cite{Sick:2018fzn}:
%
%
\be
\frac{R(r_{\rm cut})}{R}
=\sqrt{1-e^{-X}\left[
1+X+\frac{1}{2}X^2+\frac{1}{6}X^3+\frac{1}{24}X^4
\right]},
\ee
where $X=r_{\rm cut}\Lambda=\sqrt{12}r_{\rm cut}/R$.
To get more than 98\% of $R$, the value of $r_{\rm cut}$ must be
greater than $2.75R$, which is remarkably large value~\cite{Sick:2018fzn}.
If we take it seriously, this intuitive argument suggests that the spatial extent   
of $2.75R\times 4$~\footnote{In a $L^3$ spatial box with periodic boundary
conditions, one may consider that a half size of $L$ corresponds to $2 r_{\rm cut}$.}, 
which is roughly 10 fm,  is required for precise determination of the proton
charge radius within a few-percent accuracy on a {\it periodic} hyper-cubic lattice.

This paper is organized as follows: In Sec.~\ref{sec:2}, 
we present a brief introduction of general features of nucleon form factors. 
In Sec.~\ref{sec:3}, we first summarize simulation parameters in 2+1 flavor ensembles generated
by the PACS Collaboration~\cite{Ishikawa:2015rho} and then give some basic results from the nucleon
two-point function. We also describe the lattice method (standard ratio method) for calculating 
the isovector form factors of the nucleon from relevant three-point correlation functions.  
The results of our lattice calculations for the nucleon form factors are presented in Sec.~\ref{sec:4}, which
is divided into three major subsections. We first determine four kinds of nucleon couplings --- the vector coupling
$g_V$, the axial-vector coupling $g_A$, the scalar coupling $g_S$ and the tensor coupling $g_T$ --- which are
directly accessible from the three-point correlations function at zero momentum transfer in Sec.~\ref{sec:4-A}.
Section~\ref{sec:4-B} presents the results of the {\rm isovector} electric
($G_E$) and magnetic ($G_M$) form factors, including a detailed study of 
the momentum transfer dependence and determination of the {\rm isovector} electric and 
magnetic RMS radii and also the {\rm isovector} magnetic moment. 
The last subsection (Sec.~\ref{sec:4-C}) is devoted to a discussion of the results of three form factors 
obtained in the axial-vector and pseudoscalar channels, which are related by
the axial Ward-Takahashi identity in terms of the nucleon matrix elements.
Finally, we close with a brief summary and our conclusions in Sec.~\ref{sec:5}.

%
%
\begin{table*}[ht]
\caption{
Summary of simulation parameters in 2+1 flavor QCD simulations. See Ref.~\cite{Ishikawa:2015rho} for further details.
}\label{Tab:Summary_Sim}
\begin{ruledtabular}
\begin{tabular}{cccccccccc}
$\beta$& $L^3\times T$  & $C_{\rm SW}$ &  $\kappa_{ud}$  & $\kappa_{s}$ 
& $a$ [fm] &  $a^{-1}$ [GeV] &  $(La)^3$ & $m_{\pi}$ [MeV] & $N_{\rm conf}$\cr
\hline
1.82 &$96^3 \times96$ & 1.11 & 0.126117  & 0.124790  
& 0.0846(7) 
& 2.333(18) & $\sim (8.1\; {\rm fm})^3$ & $\approx146$ & 200
\cr
\end{tabular}
\end{ruledtabular}
\end{table*}
%

%
%
\begin{table*}[ht] 
\caption{Choices for nonzero spatial momenta: ${\bm q}=\pi/48\times {\bm n}$. The bottom row denote the 
degeneracy of ${\bm n}$ due to the permutation symmetry between $\pm x$, $\pm y$, $\pm z$ directions. 
\label{tab:q}
}
\begin{ruledtabular}
\begin{tabular}{c|cccccccccc} 
label & Q0 & Q1 & Q2 & Q3 & Q4 & Q5 & Q6 & Q7 & Q8 & Q9 \\ \hline 
${\bm n}$ & (0,0,0) & (1,0,0) & (1,1,0) & (1,1,1) & (2,0,0) & (2,1,0) & (2,1,1) & (2,2,0) & (3,0,0) & (2,2,1)\\
$\vert{\bm n}\vert^2$ & 0 & 1 & 2 & 3 & 4 & 5 & 6 & 8 & 9 & 9\\
degeneracy & 1 & 6 & 12 & 8 & 6 & 24 & 24 & 12 & 6 & 24 \\ 
\end{tabular}
\end{ruledtabular} 
\end{table*} 
%

\section{General features of nucleon form factors}
\label{sec:2}

In general, four form factors appear in the nucleon matrix elements
of the weak current. Here, for example, we consider the matrix element
for neutron beta decay. In this case, the vector and axial-vector currents
are given by $V^{+}_{\alpha}(x)=\bar{u}(x)\gamma_{\alpha}d(x)$ and
$A^{+}_{\alpha}(x)=\bar{u}(x)\gamma_{\alpha}\gamma_5d(x)$,
and then the general form of the matrix element for neutron
beta decay is expressed by both the vector and axial-vector transitions:
%
%
\be
\langle p|V_{\alpha}^{+}(x)+A_{\alpha}^{+}(x)|n\rangle=\bar{u}_p \left(
{\cal O}_{\alpha}^V(q)+{\cal O}_{\alpha}^A(q)
\right) u_n e^{iq\cdot x},
\ee
where $q=P_n -P_p$ is the momentum transfer between the neutron ($n$)
and proton ($p$). The vector ($F_V$) and induced tensor ($F_T$) form
factors are introduced for the vector matrix element,
%
%
\be
{\cal O}_{\alpha}^{V}(q)=\gamma_{\alpha}F_V(q^2)+\sigma_{\alpha \beta}q_{\beta}F_T(q^2)
\ee
as well as the axial vector ($F_A$) and induced pseudoscalar ($F_P$) form factors for the axial-vector matrix element,
%
%
\be
{\cal O}_{\alpha}^{A}(q)=\gamma_{\alpha}\gamma_5 F_A(q^2)+iq_{\alpha}\gamma_5
F_P(q^2).
\ee
These matrix elements are given here in the Euclidean metric convention
(we have defined $\sigma_{\alpha \beta}=\frac{1}{2i}[\gamma_\alpha, \gamma_\beta]$)~\footnote{The sign 
of all form factors is chosen to be positive. 
Remark that our $\gamma_5$ definition $\gamma_5\equiv \gamma_1 \gamma_2 \gamma_3 \gamma_4 =-\gamma_5^{M}$ 
has the opposite sign relative to that in the Minkowski convention 
($\vec{\gamma}^M=i\vec{\gamma}$ and $\gamma_0^{M}=\gamma_4$) adopted in 
the particle data group~\cite{Tanabashi:2018oca}.}.
Thus, $q^2$ denoted in this paper, which stands for the Euclidean four-momentum
squared, corresponds to the spacelike momentum squared as $q_M^2=-q^2<0$ 
in Minkowski space.

The vector part of weak processes is related to the nucleon's electromagnetic 
matrix element through an isospin rotation if heavy-flavor contributions are ignored {\it under the exact isospin symmetry}. A simple exercise in $SU(2)$ Lie algebra leads to the following relation between the vector part of the weak matrix elements of  neutron beta decay and the difference of proton and neutron
electromagnetic matrix elements
%
%
\ben
\langle p|\bar{u}\gamma_{\alpha}d|n\rangle &=&\langle p|\bar{u}\gamma_{\alpha}u-\bar{d}\gamma_{\alpha}d|p\rangle \\
&=&\langle p|j_{\alpha}^{\rm em}|p\rangle - \langle n|j_{\alpha}^{\rm em}|n\rangle
\een
where $j_{\alpha}^{\rm em}=\frac{2}{3}\bar{u}\gamma_{\alpha} u-\frac{1}{3}\bar{d}\gamma_{\alpha} d$. 
Therefore, this relation gives a connection between
the weak vector and induced tensor form factors and the {\it isovector} part
of electromagnetic nucleon form factors
%
%
\ben
F_1^{v}(q^2)&=&F_V(q^2), \\
F_2^{v}(q^2)&=&2M_N F_T(q^2),
\een
where $M_N$ denotes the nucleon mass, which is defined as the average of the neutron
and proton masses. $F_1^{v}$ ($F_2^{v}$) is given by the {\it isovector} combination of the Dirac (Pauli) 
form factors of the proton and neutron as
%
%
\be
F^{v}_{1,2}(q^2)=F_{1,2}^p(q^2)-F_{1,2}^n(q^2),
\ee
where individual form factors $F_{1,2}^N$ ($N=p, n$) are defined by
%
%
\begin{multline}
\langle N(P^\prime)|j_{\alpha}^{\rm em}(x)|N(P)\rangle\cr
{}=\bar{u}_N(P^\prime)\left(
\gamma_{\alpha}F^N_1(q^2)+\sigma_{\alpha \beta}\frac{q_{\beta}}{2M_N}F^N_2(q^2)
\right)u_N(P)e^{iq\cdot x}.
\end{multline}

Experimental data from elastic electron-nucleon scattering is usually described 
in terms of the electric $G_E(q^2)$ and magnetic $G_M(q^2)$ Sachs
form factors, which are related to the Dirac and Pauli form factors:
%
%
\be
G^{N}_E(q^2)=F_1^N(q^2)-\frac{q^2}{4M_N^2}F_2^N(q^2),
\label{Eq:GE}
\ee
%
%
\be
G^{N}_M(q^2)=F_1^N(q^2)+F_2^N(q^2).
\label{Eq:GM}
\ee
Their normalizations at $q^2=0$ are given by the proton's (neutron's) 
electric charge and magnetic moment~\footnote{For the proton, $G_E^p(0)=1$ and $G_M^p(0)=\mu_p=+2.79285$, while $G_E^n(0)=0$ and $G_M^n(0)=\mu_n=-1.91304$ for the neutron.}. Therefore, one finds
%
%
\ben
F_V(0)=F_1^v(0)&=&G_E^v(0)=1,  \\
2M_NF_T(0)=F_2^v(0)&=&G_M^v(0)-1=3.70589,
\een
where $G^v_E$ ($G^v_M$) represents the {\it isovector} combination of the electric (magnetic) 
form factors of the proton and neutron.

Regarding the $q^2$ dependence of the Sachs form factors $G^N_E(q^2)$ 
and $G^N_M(q^2)$, it is known experimentally that the standard 
dipole parametrization $G_D(q^2)=\Lambda^2/(\Lambda^2+q^2)^2$
with $\Lambda^2 = 0.71$ $[({\rm GeV})^2]$ (or $\Lambda=0.84$ [GeV]) 
describes well the magnetic form factors of both the proton
and neutron and also the electric form factor of the proton, at least,
in the low $q^2$ region. In general, if there is no singularity around $q^2=0$ for a given form factor $G(q^2)$, 
the normalized form factor can be expanded in powers of $q^2$. 
%
%
\be
G(q^2)=G(0)\left(1-\frac{1}{6}\langle r^2\rangle q^2+\frac{1}{120}\langle
r^4\rangle q^4 + \cdots\right),
\ee
where the first coefficient determines the mean squared radius $\langle r^2\rangle$, which 
is a typical size in the coordinate space. For the dipole form, the 
root-mean-square (RMS) radius $R$ is given as $R=\sqrt{\langle r^2\rangle}=\frac{\sqrt{12}}{\Lambda}$ 
by the dipole mass parameter $\Lambda$.

For the axial-vector part of weak processes, the axial-vector form factor 
at zero momentum transfer, namely, 
the axial-vector coupling $g_A=F_A(0)$, is precisely determined by measurements
of the beta asymmetry in neutron decay. The value of $g_A=1.2724(23)$ was
quoted in the 2018 PDG~\cite{Tanabashi:2018oca}.
The reason why $g_A$ deviates from the corresponding vector coupling $g_V=F_V(0)=1$ is
that the axial-vector current is strongly affected by the spontaneous chiral symmetry breaking
in the strong interaction~\cite{{Nambu:1961tp},{Nambu:1961fr}}. 
In this sense, this particular quantity allows us to perform
a precision test of lattice QCD in the baryon sector.  

The $q^2$ dependence of the axial-vector form factor $F_A(q^2)$ 
has also been studied in experiments,
where the dipole form $F_A(q^2)=F_A(0)/(q^2+M_A^2)^2$ is a good description for
low and moderate momentum transfer $q^2<1$ $[({\rm GeV})^2]$~\cite{{Bernard:2001rs},{Bodek:2007ym}}. 
Recently, the axial mass parameter $M_A$ 
has received much attention in neutrino oscillation studies~\cite{{Bhattacharya:2011ah},{Bhattacharya:2015mpa}}.

Although the induced pseudoscalar form factor $F_P(q^2)$ is less well known experimentally~\cite{{Choi:1993vt},{Gorringe:2002xx}},
it is theoretically known that two form factors $F_A(q^2)$ and $F_P(q^2)$ in the axial-vector channel are 
not fully independent. It is because the axial Ward-Takahashi identity: $\partial_\alpha A^+_\alpha(x)=2\hat{m}P^{+}(x)$
leads to the generalized Goldberger-Treiman  (GT) relation among three form factors~\cite{{Weisberger:1966ip},{Sasaki:2007gw}}: 
%
%
\begin{align}
2M_NF_A(q^2)-q^2F_P(q^2)=2\hat{m}G_P(q^2),
\label{Eq:GTrelation}
\end{align}
where $\hat m=	m_u=m_d$ is a degenerate up and down quark mass and
the pseudoscalar nucleon form factor $G_P(q^2)$ is defined in 
the pseudoscalar nucleon matrix element
%
%
\begin{align}
\langle p|P^{+}(x)|n\rangle = \bar{u}_p \left( \gamma_5 G_P(q^2)
\right) u_n e^{iq\cdot x}
\label{Eq:PSFF}
\end{align}
with a local pseudoscalar density, $P^{+}(x)\equiv\bar{u}(x)\gamma_{5}d(x)$.
Therefore, the $q^2$ dependences of three form factors, $F_A(q^2)$, $F_P(q^2)$ and $G_P(q^2)$
are constrained by Eq.~(\ref{Eq:GTrelation}) as a consequence of the axial Ward-Takahashi identity.
Therefore, the three form factors, $F_A(q^2)$, $F_P(q^2)$ and $G_P(q^2)$
are very important for verifying the axial Ward-Takahashi identity in terms of the nucleon
matrix elements.

The latest lattice calculations of the nucleon structure have been carried out 
with increasing accuracy~\cite{{Chang:2018uxx},
{Bhattacharya:2016zcn},{Alexandrou:2017hac},{Gupta:2018qil},{Green:2014xba},{Alexandrou:2017ypw},{Yamazaki:2008py},{Yamazaki:2009zq},{Yoon:2016jzj},{Bhattacharya:2013ehc},{Capitani:2015sba},{Alexandrou:2013joa},{Ohta:2017gzg}}.
It seems that there is still a gap between experimentally known values and
the lattice results, especially for the electric-magnetic nucleon radii and the magnetic moment.   
Our preliminary results computed near the physical pion mass in very large volume 
show agreement with experimental results~\cite{{Yamazaki:2015vjn},{Kuramashi:2016lql},{Tsukamoto:2017fnm}}. 
In this paper, we present an update of our previous studies~\cite{{Yamazaki:2015vjn},{Kuramashi:2016lql},{Tsukamoto:2017fnm}}, 
including the axial-vector, induced pseudoscalar and pseudoscalar form factors.

%
%
\begin{table}[hb] 
\caption{
Fitted energies of the nucleon state with the ten lowest momenta obtained from the smear-local case 
of the nucleon two-point function. Results for the nucleon energies $E_N({\bm n}^2)$ with nonzero momenta are averaged over all possible permutations of ${\bm n}=(n_x, n_y, n_z)$, including both positive and negative directions.
The values of the corresponding momentum squared $q^2=2M_N(E_N-M_N)$ are also tabulated.
\label{tab:mass}
}
\begin{ruledtabular}
\begin{tabular}{cccc} 
label & $aE_N({\bm n})$ (Smear-Local) & fit range & $q^2$ [$({\rm GeV})^2$] \\ \hline 
Q0 & 0.4107(12)& [8:15] & 0\cr
Q1 & 0.4161(12)& [8:15] & 0.024(1)\cr
Q2 & 0.4215(12)& [8:15] & 0.048(2)\cr
Q3 & 0.4268(13)& [8:15] & 0.072(2)\cr
Q4 & 0.4320(13)& [8:15] & 0.095(3)\cr
Q5 & 0.4373(14)& [8:15] & 0.119(4)\cr
Q6 & 0.4427(15)& [8:15] & 0.143(5)\cr
Q7 & 0.4531(18)& [8:15] & 0.189(7)\cr
Q8 & 0.4575(21)& [8:15] & 0.209(8)\cr
Q9 & 0.4588(20)& [8:15] & 0.215(8)\cr
\end{tabular}
\end{ruledtabular} 
\end{table} 
%

\section{Simulation details}
\label{sec:3}

We use the 2+1 flavor QCD gauge configurations generated with 
the 6-APE stout-smeared $O(a)$-improved Wilson-clover quark action 
and the Iwasaki gauge action~\cite{Iwasaki:2011np} on a lattice 
$L^3\times T=96^3\times 96$ at $\beta=1.82$, which corresponds to a lattice cutoff of 
$a^{-1}\approx 2.3$ GeV ($a \approx 0.085$ fm)~\cite{Ishikawa:2015rho}.
Periodic boundary conditions are used for the gauge and quark fields in all four directions. 
The stout smearing parameter is set to $\rho = 0.1$~\cite{Morningstar:2003gk}.
The improvement coefficient, $c_{\rm SW}=1.11$, is determined nonperturbatively 
using the Schr{\"o}dinger functional (SF) scheme~\cite{Taniguchi:2012kk}. 
The improved factor $c_A$ for the axial-vector current becomes very small
at the nonperturbative value of $c_{\rm SW}$ and is
consistent with zero within the statistical error~\cite{Taniguchi:2012kk}. 
Therefore, we do not consider ${\cal O}(a)$ improvement of the quark bilinear current.
The hopping parameters for the light sea quarks $(\kappa_{ud}, \kappa_{s})$ = (0.126117, 0.124790) are 
chosen to be near the physical point. For the first time, a simulated pion mass reaches down to 
$m_\pi \approx 146$ MeV
in a large spatial volume of $(8.1~{\rm fm})^3$ in 2+1 flavor QCD.
A brief summary of our simulation parameters with 2+1 flavor QCD simulations is given in Table~\ref{Tab:Summary_Sim}.

The degenerated up-down quarks are simulated with the DDHMC algorithm~\cite{{Luscher:2003vf},{Luscher:2005rx}}
using the even-odd preconditioning and the twofold mass preconditioning~\cite{{Hasenbusch:2001ne},{Hasenbusch:2002ai}}.
The strange quark is simulated with the UVPHMC algorithm~\cite{{deForcrand:1996ck},{Frezzotti:1997ym},{Frezzotti:1998eu},{Frezzotti:1998yp},{Aoki:2001pt},{Ishikawa:2006pb}}
where the action is UV filtered~\cite{{deForcrand:1998sv},{Alexandrou:1999ii}}
after the even-odd preconditioning without domain decomposition.
The total number of gauge configurations reaches 200 which corresponds to 2000 trajectories after thermalization.
Each measurement is separated by 10 trajectories. The results for the hadron spectrum and other physical quantities 
were already presented in Ref.~\cite{Ishikawa:2015rho}. We use the jackknife method with a bin size of 50 trajectories 
to estimate the statistical errors. 
Our preliminary results of the nucleon vector form factors with less number of measurements 
were first reported in Refs.~\cite{{Yamazaki:2015vjn},{Kuramashi:2016lql}}.

\subsection{Nucleon two-point functions}
\label{sec:3-A}

Let us first examine the nucleon rest mass and the dispersion relation, which 
are obtained from the nucleon two-point functions. 
In order to compute nucleon energies or matrix elements, we define the nucleon (proton) operator as
%
%
\begin{multline}
N_X(t,{\bm p})\\
{}=\sum_{{\bm x},{\bm x}_{1},{\bm x}_{2},{\bm x}_{3}}e^{-i{\bm p}\cdot{\bm x}}\varepsilon_{abc}[u_a^T({\bm x}_1,t)C\gamma_5d_b({\bm x}_2, t)]u_c({\bm x}_3, t)\cr
{}\times\phi({\bm x}_1-{\bm x})\phi({\bm x}_2-{\bm x})\phi({\bm x}_3-{\bm x}),
\label{eq:nuc_ope}
\end{multline}
where the superscript $T$ denotes a transposition and $C$ is the charge-conjugation matrix defined as $C=\gamma_4\gamma_2$.
The indices $abc$ and $ud$ label color and flavor, respectively.
The subscript $X$ of the nucleon operator specifies the type of the smearing for the quark propagators. 
In this study, we use 
two types of smearing function $\phi$: the local function as $\phi({\bm x}_i-{\bm x})=\delta({\bm x}_i -{\bm x})$ (denoted as $X=L$) and
the exponential smearing function: $\phi({\bm x}_i-{\bm x})=A\exp[-B|{\bm x}_i-{\bm x}|]$ 
with $A=1.2$ and $B=0.11$ (denoted as $X=S$). 
For simplicity, ${\bm x}_1={\bm x}_2={\bm x}_3$ is chosen.

We then construct two types of two-point functions with the exponential smearing source (the source-time location denoted as $t_{\rm src}$) as
%
%
\begin{multline}
C_{XS}(t_{\rm sink}-t_{\rm src}, {\bm p})\\
{}=\frac{1}{4}{\rm Tr}\left\{{\cal P_+}\langle N_X(t_{\rm sink},{\bm p}){\bar N}_S(t_{\rm src},-{\bm p})\rangle
\right\}, \label{eq:2pt}
\end{multline}
where $X=L$ (local) or $S$ (smear) stands for the type of smearing at the sink operator (with 
the sink-time location denoted as $t_{\rm sink}$).
A projection operator ${\cal P}_+=\frac{1+\gamma_4}{2}$ can eliminate contributions from the opposite-parity
state for $|{\bm p}|=0$~\cite{{Sasaki:2001nf}, {Sasaki:2005ug}}.
In this study, for nonzero spatial momentum, we use the nine lowest momenta: ${\bm p}=2\pi/L\times {\bm n}$ 
with the vector of integers ${\bm n}\in Z^3$. All choices of ${\bm n}$ are listed in Table~\ref{tab:q}.

%
%
\begin{figure}[ht!]
\begin{minipage}[t]{0.50\textwidth}
\includegraphics[height=6.0cm,keepaspectratio,clip]{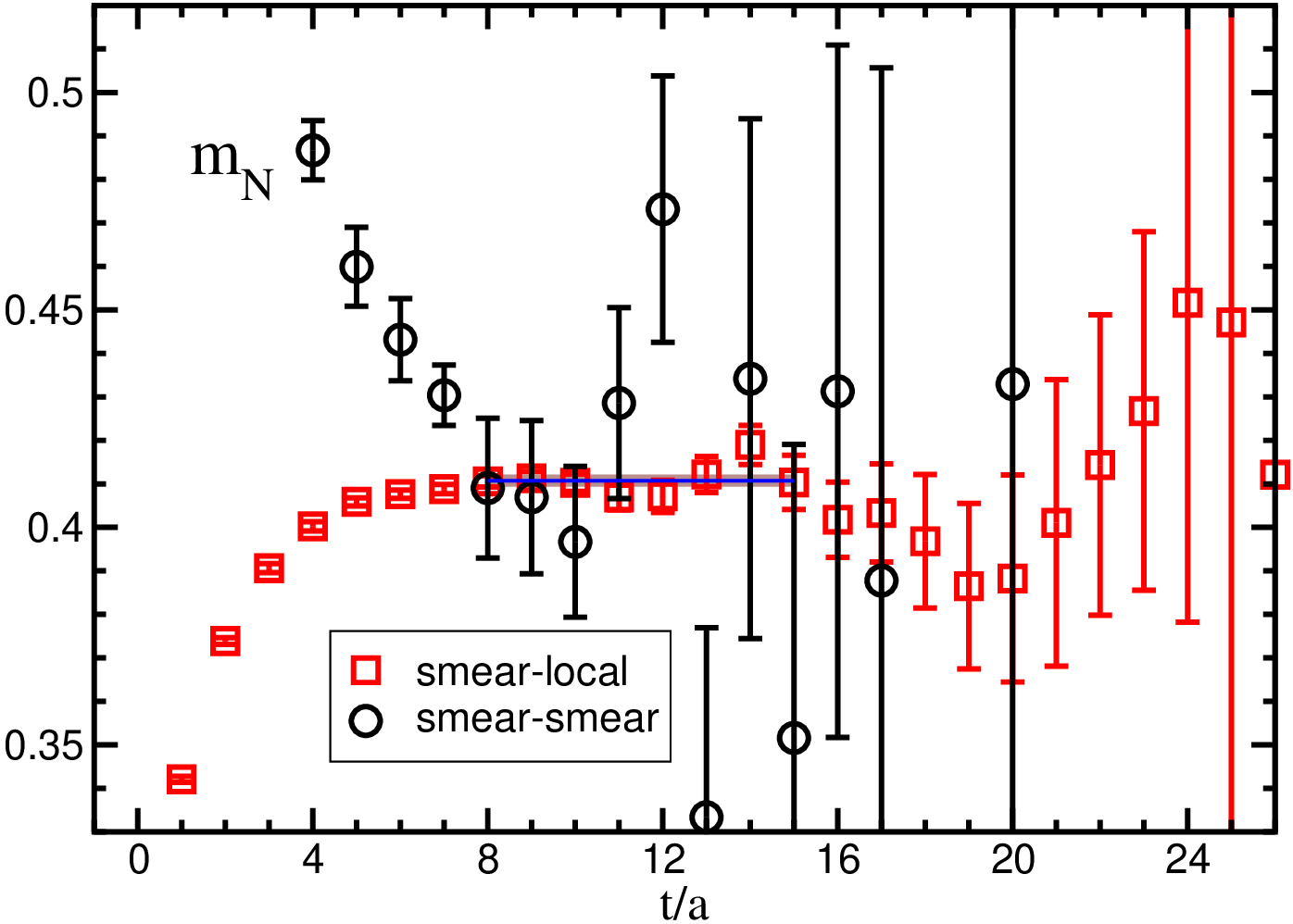}
\caption{Effective mass for the nucleon from the smear-local (squared symbols) and smear-smear (circle symbols)
cases of the nucleon two-point functions.}
\label{fig:m_N}
\end{minipage}
\end{figure}
%

\subsection{Nucleon spectra and dispersion relation}
\label{sec:3-B}

In Fig.~\ref{fig:m_N} we plot the nucleon effective mass with $|{\bm p}|=0$
for two cases: smear-local denotes the nucleon two-point function with the 
smeared source and the local sink operators 
and smear-smear denotes the smeared source and the smeared sink ones. 
The values of the smearing parameters ($A=1.2$ and $B=0.11$) are chosen to optimize 
the effective mass plateau for the smear-local case.
We thus observe that the smear-local case shows a good plateau for $t/a \ge 6$
with our choice of smearing parameters. 
On the other hand, the signal becomes noisier for the smear-smear case since
the extra smearing in general causes statistical noise.

We also measure the nucleon energies $E_N({\bm p})$ from the smear-local 
case of the nucleon two-point function for nine different cases with nonzero 
spatial momenta. All fit results for $E_N({\bm p})$ obtained from the 
single-exponential fit, where we take into account the full covariance matrix of the data
during the fitting process, are tabulated in Table~\ref{tab:mass}. 

\begin{figure}[t]
\begin{minipage}[t]{0.50\textwidth}
\includegraphics[height=6.0cm,keepaspectratio,clip]{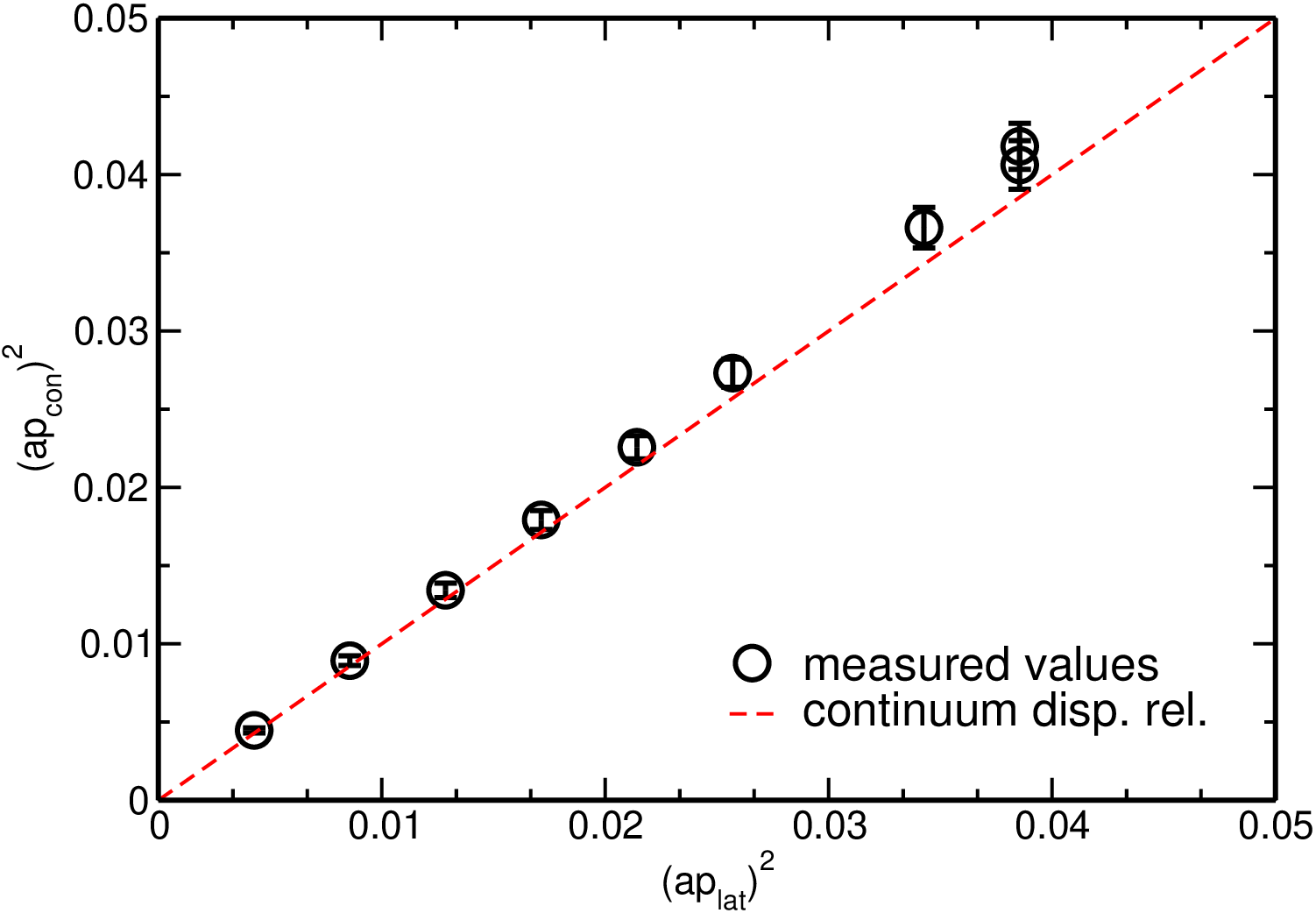}
\caption{Check of the dispersion relation for the nucleon. The variables $p_{\rm con}^2$ and $p_{\rm lat}^2$ 
appearing on the x-axis and y-axis are defined in the text. For comparison, the relativistic continuum dispersion relation is
denoted as a dashed line.}
\label{fig:disp_N}
\end{minipage}
\end{figure}

Figure~\ref{fig:disp_N} shows a check of the dispersion relation for the nucleon. 
The vertical axis shows the momentum squared defined through the relativistic continuum 
dispersion relation as $p_{\rm con}^2=E^2_N({\bm p})- M_N^2$, while the horizontal axis shows
the momentum squared given by the lattice momentum $p^2_{\rm lat}=(2\pi/L)^2 \times |{\bm n}|^2$ 
in lattice units.
As can be seen in Fig.~\ref{fig:disp_N}, the relativistic continuum dispersion relation 
is well satisfied up to $|{\bm n}|^2=9$.

%
%
%

\subsection{Three-point correlation functions for nucleon form factors}
\label{sec:3-C}

The nucleon form factors are extracted from the three-point correlation functions consisting of the nucleon source and sink operators with a given local current
$J^{O}_\alpha$ ($O=S, P, V, A$ and $T$) located at the time slice $t$: 
%
%
\begin{multline}
C_{O,\alpha}^{{\cal P}_k}(t,{\bm p}^\prime,{\bm p})\cr
{}=\frac{1}{4}{\rm Tr}\left\{{\cal P}_k\langle N(t_{\rm sink},{\bm p}^\prime)
J^O_\alpha(t,{\bm q}){\bar N}(t_{\rm src},-{\bm p})\rangle
\right\}, \label{eq:3pt_J}
\end{multline}
where ${\cal P}_k$ denotes an appropriate projection operator to extract the form factors and ${\bm q}={\bm p}-{\bm p}^\prime$ represents the three-dimensional momentum transfer. The local current is given by $J^{O}_\alpha=\bar{u}(\Gamma^O)_{\alpha} d$ where $\Gamma^O$ is a Dirac matrix appropriate for the channel $O$.

We then calculate the following ratio constructed from the three-point correlation function $C_{O,\alpha}^{\cal P}$ 
with the nucleon two-point functions $C_{XS}$:
%
%
\begin{widetext}
\be
{\cal R}^{k}_{O,\alpha}(t,{\bm p}^\prime,{\bm p})=
\frac{C_{O,\alpha}^{{\cal P}_k}(t,{\bm p}^\prime,{\bm p})}{C_{SS}(t_{\rm sink}-t_{\rm src}, {\bm p}^\prime)}
\sqrt{
\frac{C_{LS}(t_{\rm sink}-t, {\bm p})C_{SS}(t-t_{\rm src}, {\bm p}^\prime)C_{LS}(t_{\rm sink}-t_{\rm src}, {\bm p}^\prime)}
{C_{LS}(t_{\rm sink}-t, {\bm p}^\prime)C_{SS}(t-t_{\rm src}, {\bm p})C_{LS}(t_{\rm sink}-t_{\rm src}, {\bm p})}
},
\label{Eq:RatioQ}
\ee
\end{widetext}
which is a function of the current operator insertion time $t$ at the given values of 
momenta ${\bm p}$ and ${\bm p}^\prime$ for the initial and final states of the nucleon.

We consider only the rest frame of the final state with ${\bm p}^\prime={\bm 0}$, which leads to $q^2=2M_N(E_N({\bm q})-M_N)$ for the squared four-momentum transfer. 
Hereafter, the nucleon energy $E_N({\bm q})$ is simply abbreviated as $E_N$.
In this kinematics, ${\cal R}_{O,\alpha}(t,{\bm p}^\prime,{\bm p})$ is represented by a simple notation
${\cal R}_{O,\alpha}(t,{\bm q})$. We calculate only the connected diagrams to concentrate 
on the {\it isovector} part of the nucleon form factors. 
The current insertion points $t$ are moved between the nucleon source and sink operators, both of 
which are exponentially smeared in the Coulomb gauge, separated by 15 time slices.
In the physical units, the source-sink separation of $t_{\rm sep}/a=15$ (denoted as $t_{\rm sep}=t_{\rm sink}-t_{\rm src}$)
corresponds to 1.27 fm. In previous calculations~\cite{{Bhattacharya:2013ehc},{Green:2014xba}, {Capitani:2015sba}, {Alexandrou:2017ypw},{Yamazaki:2009zq},{Ohta:2017gzg}}, the maximum values of the source-sink separation were
typically as large as 1.3-1.4 fm, where the excited-state effect is marginally insignificant.

The three-point correlation functions are calculated using the sequential source method with a fixed 
source location~\cite{{Martinelli:1988rr},{Sasaki:2003jh}}. 
This method requires the sequential quark propagator for each choice of a projection operator ${\cal P}$ 
regardless of the types of current $J^{O}_\alpha$. 
To minimize the numerical cost, we consider two types of the projection operators 
${\cal P}_t={\cal P}_{+}\gamma_4$ and ${\cal P}_{5z}={\cal P}_{+}\gamma_5\gamma_3$ in this study. 

The ratio (\ref{Eq:RatioQ}) with appropriate combinations of the projection operator ${\cal P}_k$ ($k=t, 5z$) and
the $\alpha$ component of the current $j_\alpha^{O}$ gives the asymptotic values~\cite{Sasaki:2007gw}:
%
%
\be
{\cal R}^{t}_{V,4}(t,{\bm q})\rightarrow \sqrt{\frac{E_N+M_N}{2E_N}}G^v_E(q^2)
\label{Eq:GE}
\ee
and
%
%
\be
{\cal R}^{5z}_{V,i}(t,{\bm q})\rightarrow \frac{-i\varepsilon_{ij3}q_j}{\sqrt{2E_N(E_N+M_N)}}G^v_M(q^2)
\label{Eq:GM}
\ee
%
for the vector current ($O=V$) in the limit when the Euclidean time separation between 
all operators is large, $t_{\rm sink}\gg t \gg t_{\rm src}$ with fixed $t_{\rm src}$ and
$t_{\rm sink}$. Similarly, we get
%
%
\begin{multline}
{\cal R}^{5z}_{A,i}(t,{\bm q}) \cr
{}\rightarrow\sqrt{\frac{E_N+M_N}{2E_N}}\left[F_A(q^2)\delta_{i3}-\frac{q_iq_3}{E_N+M_N}F_P(q^2)\right]
\label{Eq:FA_FP}
\end{multline}
for the axial-vector current ($O=A$), and
%
%
\be
{\cal R}^{5z}_{P}(t,{\bm q})\rightarrow \frac{iq_3}{\sqrt{2E_N(E_N+M_N)}}G_P(q^2)
\label{Eq:GP}
\ee
for the pseudoscalar ($O=P$), respectively.
Here, we recall that the lattice operators receive finite renormalizations relative to their continuum
counterparts in general. Thus, all of the above form factors $G^v_E$, $G^v_M$, $F_A$, $F_P$ 
and $G_P$ are not renormalized yet. Their renormalized values require the renormalization 
factor $Z_O$ ($O=V, A, P$), which is defined through the renormalization of 
the quark bilinear currents
%
%
\be
[\bar{u}\Gamma^O d]^{\rm ren}=Z_{O}[\bar{u}\Gamma^O d]^{\rm lattice}.
\ee
The renormalization factors $Z_V$, $Z_A$ and $Z_P$ are independently obtained using
the SF scheme at the vanishing quark mass defined 
by the partially conserved axial-vector current (PCAC) relation~\cite{Ishikawa:2015fzw}.
Hereafter, the {\it isovector} electric $G^v_E$ and magnetic $G^v_M$ form factors are simply 
abbreviated as $G_E$ and $G_M$.

%
%
\begin{figure*}[ht!]
\begin{tabular}{ll}
\begin{minipage}[t]{0.50\textwidth}
\includegraphics[height=6.0cm,keepaspectratio,clip]{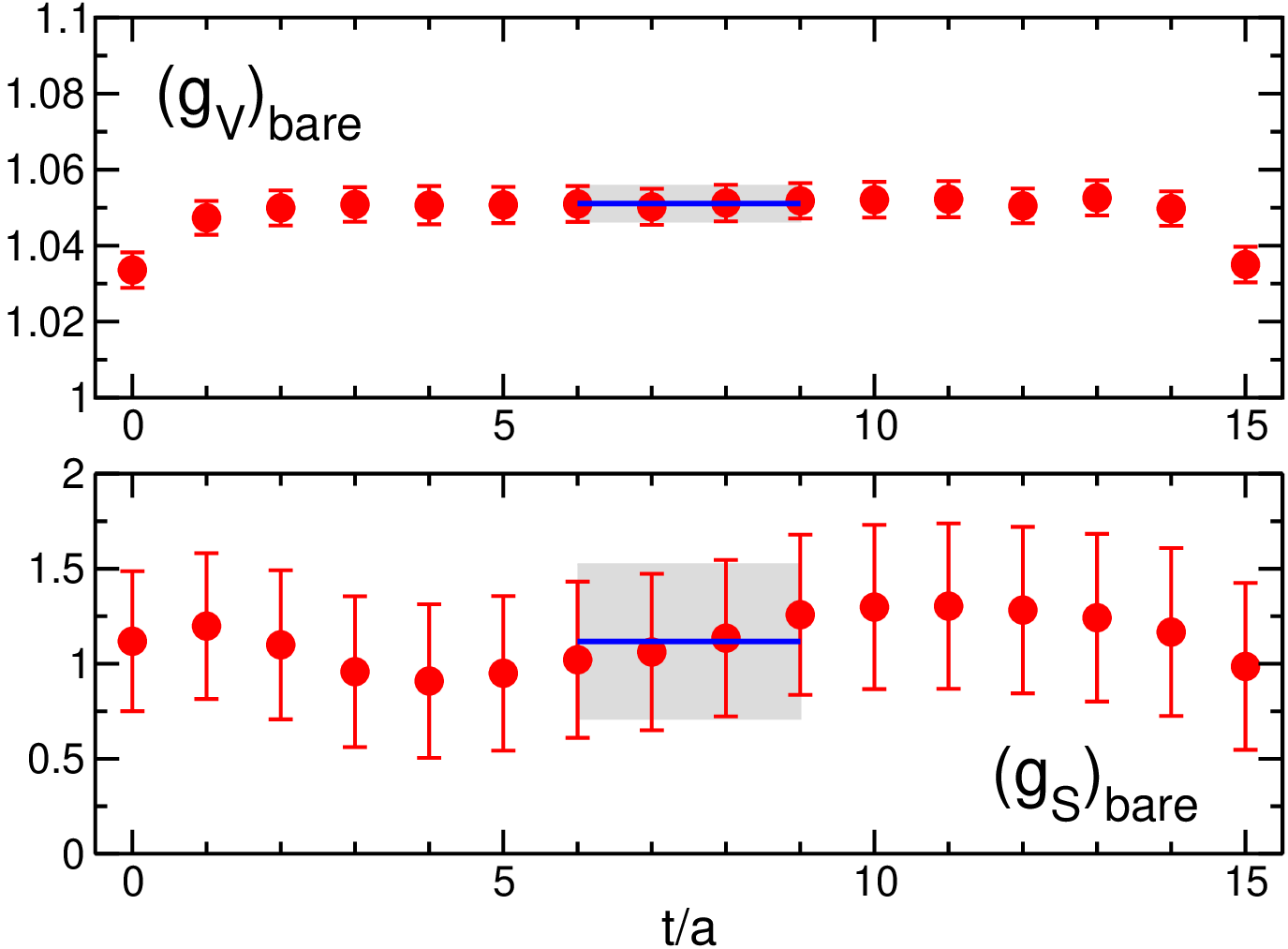}
\end{minipage} &
\begin{minipage}[t]{0.50\textwidth}
\includegraphics[height=6.0cm,keepaspectratio,clip]{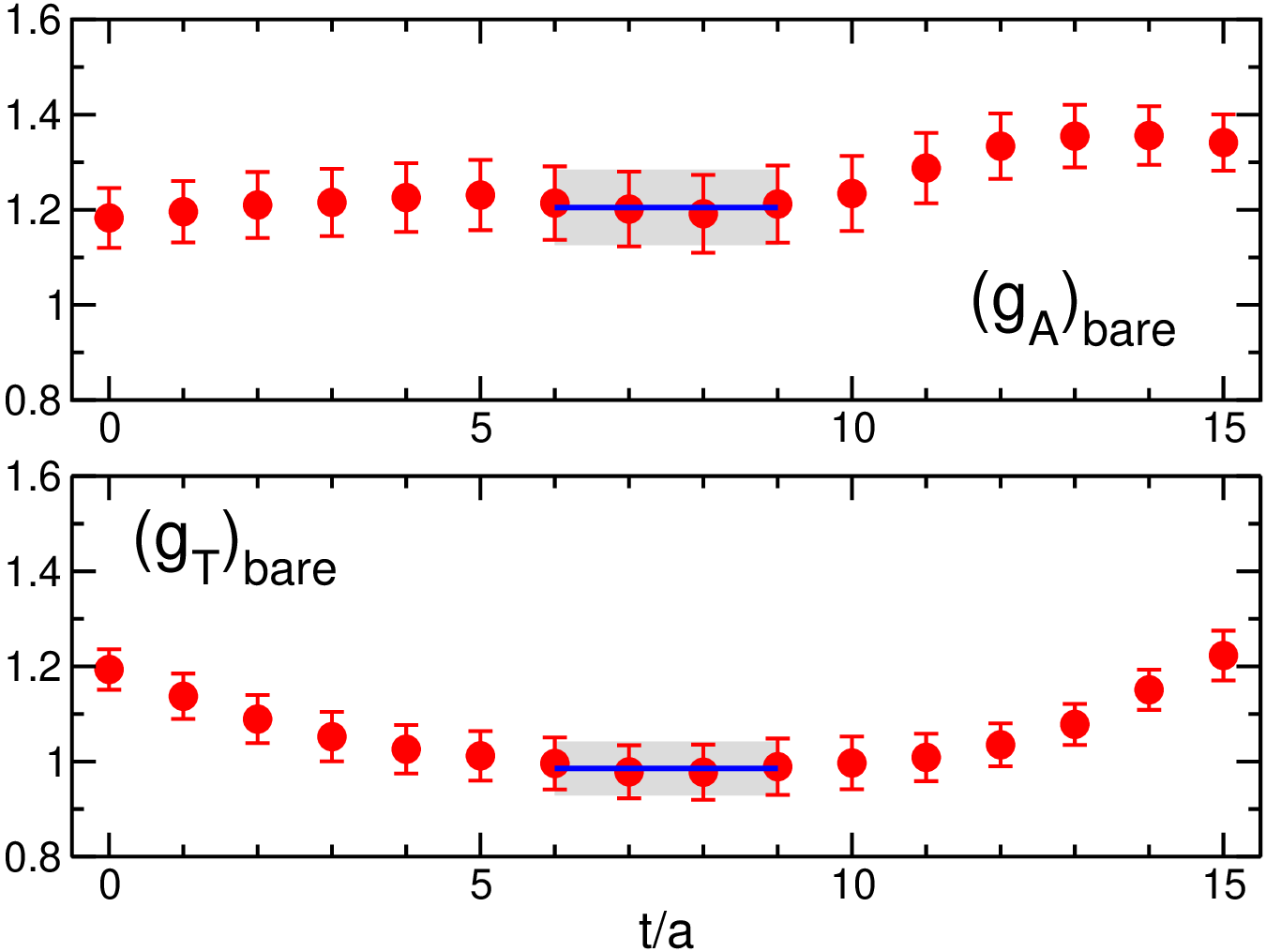}
\end{minipage} 
\end{tabular}
\caption{Bare coupling $g_{O}$ for $O=V$ (left upper), $A$ (right upper), $S$ (left lower) and $T$ (right lower) 
as a function of the current insertion time slice.
}
\label{fig:bare_couplings}
\end{figure*}
%

\section{Results of nucleon form factors}
\label{sec:4}

All the results presented in this work are obtained with 200 configurations. To reduce the
statistical uncertainties, we perform 64 measurements of the two-point and three-point correlation functions on each configuration.
For 64 measurements, we use eight different source locations, four choices for a temporal direction
on a $96^4$ lattice, and two choices of both forward and backward directions in time.
In the analysis, all 64 sets of three-point correlation functions and nucleon two-point functions
are folded together to create the single-correlation functions, respectively. 
It can reduce possible autocorrelation among measurements. 
The statistical errors are evaluated using the jackknife analysis with a bin size of five configurations. 

We employ nine cases of spacial momentum transfer ${\bm q}=\pi/48\times{\bm n}$ with $\vert {\bm n}\vert^2\le 9$
as described in Table~\ref{tab:q}. The minimum momentum transfer is about 152 MeV, which is as small as the pion mass. 
A difference between the results for Q8 and Q9 cases could probe the possible lattice discretization error.

%
%
\begin{table}[ht]
\caption{
Summary of bare couplings for the vector, axial vector, scalar and tensor.
}\label{Tab:Summary_Charges}
\begin{ruledtabular}
\begin{tabular}{cccc}
\hline
$g_V^{\rm bare}$ & $g_A^{\rm bare}$ & $g_S^{\rm bare}$ & $g_T^{\rm bare}$ \cr
\hline
1.0511(47) & 1.205(78) & 1.117(407) & 0.985(55)
\cr
\hline
\end{tabular}
\end{ruledtabular}
\end{table}
%

\subsection{Nucleon three-point function with zero momentum transfer}
\label{sec:4-A}

For zero momentum transfer ${\bm q}={\bm 0}$, the ratio (\ref{Eq:RatioQ}) becomes
%
%
\be
{\cal R}^k_{O,\alpha}(t,{\bm 0})=\frac{C_{O,\alpha}^{{\cal P}_k}(t,{\bm 0})}{C_{SS}(t_{\rm sink}-t_{\rm src}, {\bm 0})},
\label{Eq:Ratio0}
\ee
which vanishes unless $\Gamma^O=1(S), \gamma_4(V), \gamma_i\gamma_5(A)$, and $\sigma_{ij}(T)$ 
with $i,j=1,2,3$~\cite{Sasaki:2003jh}. 
The nonvanishing ratio gives an asymptotic plateau corresponding to the bare value of the coupling $g_O$ relevant for the $O$ channel in the middle region between the source and sink points, when the condition $t_{\rm sink}\gg t \gg t_{\rm src}$ is satisfied.

With our choice of the projection operators $k=t, 5z$, we can determine four different couplings:  
the vector coupling $g_V$, the axial-vector coupling $g_A$, the scalar coupling $g_S$ and the tensor 
coupling $g_T$, from the four ratios,
%
%
\ben
{\cal R}^t_{V,4}(t,{\bm 0})&\rightarrow& G_E(0)=F_V(0)=g_V ,\\
{\cal R}^{5z}_{A,3}(t,{\bm 0})&\rightarrow& F_A(0)=g_A  ,\\
{\cal R}^t_{S}(t,{\bm 0})&\rightarrow& g_S ,\\
{\cal R}^{5z}_{T,12}(t,{\bm 0})&\rightarrow& g_T , 
\een
whose values are not yet renormalized.

In Fig.~\ref{fig:bare_couplings}, we plot the above four ratios 
as a function of the current insertion time slice $t$ for
the vector (left upper panel), axial-vector (right upper panel),
scalar (left lower panel) and tensor (right lower panel) channels. 
The best plateau is clearly observed in the vector channel since
the vector coupling $g_V$ corresponds to the conserved charge
associated to the isospin symmetry. The exact symmetry would
suppress the statistical fluctuations and hinder parts of the 
excited-state contamination. Therefore, a very long plateau 
indeed appears in the ratio of ${\cal R}^t_{V,4}(t,{\bm 0})$. 
It is also worth recalling that the time-reversal average was implemented
in our averaging procedure on each configuration with multiple measurements, which include
both forward and backward propagations in time, as described previously.

In the vector channel (left upper panel), $t$ dependence
of ${\cal R}^t_{V,4}(t,{\bm 0})$ is symmetric between the source ($t/a=0$) 
and sink ($t/a=15$) points. 
Although this symmetric behavior with respect to $t$ is hindered by larger statistical fluctuations
in both the scalar (left lower panel) and the axial-vector (right upper panel) channels,
its behavior is clearly seen in the tensor channel (right lower panel) with relatively small errors. 
In the case of the tensor, good plateau is observed only in the middle region 
between the source and sink points. Therefore, we choose the range $6\le t/a \le 9$,
where an asymptotic plateau behavior appears in the ratio of ${\cal R}^k_{O,\alpha}(t,{\bm 0})$
with our choice of source-sink separation.

In each plot, a solid line indicates the average value over range $6\le t/a \le 9$ and 
a shaded band displays one standard deviation estimated using the jackknife method.
The obtained values of the bare couplings $g_{O}$, which are not yet
renormalized, are summarized in Table~\ref{Tab:Summary_Charges}.

We next evaluate the renormalization factor of the local vector current $Z_V$ through the property
of $g^{\rm ren}_V=1$ for the renormalized value of the vector charge $g^{\rm ren}_V$ under the exact 
isospin symmetry. Therefore, the value of $Z_V$ can be evaluated using $1/g_V^{\rm bare}$ since $g_V^{\rm ren}=Z_V g_V^{\rm bare}$.
 
\begin{figure}[ht]
\begin{minipage}[t]{0.50\textwidth}
\includegraphics[height=6.0cm,keepaspectratio,clip]{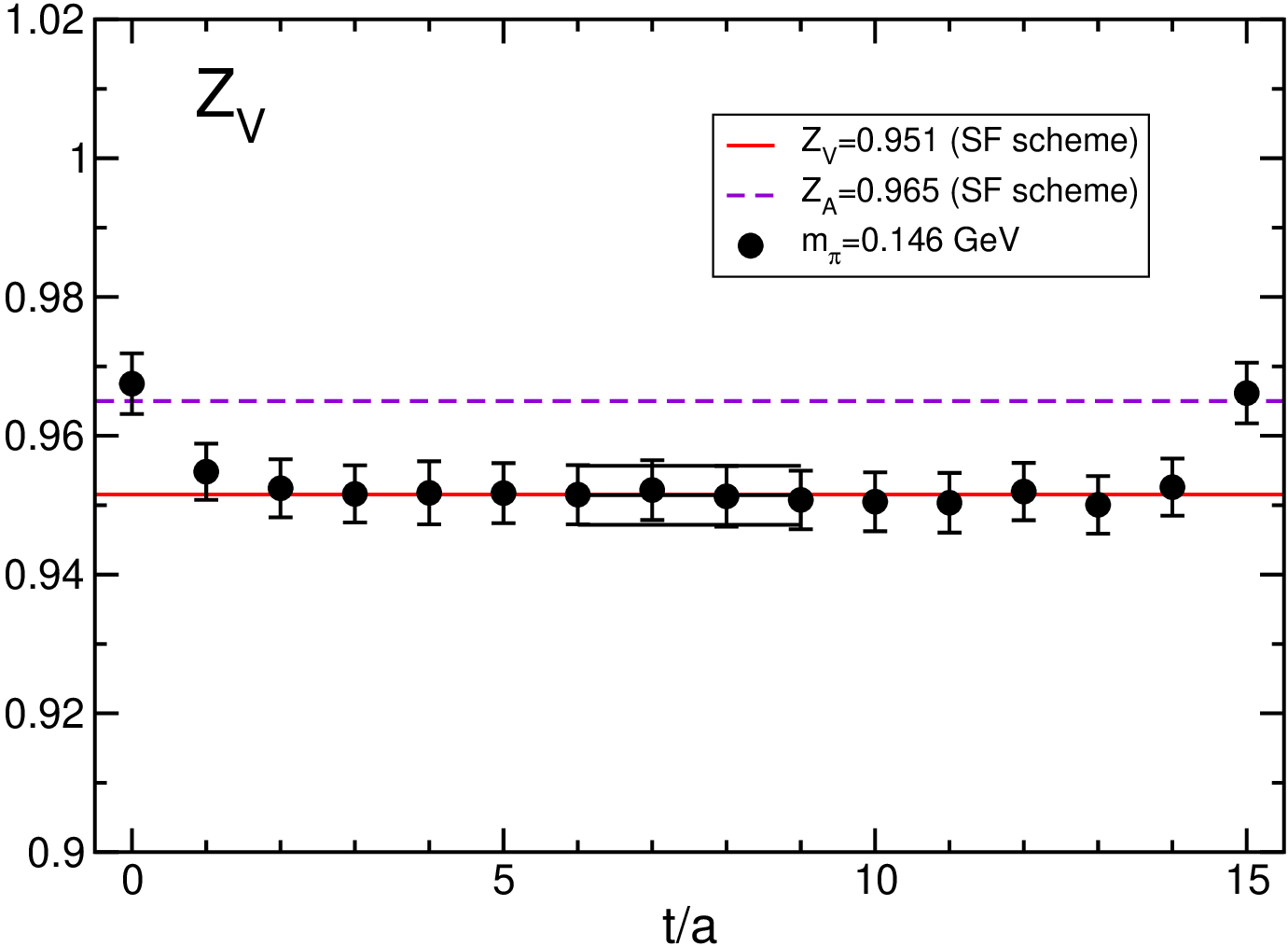}
\caption{Renormalization factor of the local vector current determined 
by $Z_V=1/g_V^{\rm bare}$.}
\label{fig:zv}
\end{minipage}
\end{figure}

Figure~\ref{fig:zv} plots the result of $Z_V$, which shows a good plateau in the time range $2\le t/a\le 14$. The 
constant-fit result with one-standard-error band denoted by three solid lines shows good consistency with the value of $Z_V$ (red line) obtained using the SF scheme at the vanishing PCAC quark mass~\cite{Ishikawa:2015fzw}. This consistency may suggest that the excited-state contamination in three-point functions is not 
serious with our choice of source-sink separation.
We also draw the value of $Z_A$ with the SF scheme for comparison in the same figure. The difference 
between $Z_V$ and $Z_A$ is 1.5\%, which indicates a fairly small chiral symmetry breaking effect in our simulations.

%
%
%

%
%
\begin{figure}[ht]
\begin{minipage}[t]{0.50\textwidth}
\includegraphics[height=6.0cm,keepaspectratio,clip]{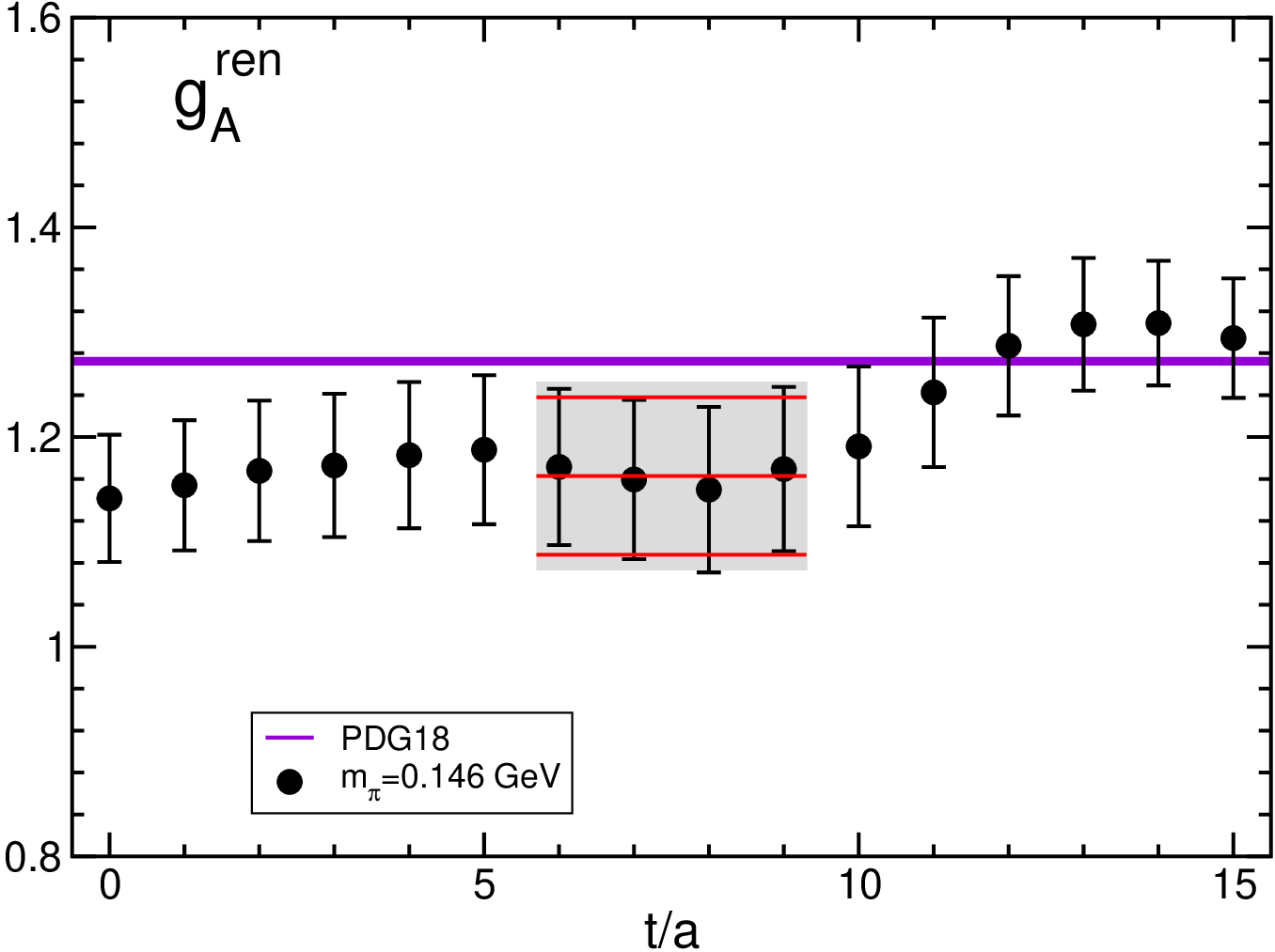}
\caption{Renormalized axial charge with $Z_A=0.9650(68)(95)$ 
in the SF scheme.}
\label{fig:ga}
\end{minipage}
\end{figure}
%

%
%
\begin{figure*}[t]
\begin{tabular}{ll}
\begin{minipage}[t]{0.50\textwidth}
\includegraphics[height=6.0cm,keepaspectratio,clip]{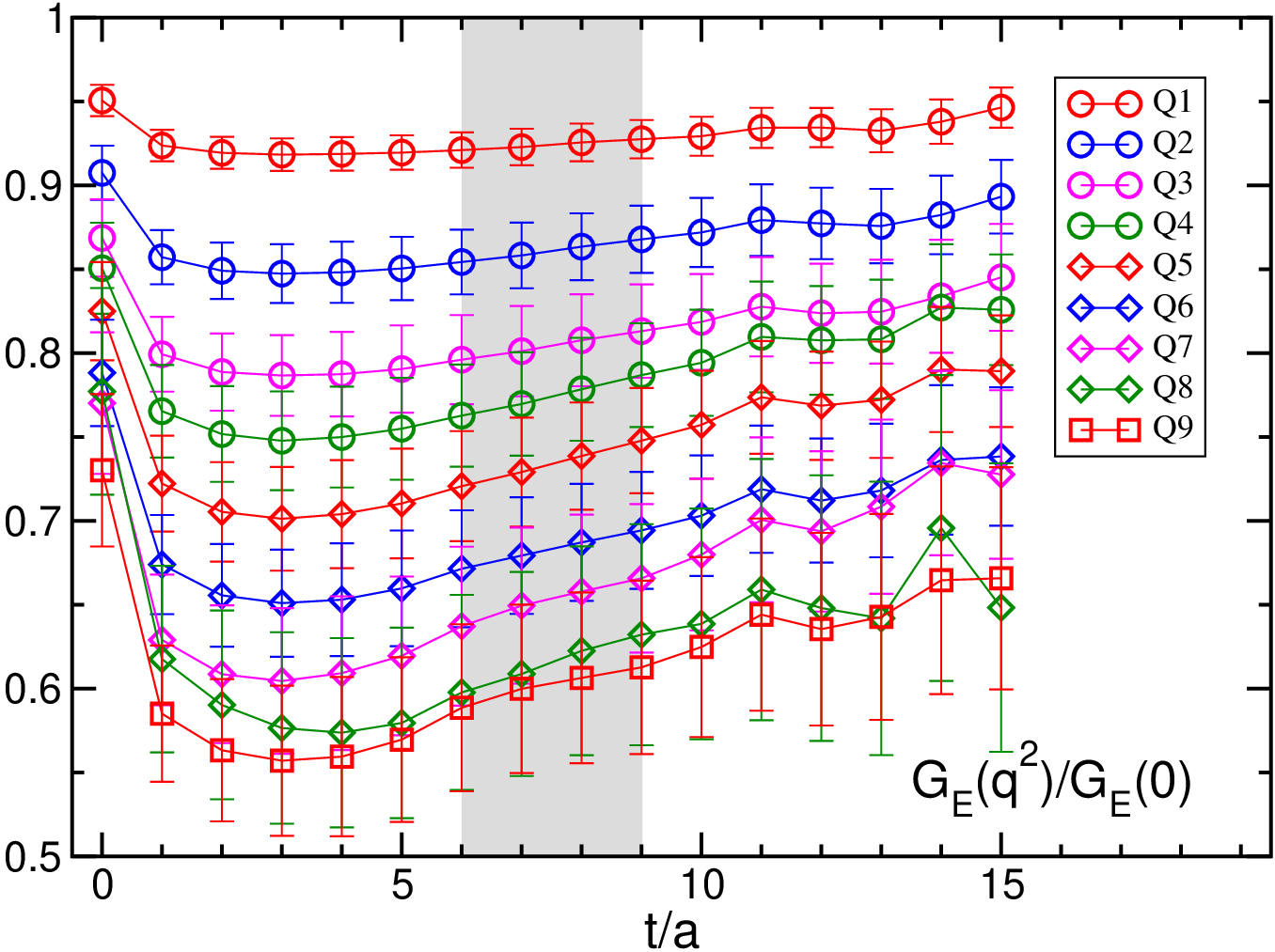}
\end{minipage}
&
\begin{minipage}[t]{0.50\textwidth}
\includegraphics[height=6.0cm,keepaspectratio,clip]{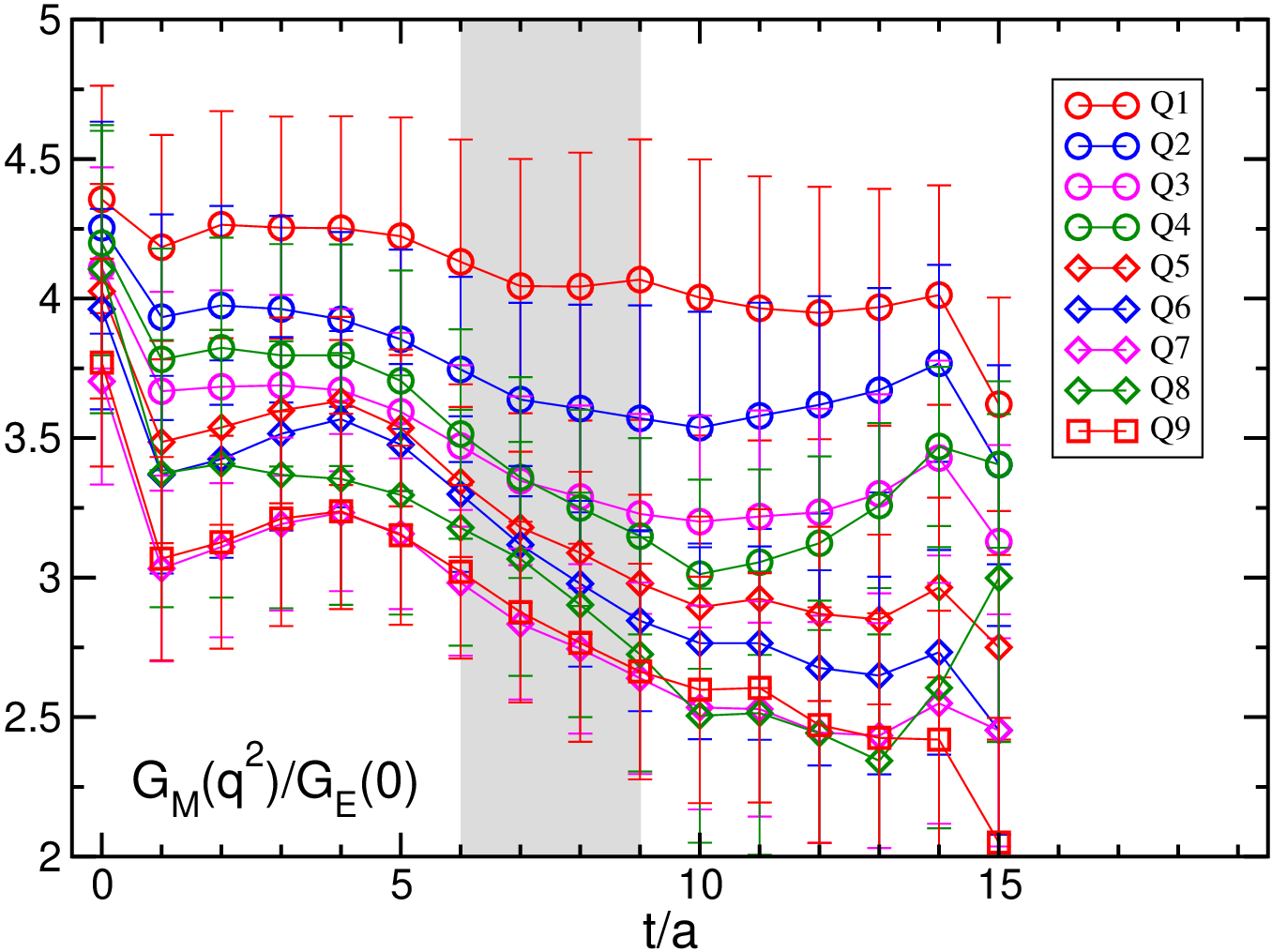}
\end{minipage}
\end{tabular}
\caption{Ratios of the three- and two-point functions (\ref{Eq:GE}) and (\ref{Eq:GM})
for the {\it isovector} electric form factor $G_E$ (left) and magnetic form factor $G_M$ (right)
at the nine lowest nonzero momentum transfers after extracting the relevant kinematical factors.
The gray shaded area ($6\le t/a \le 9$) in each panel shows the region where 
the values of the corresponding form factor are estimated.}
\label{fig:tdep_GeGm}
\end{figure*}
%

%
%
\begin{figure*}[t]
\begin{tabular}{ll}
\begin{minipage}[t]{0.50\textwidth}
\includegraphics[height=6.0cm,keepaspectratio,clip]{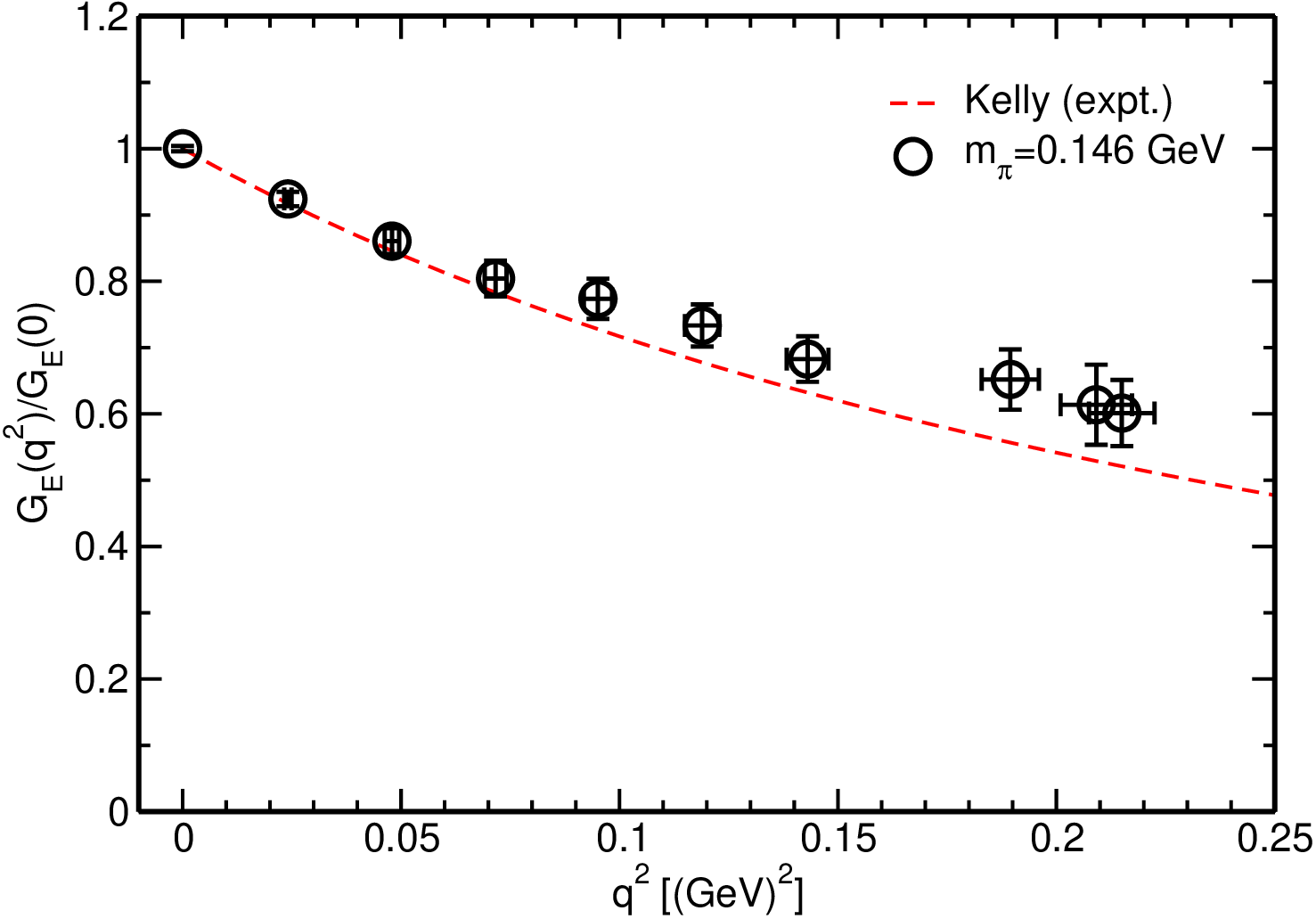}
\end{minipage}
&
\begin{minipage}[t]{0.50\textwidth}
\includegraphics[height=6.0cm,keepaspectratio,clip]{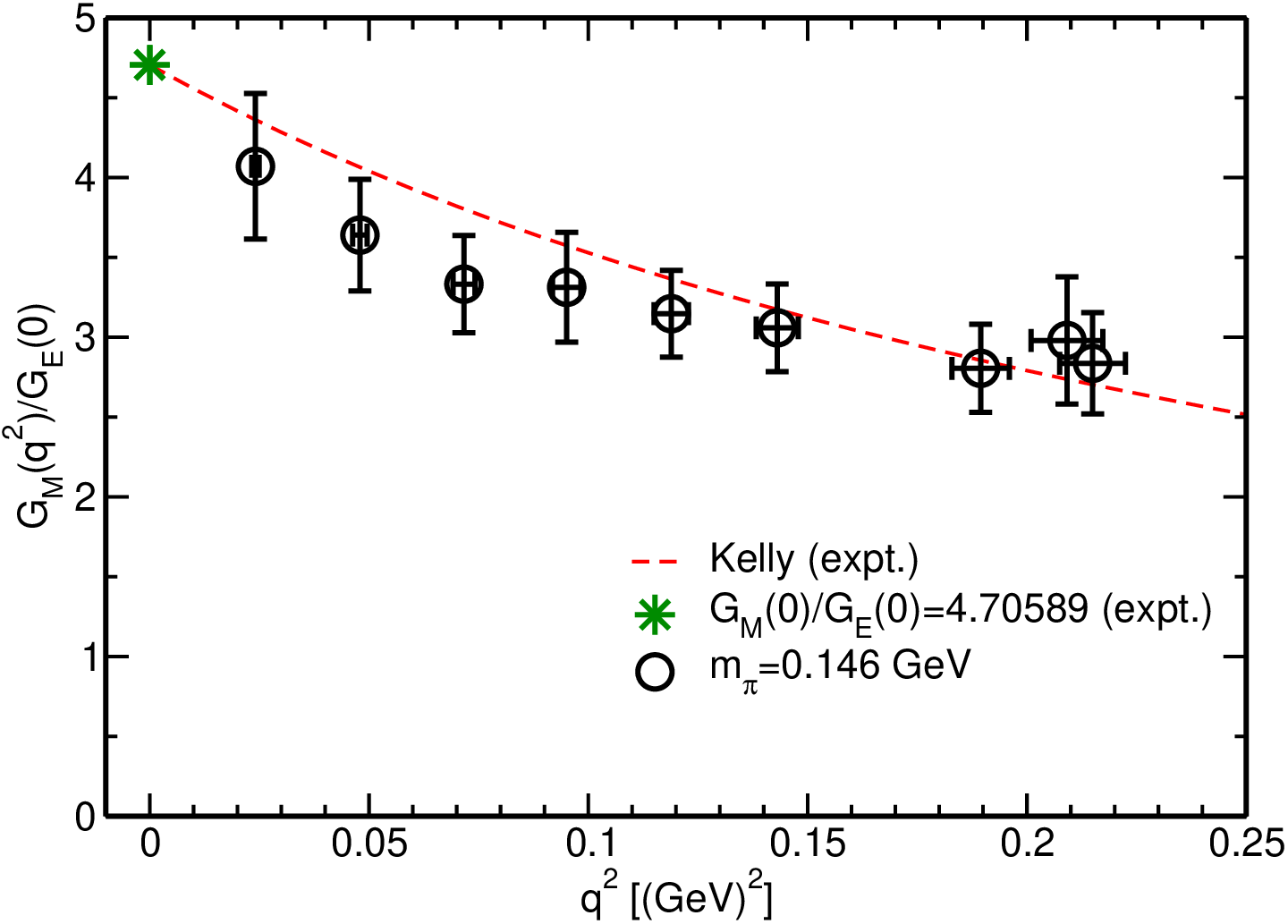}
\end{minipage}
\end{tabular}
\caption{Momentum-squared dependence of the {\it isovector} electric form factor $G_E$ (left) and 
magnetic form factor $G_M$ (right). We also plot Kelly's fit~\cite{Kelly:2004hm} as their experimental data. 
}
\label{fig:gem_q2}
\end{figure*}

The local axial-vector current is renormalized with the value of $Z_A=0.9650(68)(95)$ obtained 
with the SF scheme~\cite{Ishikawa:2015fzw} shown in Fig.~\ref{fig:zv}. 
We then plot the renormalized value of the axial charge $g_A^{\rm ren}=Z_A g_A^{\rm bare}$ 
as a function of the current insertion time slice $t$ in Fig.~\ref{fig:ga}. 
The three solid lines represent the fit result with one-standard error band in the time 
region of $6\le t/a \le 9$, while the uncertainty in the determination of the value of $Z_A$ 
is taken into account by the shaded band. We finally obtain the renormalized value of the axial charge 
%
%
\be
g_A^{\rm ren}=1.163(75)_{\rm stat}(14)_{Z_A},
\ee
which slightly underestimates the experimental value of 1.2724(23)
by less than 10\%. However, the dominant statistical error is of the same order. 
We also recall that the $t$ dependence of ${\cal R}^{5z}_{A,3}(t,{\bm 0})$ 
in Fig.~\ref{fig:bare_couplings}
shows large wiggles, 
which seem to break time reversal between the source and sink points. However the 
time-reversal feature should eventually be restored as observed in the vector and tensor channels. 
This suggests that the ratio of ${\cal R}^{5z}_{A,3}(t,{\bm 0})$ still has large statistical fluctuations, which are not well under control. 
A new class of statistical error reduction techniques such as low-eigenmode deflation 
and all-mode-averaging (AMA)~\cite{Blum:2012uh} should be useful and efficient to improve the statistical accuracy of 
${\cal R}^{5z}_{A,3}(t,{\bm 0})$ with the limited number of configurations. 
We intend to examine this possibility in future research.

\subsection{Results of nucleon form factors in the vector channel}
\label{sec:4-B}

\subsubsection{Electric $G_E$ and magnetic $G_M$ form factors}
\label{sec:4-B1}

The {\it isovector} electric form factor $G_E$ and magnetic form factor $G_M$ are separately 
obtained from the ratios (\ref{Eq:GE}) and (\ref{Eq:GM}). 
Figure~\ref{fig:tdep_GeGm} shows the
ratios of the three- and two-point functions (\ref{Eq:GE}) and (\ref{Eq:GM}) as a function of 
the current insertion time slice $t$ for the {\it isovector} electric form factor $G_E$ (left) and 
magnetic form factor $G_M$ (right) at the nine lowest nonzero momentum transfers after 
extracting the relevant kinematical factors.
Although the time region $6\le t/a \le 9$ certainly exhibits an asymptotic plateau 
at low momentum transfers with our choice of source-sink separation, the plateau region is not 
clearly defined at the higher momentum transfers especially for the magnetic form factor $G_M$. 

This observation suggests that
the excited-state contamination could not be well under control for the higher momentum transfers
with the source-sink separation $t_{\rm sep}/a=15$, since our choice of source-sink separation 
is determined with a set of the smearing parameters ($A=1.2$ and $B=0.11$
for the exponential smearing function as described previously) that is optimized 
for the case of the nucleon operator with zero momentum. However, the remaining excited-state contamination
is relatively smaller than the statistical uncertainties on both electric and magnetic form factors 
at the higher momentum transfers.
Therefore, we simply evaluate the values of both $G_E$ and $G_M$ form factors by constant fits
to the data points in the range $6\le t/a \le 9$ denoted by the gray shaded area
as in the cases of the coupling $g_O$ ($O=V, A, S, T$).

In Table~\ref{tab:formfactor}, we compile the results for $G_E$ and $G_M$ form 
factors together with the results of $F_1$ and $F_2$ form factors, which are evaluated 
using linear combinations of $G_E$ and $G_M$ with appropriate coefficients 
determined by Eqs.~(\ref{Eq:GE}) and (\ref{Eq:GM})
with measured values of the momentum squared $q^2$ and the nucleon mass $M_N$.
The $q^2$ dependences of $G_E$ (left panel) and $G_M$ 
(right panel) are plotted in Fig.~\ref{fig:gem_q2}, respectively. 
For the finite three-momentum ${\bm q}$, we use the nine lowest nonzero momenta: 
${\bm q}=\pi/48 \times {\bm n}$ 
with $|{\bm n}|^2\le9$. The lowest nonzero momentum transfer $q^2$ in the present work 
reaches the value of 0.024(1) [$({\rm GeV})^2$], which is much smaller than $4m_\pi^2$ 
even at $m_\pi \approx 146$ MeV. 
In each panel, we also plot Kelly's fit~\cite{Kelly:2004hm} (red dashed curves) 
as their experimental data.
The simulated electric form factor $G_E$ is fairly consistent with the experimental results, 
especially at low $q^2$. 
The magnetic form factor $G_M$ agrees with Kelly's fit albeit with rather large errors.

This feature suggests that our results successfully reproduce the experimental value for 
the electric RMS radius.
On the other hand, noisier signals for $G_M$ would prevent the precise 
determination of the magnetic RMS radius.
The electric (magnetic) RMS radius, $\sqrt{\la r^2_{E}\ra}$ ($\sqrt{\la r^2_{M}\ra}$), 
can be evaluated from the slope of the respective form factor at zero momentum transfer 
%
%
\ben
\la r^2_l\ra&=&-\left.\frac{6}{G_l(0)}\frac{dG_l(q^2)}{dq^2}\right\vert_{q^2=0}
\een
with $l=E$ ($M$). 
Recall that $G_M$ at zero momentum transfer, whose value corresponds to
the isovector magnetic moment $\mu_v=G_M(0)$, cannot be directly measured for 
kinematical reasons~\cite{Sasaki:2007gw}. Therefore, a $q^2$-extrapolation is 
necessary to evaluate the value of $G_M(0)$.

In principle, low-$q^2$ data points are crucial for determining both the RMS radii and 
magnetic moment. However, the accessible momentum transfer is limited on the lattice since 
the components of the three momentum are discrete in units of $2\pi/L$.
In this sense, a large spatial size $L$ is required to precisely determine the 
RMS radii ($\sqrt{\la r^2_{E}\ra}$ and $\sqrt{\la r^2_{M}\ra}$) 
and magnetic moment ($\mu_v$).

%
%
\begin{table}[hb] 
\caption{
Results for the nucleon form factors in the vector channel.
The values are all renormalized. 
\label{tab:formfactor}
}
\begin{ruledtabular}
\begin{tabular}{llllll} 
\multicolumn{1}{c}{$q^2$ [$({\rm GeV})^2$]}
& \multicolumn{1}{c}{$G^{\rm ren}_E(q^2)$} & \multicolumn{1}{c}{$G^{\rm ren}_M(q^2)$} 
& \multicolumn{1}{c}{$F^{\rm ren}_1(q^2)$} & \multicolumn{1}{c}{$F^{\rm ren}_2(q^2)$} \\ \hline 
0.000& 1.000(4) & N/A & 1.000(4) & N/A\cr
0.024(1)& 0.924(11) & 4.071(456)& 0.944(12)& 3.127(449)\cr
0.048(2)& 0.861(19) & 3.640(350)& 0.895(21)& 2.744(341)\cr
0.072(2)& 0.804(27) & 3.333(305)&0.851(28)& 2.482(295)\cr
0.095(3)& 0.774(30) & 3.313(344)&0.836(33)& 2.478(327)\cr
0.119(4)& 0.733(32) & 3.148(272)&0.806(33)& 2.342(259)\cr
0.143(5)& 0.683(34) & 3.059(275)&0.767(36)& 2.292(263)\cr
0.189(7)& 0.652(46) & 2.806(276)&0.751(44)& 2.054(270)\cr
0.209(8)& 0.614(60) & 2.980(398)&0.736(65)& 2.245(370)\cr
0.215(8)& 0.601(50) & 2.837(317)&0.716(50)& 2.121(304)\cr
\end{tabular}
\end{ruledtabular} 
\end{table} 
%

%
%
\begin{figure*}[ht]
\begin{tabular}{ccc}
\begin{minipage}[t]{0.33\textwidth}
\includegraphics[height=4.0cm,keepaspectratio,clip]{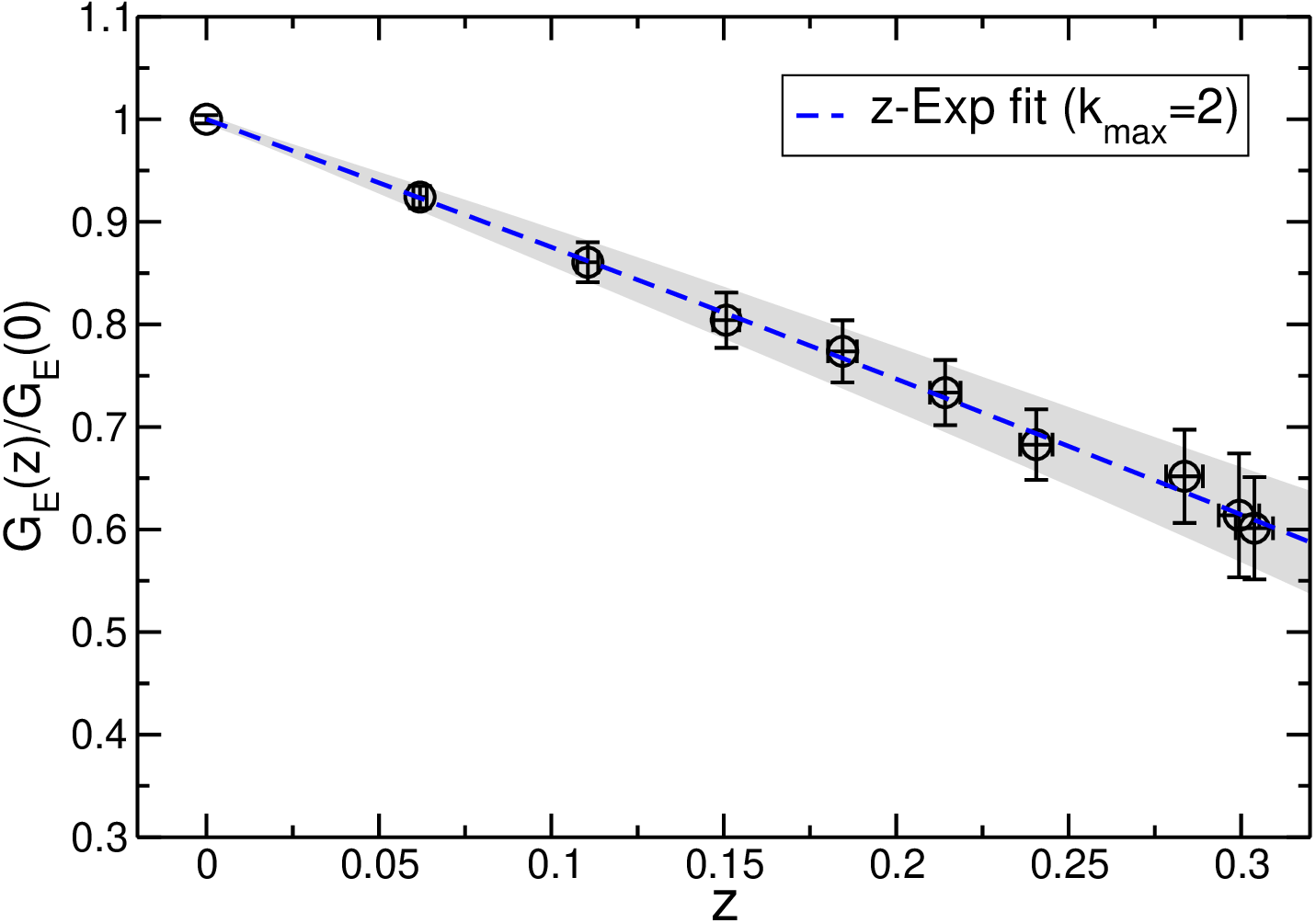}
\end{minipage}
&
\begin{minipage}[t]{0.33\textwidth}
\includegraphics[height=4.0cm,keepaspectratio,clip]{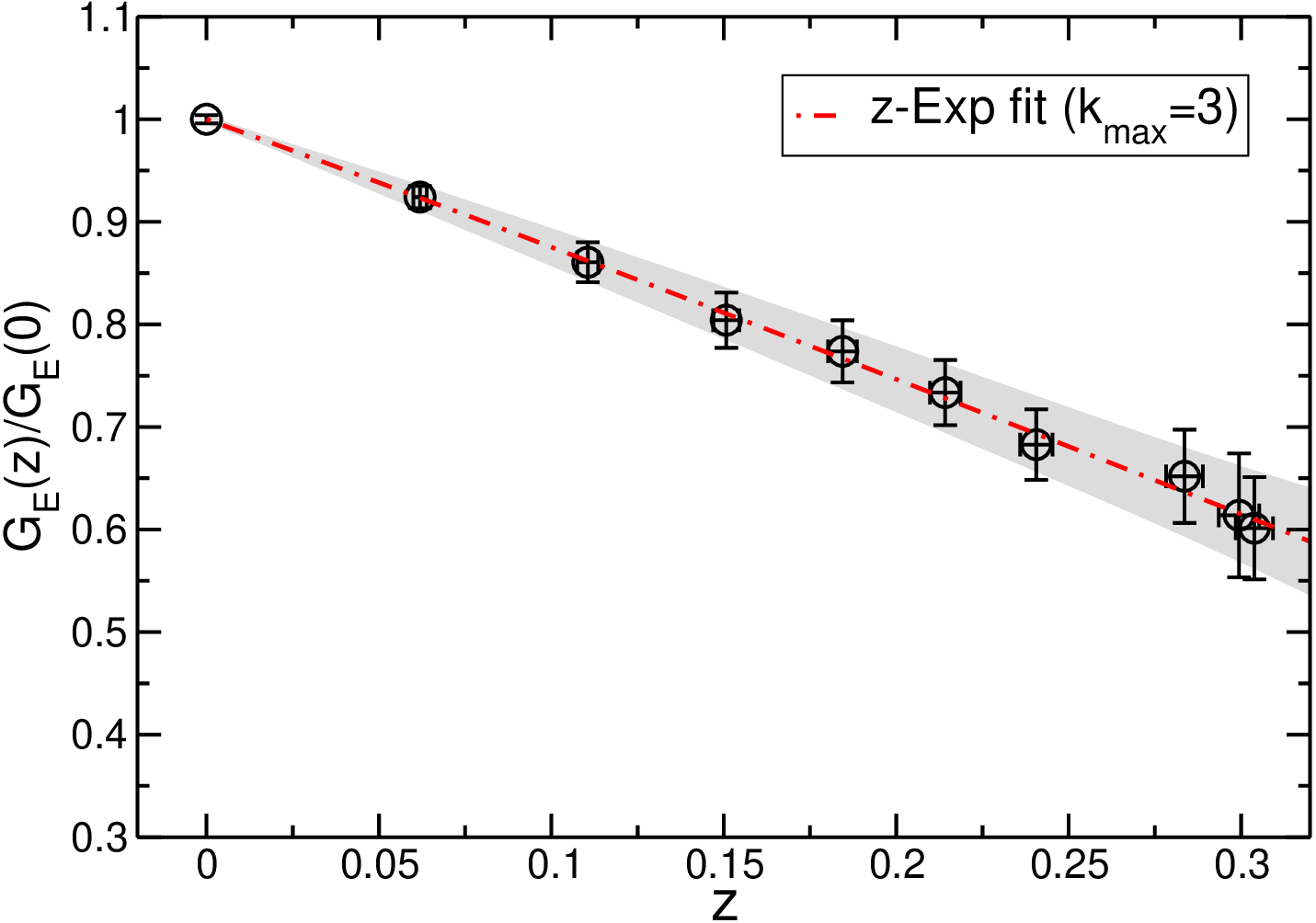}
\end{minipage}
&
\begin{minipage}[t]{0.33\textwidth}
\includegraphics[height=4.0cm,keepaspectratio,clip]{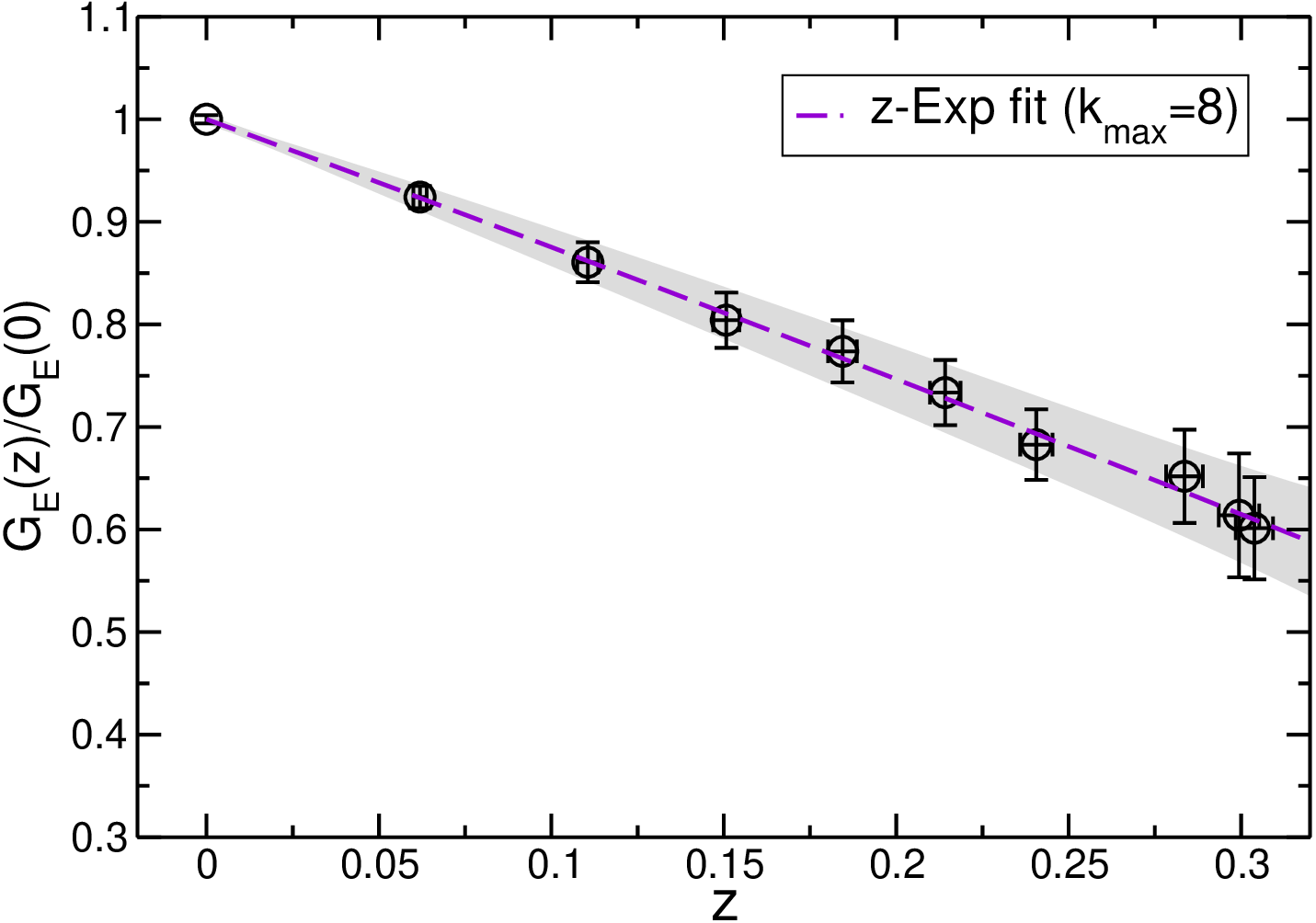}
\end{minipage}
\end{tabular}
\caption{Results for $z$-Exp fit of $G_E$ with $k_{\rm max}=2$ (left), 3 (middle) and 8 (right) using all ten data points.}
\label{fig:fit_e_zexp}
\end{figure*}
%

%
%
\begin{figure*}[ht]
\begin{tabular}{ccc}
\begin{minipage}[t]{0.33\textwidth}
\includegraphics[height=4.0cm,keepaspectratio,clip]{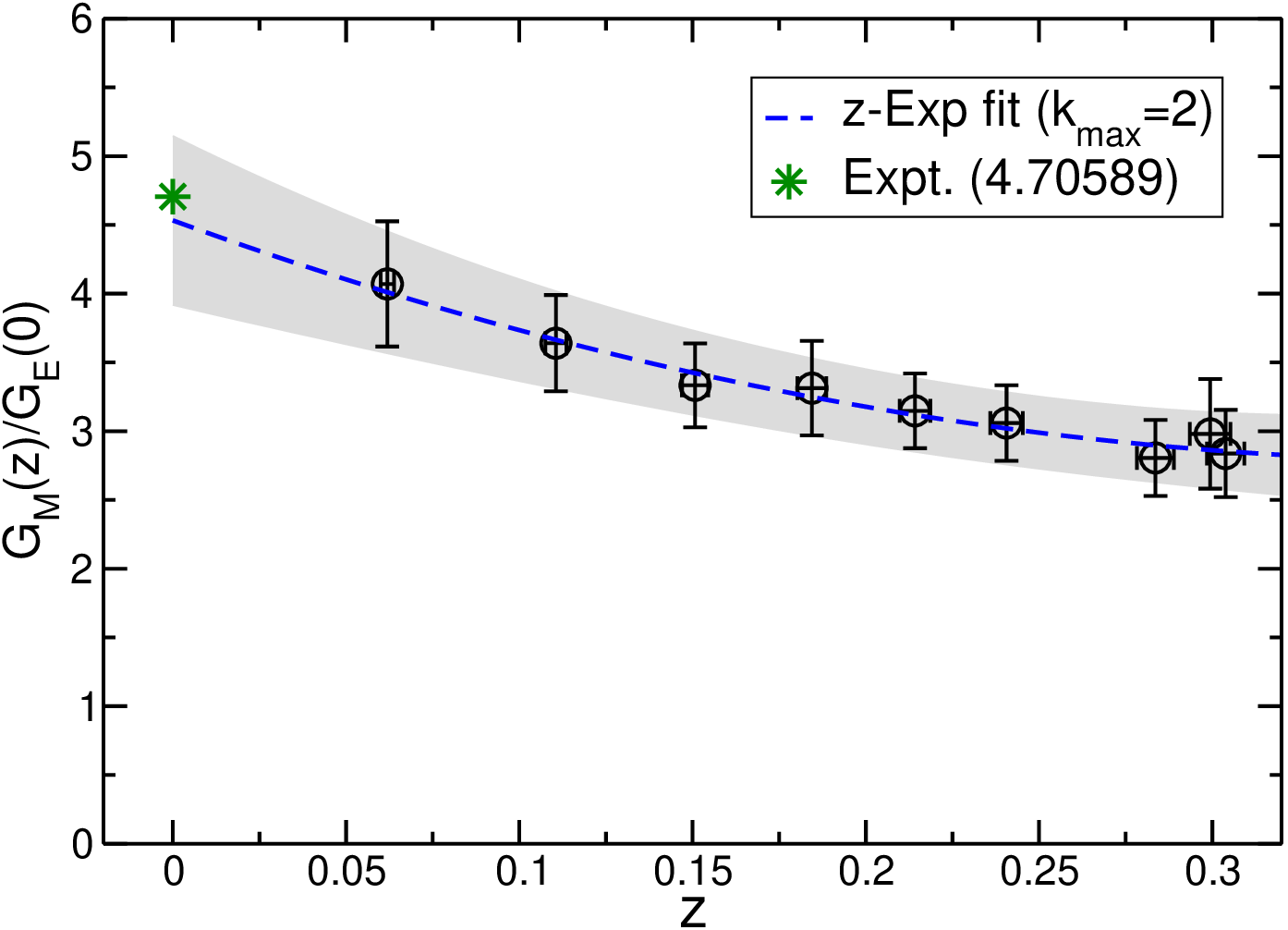}
\end{minipage}
&
\begin{minipage}[t]{0.33\textwidth}
\includegraphics[height=4.0cm,keepaspectratio,clip]{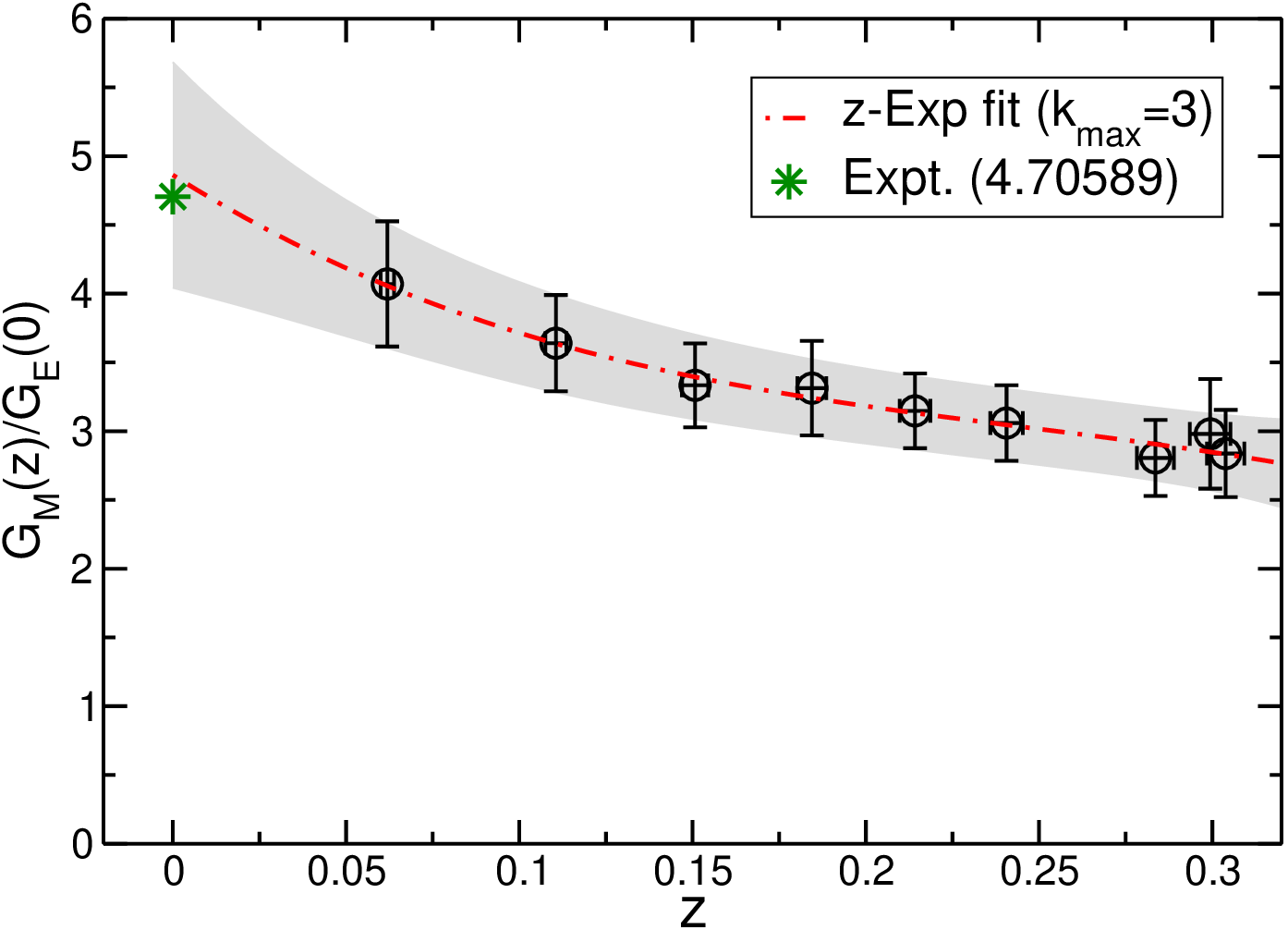}
\end{minipage}
&
\begin{minipage}[t]{0.33\textwidth}
\includegraphics[height=4.0cm,keepaspectratio,clip]{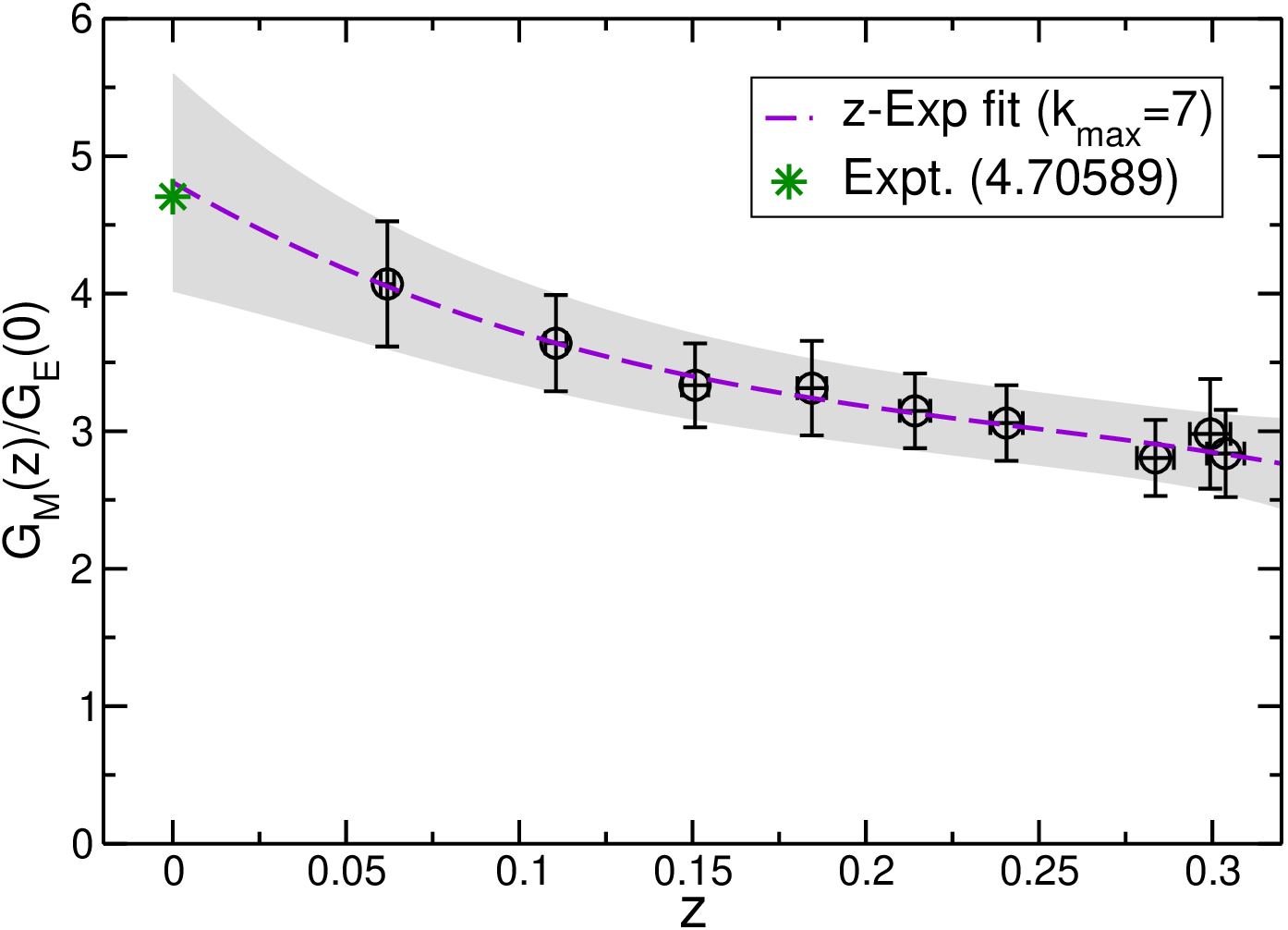}
\end{minipage}
\end{tabular}
\caption{Results for the $z$-Exp fit of $G_M$ with $k_{\rm max}=2$ (left), 3 (middle) and 7 (right) using all nine data points.}
\label{fig:fit_m_zexp}
\end{figure*}

The determination of the slope (or $q^2$ extrapolation) is highly sensitive to how we fit the $q^2$-dependence 
of the form factors. In the previous studies~\cite{{Bhattacharya:2013ehc},{Green:2014xba},{Yamazaki:2009zq},{Capitani:2015sba},{Alexandrou:2013joa}}, 
the dipole form $G(q^2) = a_0/(1+a_1 q^2)^2$, and the polynomial form 
$G(q^2)  = \sum_{k=0} d_kq^{2k}$ were often adopted for $G_E$ and $G_M$. 
However, the fitting form ansatz may tend to constrain an interpolation (or extrapolation) 
and introduce a model dependence into the final result for the RMS radius (or the value of $G(0)$).
In order to reduce systematic errors associated with the interpolation or extrapolation
of the form factor in momentum transfer, 
we use the model-independent $z$-expansion method~\cite{{Boyd:1995cf},{Hill:2010yb}}.

\subsubsection{Analysis with $z$-expansion method}
\label{sec:4-B2}

Let us recall the analytic structure of the form factors in the complex $q^2$ plane.
There is a nonanalytic region for $G(q^2)$ due to threshold of two or more particles, \textit{e.g.} 
the $2\pi$ continuum. Hence the $q^2$ expansion, $G(q^2) = \sum_{k=0}d_kq^{2k}$, does not converge if
$|q^2| > 4m_\pi^2$ where $q^2 = -4m_\pi^2$ is a branch point associated 
with the pion pair creation for $G=G_E$ or $G_M$~\cite{Hill:2010yb}.

The $z$-expansion (denoted as z-Exp) makes use
of a conformal mapping from $q^2$ to a new variable $z$.
Since this transformation maps the analytic domain mapped inside a unit-circle
$|z| < 1$ in the $z$-plane, the form factors can be described by 
a convergent Taylor series in terms of $z$:
%
%
\be
G(z) = \sum_{k=0}^{k_{\rm max}} c_k z(q^2)^k
\ee
with
%
%
\be
z(q^2)=\frac{\sqrt{t_{\rm cut}+q^2}-\sqrt{t_{\rm cut}}}{\sqrt{t_{\rm cut}+q^2}+\sqrt{t_{\rm cut}}},
\label{Eq:z-value}
\ee
where $k_{\rm max}$ truncates an infinite series expansion in $z$~\footnote{We use 
the singular value decomposition (SVD) algorithm to solve the least squares problem
for high degree polynomials.}.
Here, $t_{\rm cut}=4m_{\pi}^2$, where $m_\pi$ corresponds 
to the simulated pion mass, is chosen for the vector channel ($G=G_E$ or $G_M$), while
$t_{\rm cut}=9m_{\pi}^2$, which is associated with the $3\pi$ continuum, 
will be chosen later for the axial-vector channel ($G=F_A$)~\cite{Bhattacharya:2011ah}.

%
%
\begin{figure}[hb]
\begin{minipage}[t]{0.5\textwidth}
\includegraphics[height=6.0cm,keepaspectratio,clip]{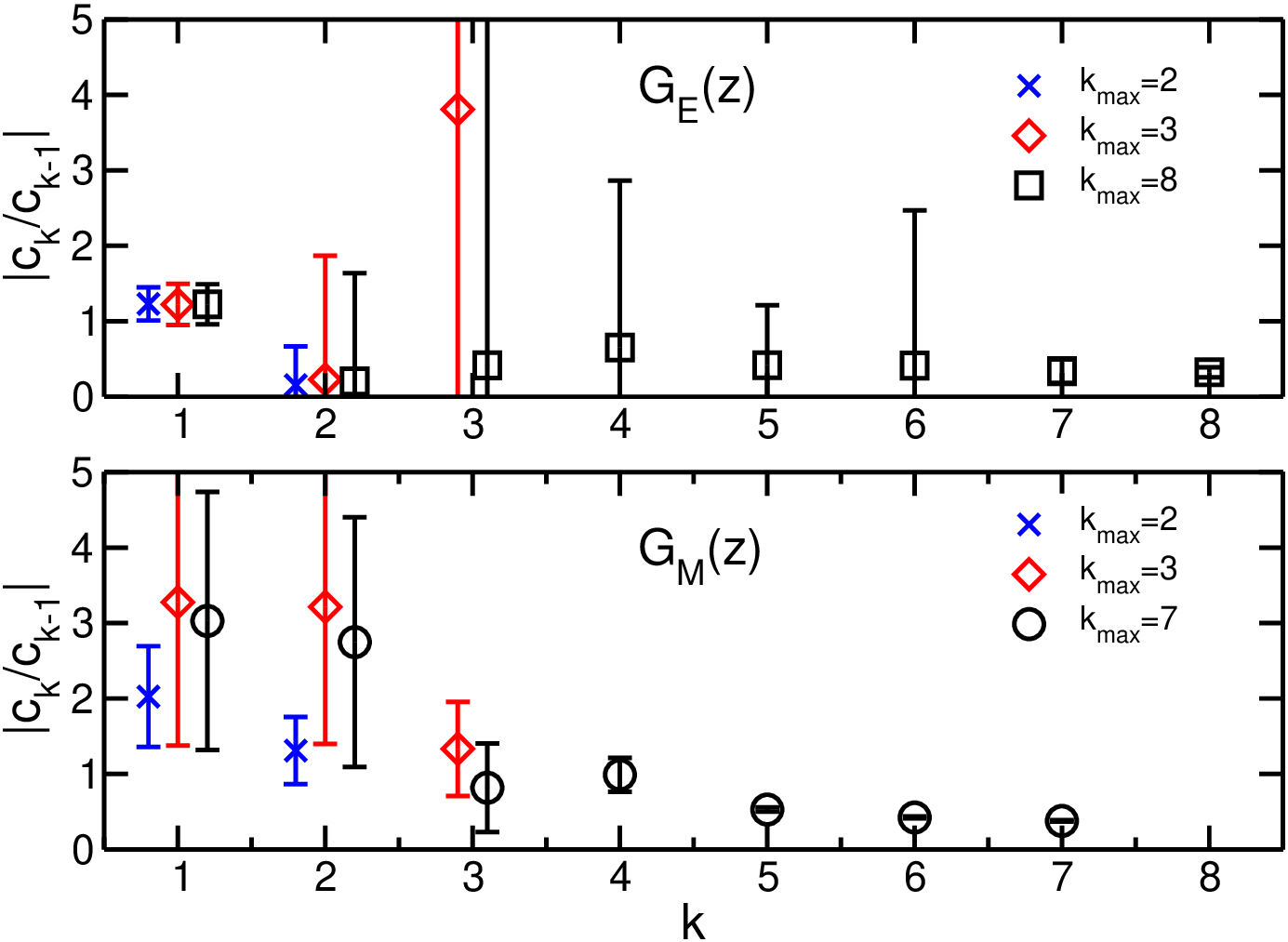}
\end{minipage}
\caption{Convergence behavior of the z-Exp fits for $G_E$ (upper panel) and 
$G_M$ (lower panel). 
The values of $|c_1/c_0|$ are insensitive to the choice of $k_{\rm max}$ if $k_{\rm max}\ge 3$.
For $G_E$, the ratios of $|c_k/c_{k-1}|$ with $k\ge 2$ become zero within the statistical error, 
while the ratios of $|c_k/c_{k-1}|$ for $G_M$ reach a convergence value less than unity at $k\approx 5$ or 6.
}
\label{fig:RCoeff}
\end{figure}
%

%
%
\begin{figure*}[ht]
\begin{tabular}{ll}
\begin{minipage}[t]{0.50\textwidth}
\includegraphics[height=6.0cm,keepaspectratio,clip]{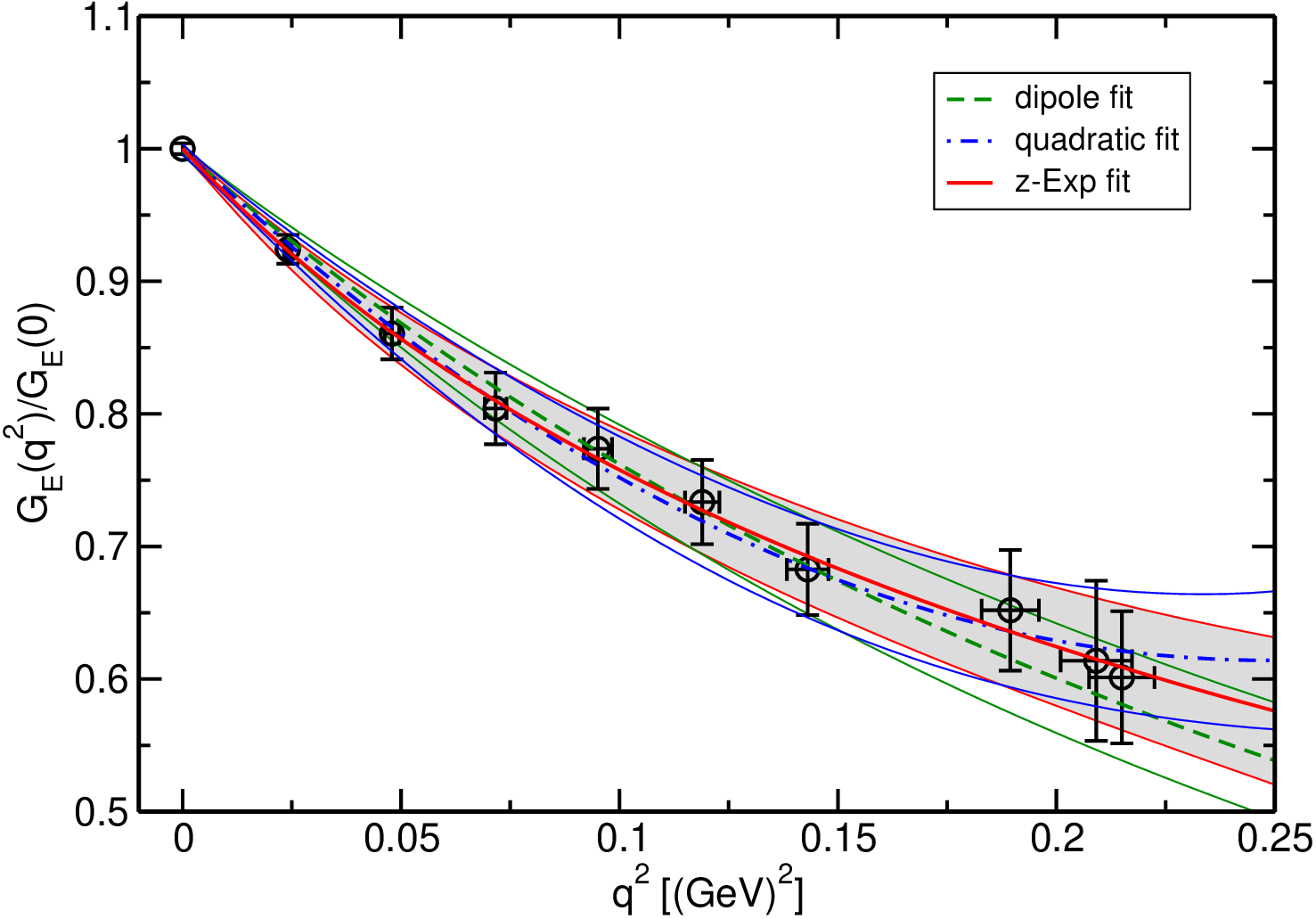}
\end{minipage}
&
\begin{minipage}[t]{0.50\textwidth}
\includegraphics[height=6.0cm,keepaspectratio,clip]{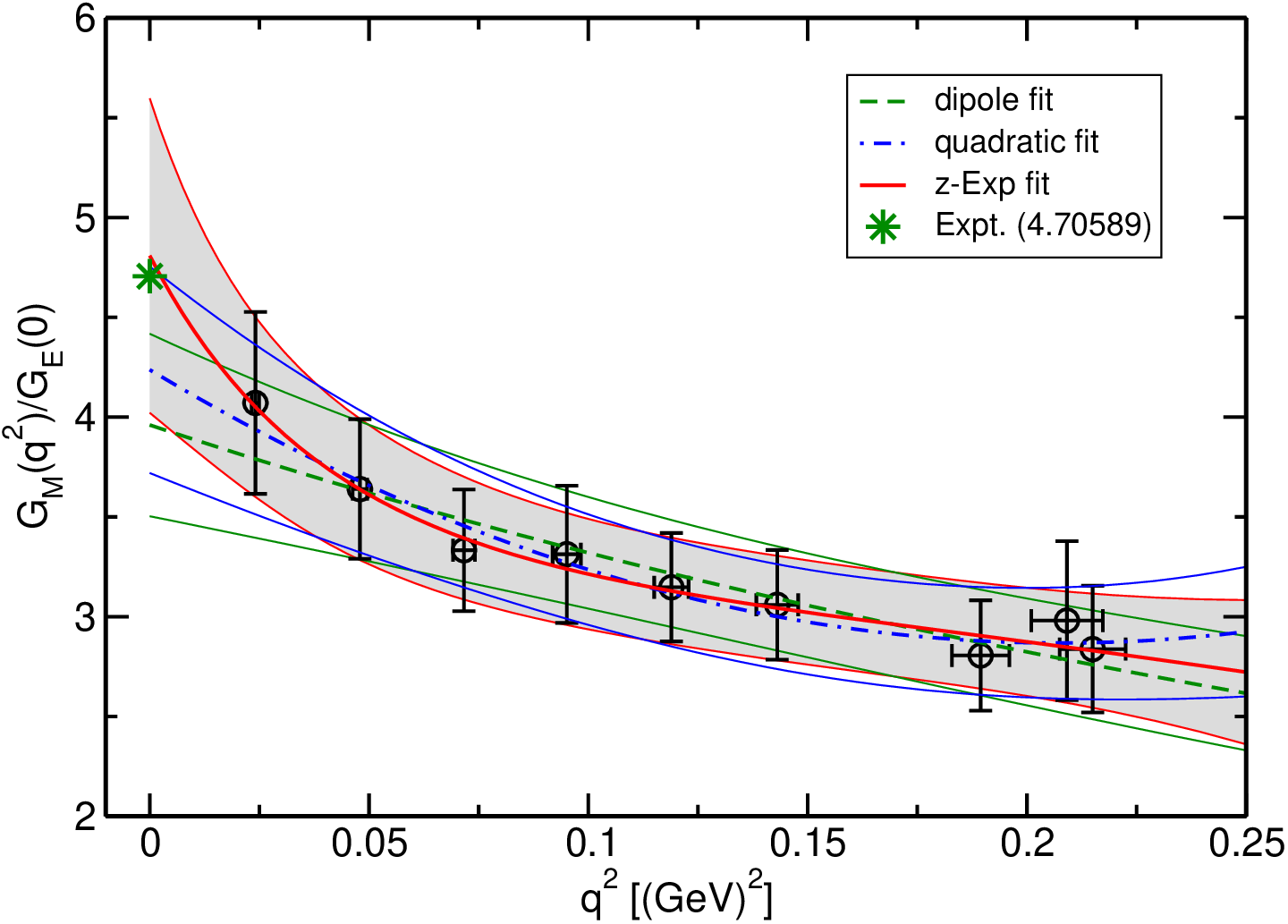}
\end{minipage}
\end{tabular}
\caption{
Results for $G_E$ (left panel) and $G_M$ (right panel)
with three types of fitting form ansatz: dipole (green),
quadratic (blue) and z-Exp (red) fits.
All fits are performed with all ten (nine) data points for $G_E$ ($G_M$).\label{fig:gegmfit}}
\end{figure*}
%

%
%
\begin{table*}[h!]
\begin{ruledtabular}
\caption{
Results for the {\it isovector} electric RMS radius $\sqrt{\la r^2_{E}\ra}$, magnetic moment $\mu_v$ and magnetic 
RMS radius $\sqrt{\la r^2_{M}\ra}$ obtained from various uncorrelated fits.
\label{tab:RMSemMag}}
\begin{tabular}{ c  c c  c c c c c c c}
& & & \multicolumn{3}{ c }{z-Exp fit}  & \multicolumn{2}{ c }{Experimental value}  \cr
\cline{4-6}\cline{7-8}
Observable & dipole fit & quadratic fit & $k_{\rm max}=2$ & $k_{\rm max}=3$ & $k_{\rm max}=8$  & $ep$ scatt. &  $\mu$-$H$ atom\cr
\hline
$\sqrt{\la r^2_{E}\ra}$ [fm] 
& 0.822(63) & 0.851(62) & 0.917(81)& 0.914(101)& 0.915(99) 
&0.939(6) & 0.907(1)
\cr
$\chi^2/{\rm dof}$ & 2.60/8 & 0.99/7 & 0.42/7 & 0.41/6 & 0.41/1 & --- & ---\cr
\hline\hline
& & & \multicolumn{3}{ c }{z-Exp fit}  &  \cr
\cline{4-6} 
Observable & dipole fit & quadratic fit & $k_{\rm max}=2$ & $k_{\rm max}=3$ & $k_{\rm max}=7$  &  \multicolumn{2}{ c }{Experimental value}\cr\hline
$\mu_v\equiv\mu_p-\mu_n$ 
& 3.96(46)& 4.24(52) & 4.53(61) & 4.86(82) & 4.81(79) & \multicolumn{2}{ c }{4.70589}\cr
$\sqrt{\la r^2_{M}\ra}$ [fm] 
& 0.656(133)& 0.852(130)& 1.177(195) & 1.495(437)&1.437(409)
& \multicolumn{2}{ c }{0.862(14)}\cr
$\chi^2/{\rm dof}$ & 1.04/7 & 0.52/6&  0.41/6 & 0.36/5 & 0.37/1 & \multicolumn{2}{ c }{---}\cr
\end{tabular}
\end{ruledtabular}
\end{table*}
%

%
%
\begin{table*}[h!]
\begin{ruledtabular}
\caption{
Results for the {\it isovector} Dirac RMS radius $\sqrt{\la r^2_{1}\ra}$, anomalous magnetic moment $F^v_2(0)$ and Pauli RMS radius $\sqrt{\la r^2_{2}\ra}$ obtained from various uncorrelated fits.
\label{tab:RMS_DP}}
\begin{tabular}{ c  c c  c c c c c c}
& & & \multicolumn{3}{ c }{z-Exp fit}  & \multicolumn{2}{ c }{Experimental value}  \cr
\cline{4-6}\cline{7-8}
Observable & dipole fit & quadratic fit & $k_{\rm max}=2$ & $k_{\rm max}=3$ & $k_{\rm max}=8$  
& $ep$ scatt. &  $\mu$-$H$ atom 
& LHPC'14~\cite{Green:2014xba}\cr
\hline
$\sqrt{\la r^2_{1}\ra}$ [fm] 
&0.668(68)  & 0.740(75) &  0.807(100)  & 0.782(129) &  0.784(125)  
& 0.798(7)  & 0.760(2)
& 0.706(40)
\cr
$\chi^2/{\rm dof}$ & 0.48/8 & 0.65/7 & 0.42/7 & 0.38/6 & 0.39/1 & --- & --- & ---\cr
\hline\hline
& & & \multicolumn{3}{ c }{z-Exp fit}  &  \cr
\cline{4-6}
Observable & dipole fit & quadratic fit & $k_{\rm max}=2$ & $k_{\rm max}=3$ & $k_{\rm max}=7$  &  \multicolumn{2}{ c }{Experimental value}
& LHPC'14~\cite{Green:2014xba}
\cr\hline
$\kappa_v\equiv F_2^v(0)$ 
& 3.01(45) &  3.25(51) & 3.52(60) &  3.84(82) & 3.79(79) & \multicolumn{2}{ c }{3.70589} & 3.89(39)\cr
$\sqrt{\la r^2_{2}\ra}$ [fm] 
& 0.677(173) & 0.896(157) & 1.254(227) & 1.606(499) & 1.542(469)
& \multicolumn{2}{ c }{0.879(17)\footnotemark[1]\footnotetext[1]{Input of $ep$ scatt.} or 0.888(17)\footnotemark[2]\footnotetext[2]{Input of $\mu$-$H$ atom}} & 0.844(144)\cr
$\chi^2/{\rm dof}$ & 1.01/7 & 0.54/6&  0.44/6 & 0.40/5 & 0.40/1 & \multicolumn{2}{ c }{---} & ---\cr
\end{tabular}
\end{ruledtabular}
\end{table*}

We note here that $k_{\rm max}$ must ensure that terms $c_k z^k$ become 
numerically negligible for $k>k_{\rm max}$ for {\it a model-independent fit}.
Although $|c_k/c_{k-1}|<1$ is expected for sufficiently large $k$, the range of 
possible values of $k_{\rm max}$ is limited by the condition $k_{\rm max}\le 8$ (7) 
for the $G_E$ ($G_M$) form factor that are calculated at totally ten (nine) data points 
in this study. We then first check the stability of the fit results with a given $k_{\rm max}$. 

Figure~\ref{fig:fit_e_zexp} shows the fit results for all ten data points of $G_E(q^2)$ 
with $k_{\rm max}=2, 3$ and 8 in the $z$-expansion. Similarly, we also plot the fit results 
for all nine data points of $G_M(q^2)$ with $k_{\rm max}=2, 3$ and 7 
in Fig.~\ref{fig:fit_m_zexp}. At a glance, one can see from Fig.~\ref{fig:fit_e_zexp} 
and Fig.~\ref{fig:fit_m_zexp} that the curvature becomes smaller in the $z$ variable 
than the $q^2$ variable for both cases of $G_E$ and $G_M$. 
This indicates that 
a power series in terms of $z$ is clearly more 
efficient than one in $q^2$.
It is also observed that both fit results are not sensitive to the choice of $k_{\rm max}$. 
This suggests that the $z$ expansion gives a rapid convergence series for both cases. 
Indeed, as shown in Fig.~\ref{fig:RCoeff}, the values of $|c_1/c_{0}|$ are insensitive 
to the choice of $k_{\rm max}$ if $k_{\rm max}\ge 3$. 
For $G_E$, the ratios of $|c_k/c_{k-1}|$ with $k\ge 2$ become zero within the statistical error, while the ratios of $|c_k/c_{k-1}|$ for $G_M$ reach a convergence value less than unity at $k\approx 5$ or 6.
Thus, the value of $k_{\rm max}\le 7$ is large enough to guarantee that the z-Exp analysis makes 
{\it a model-independent fit} and reduces systematic uncertainties to determine the RMS radii 
and the value of $G_M(0)$. For these reasons, $k_{\rm max}=8$ (7) is hereafter chosen 
for the $G_E$ ($G_M$) form factor in the z-Exp method. 

In Fig.~\ref{fig:gegmfit}, we show the fit results for $G_E(q^2)$ (left panel) and $G_M(q^2)$ (right panel)
with three types of fits: dipole (green dashed curve), quadratic (blue dot-dashed curve) and z-Exp (red solid curve) fits.
All the fits reasonably describe all ten (nine) data points for $G_E$ ($G_M$) 
with $\chi^2/{\rm dof}<1$.
However, if one takes a closer look at low $q^2$, the fit curve given by 
the $z$-expansion fit goes nicely through the data points in the low-$q^2$ region. 
This implies that the z-Exp fit is relatively insensitive on the higher $q^2$ data points as was expected. 
The RMS radius is determined to be $r_{\rm RMS}=\sqrt{-6(c_1/c_0)/(4t_{\rm cut})}$ (z-Exp fit),
$\sqrt{-12 a_1}$ (dipole fit) and $\sqrt{-6d_2/d_0}$ (quadratic fit), while the coefficients of 
$c_0$, $a_0$ and $d_0$ for $G_M$ correspond to the value of the magnetic moment, respectively.
All results obtained from various uncorrelated fits, where we do not take into account correlations 
between the form factor data at different $q^2$, are compiled in Table~\ref{tab:RMSemMag}.

\begin{figure*}[hb]
\begin{tabular}{ll}
\begin{minipage}[t]{0.50\textwidth}
\includegraphics[height=5.5cm,keepaspectratio,clip]{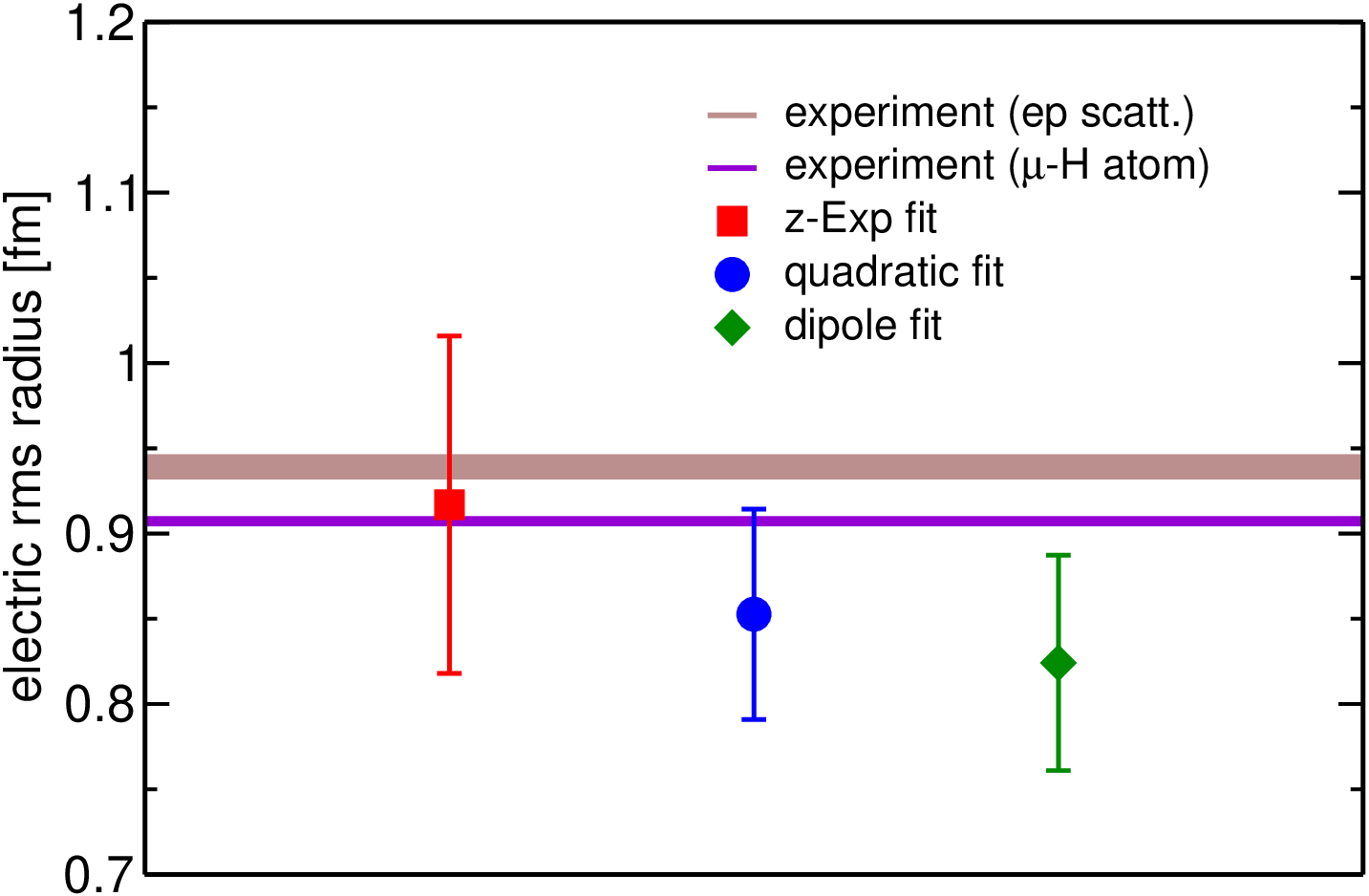}
\end{minipage}
&
\begin{minipage}[t]{0.50\textwidth}
\includegraphics[height=5.5cm,keepaspectratio,clip]{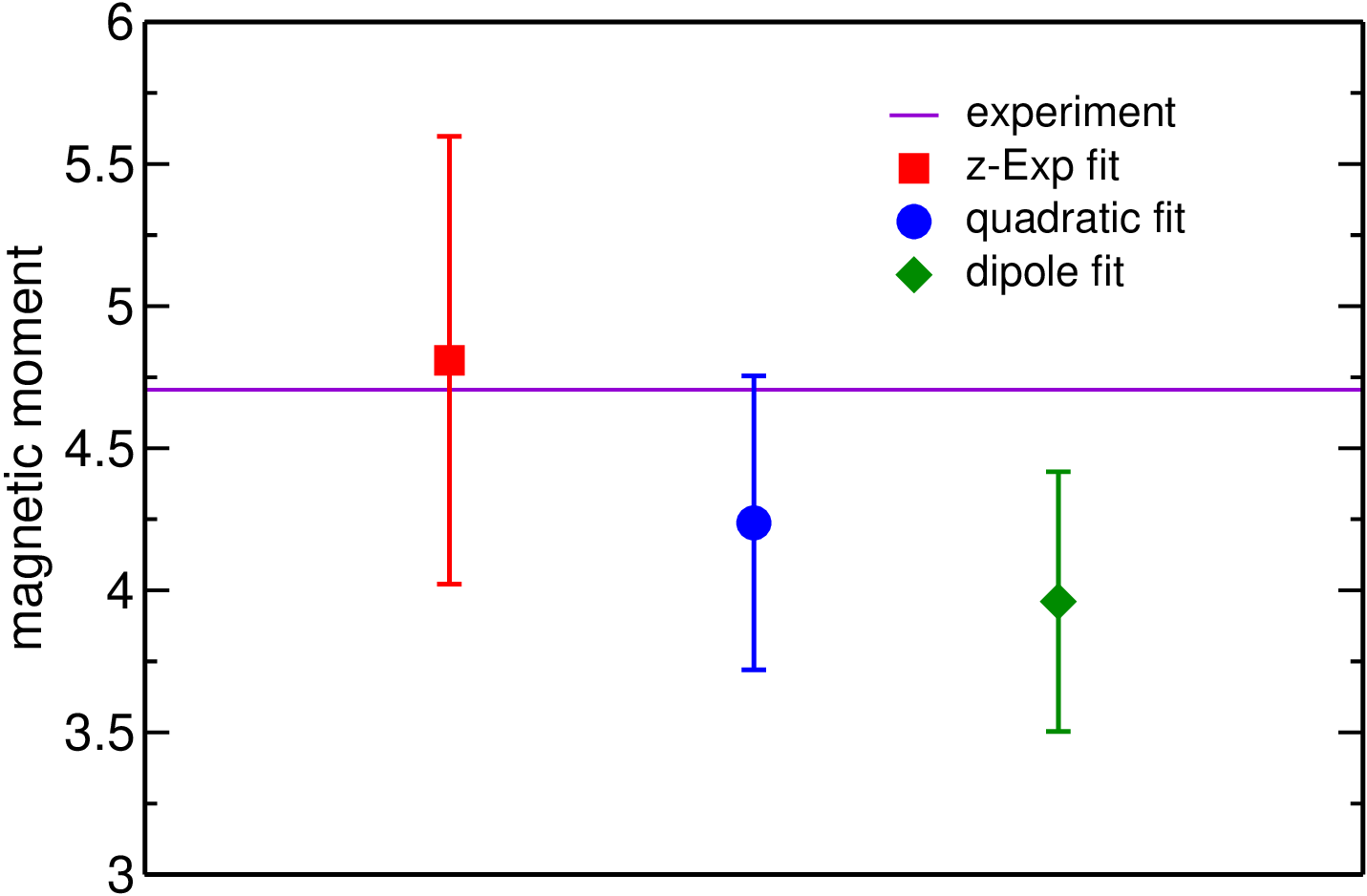}
\end{minipage}
\end{tabular}
 \caption{Comparison of results obtained with three types of fitting form ansatz
 for the electric RMS radius $\sqrt{\la r^2_{E}\ra}$ (left panel) and magnetic moment $\mu_v$ (right panel).
 In the left panel, two horizontal bands represent experimental results from 
 $ep$ scattering (upper) and $\mu$-$H$ atom spectroscopy (lower).\label{fig:RMSaMM}}
\end{figure*}

In Fig.~\ref{fig:RMSaMM}, we compare the results for $\sqrt{\la r^2_{E}\ra}$ (left panel) 
and $\mu_v$ (right panel) from the z-Exp fit (red square) with those from the quadratic 
(blue circle) and dipole (green diamond) fits. In the left panel, the two horizontal bands 
represent experimental results from $ep$ scattering (upper) and muonic hydrogen ($\mu$-$H$) atom (lower).
The dipole results are underestimated in comparison to the corresponding experimental values.
Although each z-Exp fit result has relatively larger error than the other results, its error may 
include both statistical and systematic uncertainties without any model dependence. 
Moreover each result of $\sqrt{\la r^2_{E}\ra}$ and $\mu_v$ from the z-Exp fit is 
most consistent with its respective experiment. As our final results, we quote the value 
of the (isovector) electric RMS radius:
%
%
\be
\sqrt{\la r^2_{E}\ra} = 0.915\pm0.099\;\mathrm{fm}
\ee
and the value of the (isovector) magnetic moment:
%
%
\be 
\mu_v = 4.81\pm0.79, 
\ee
which are obtained from the z-Exp fits. The former is consistent with both the experimental values from $ep$ 
scattering and $\mu$-H atom spectroscopy, though 
statistical uncertainties should be reduced down to a few percent
so as to resolve the experimental puzzle on the proton size.

%
%
%

\subsubsection{Comparison with previous results}
\label{sec:4-B4}

\begin{figure*}[hb]
\begin{tabular}{cc}
\begin{minipage}[t]{0.5\textwidth}
\includegraphics[height=6.0cm,keepaspectratio,clip]{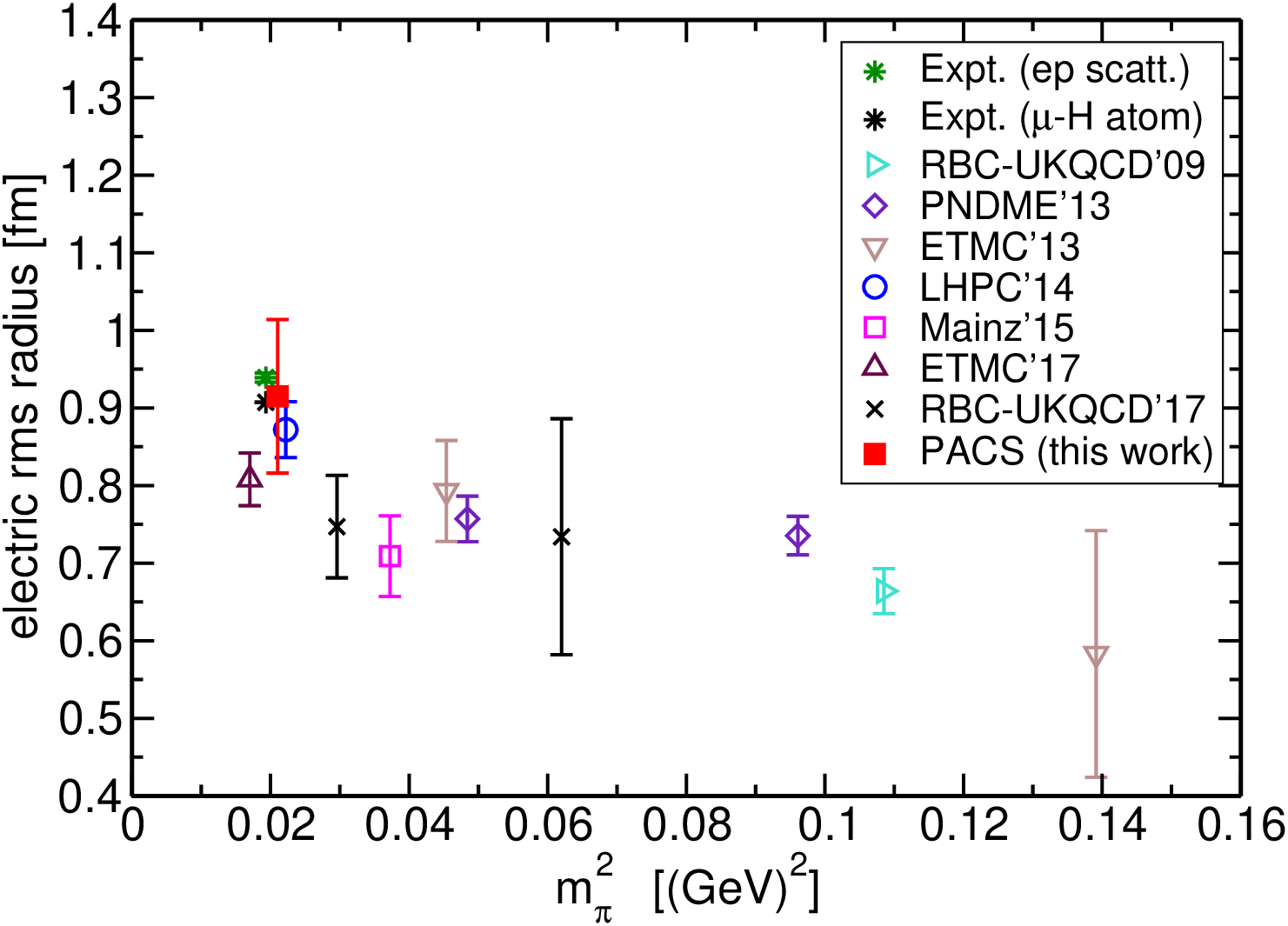}
\end{minipage}
&
\begin{minipage}[t]{0.5\textwidth}
\includegraphics[height=6.0cm,keepaspectratio,clip]{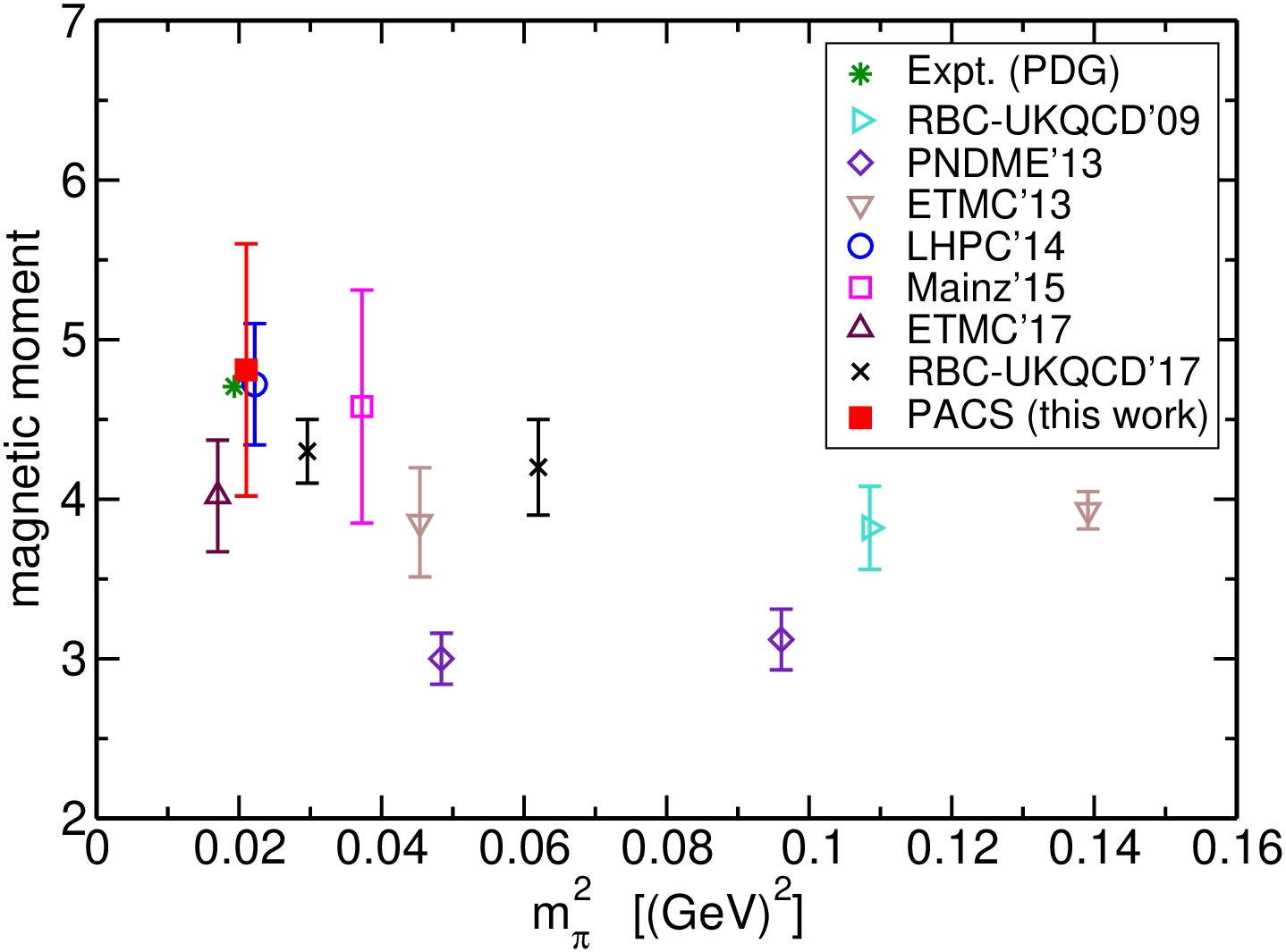}
\end{minipage}
\end{tabular}
\caption{Our results for $\sqrt{\la r^2_{E}\ra}$ (left) and $\mu_v$ (right) 
at $m_\pi\approx146$ MeV
 (filled square) compared to recent lattice results~\cite{Bhattacharya:2013ehc,Green:2014xba,Capitani:2015sba,Alexandrou:2013joa,Alexandrou:2017ypw,Ohta:2017gzg}.
The asterisks represent the experimental results~\cite{Tanabashi:2018oca}.
\label{fig:Comp_RMS_Gm0}}
\end{figure*}

We finally compare our results for $\sqrt{\la r^2_{E}\ra}$ and $\mu_v$ with recent lattice results 
from LHPC~\cite{Green:2014xba}, PNDME~\cite{Bhattacharya:2013ehc},
the Mainz group~\cite{Capitani:2015sba}, ETMC (denoted as ETMC'13~\cite{Alexandrou:2013joa} 
and ETMC'17~\cite{Alexandrou:2017ypw}) and the RBC-UKQCD (denoted as RBC-UKQCD'09~\cite{Yamazaki:2009zq} and
RBC-UKQCD'17~\cite{Ohta:2017gzg}) as shown in Fig.~\ref{fig:Comp_RMS_Gm0}. See also 
Table~\ref{tab:comp_previous_studies} for a summary of previous lattice calculations in comparison with this study. 
Although our results are compatible with both experiments albeit with large errors, many results for 
both the electric RMS radius and magnetic momentum
are often underestimated relative to their respective experimental values in the previous calculations as can be seen in Fig.~\ref{fig:Comp_RMS_Gm0}.

The following are major points of concern for this issue:
1) both quantities are sensitive to the simulated pion mass, 2) their finite size effects become significant as the pion mass decreases, 3) the longer $q^2$ interpolations or extrapolations to estimate them cause larger systematic uncertainties. 
The last point is connected to the second point since the larger spatial volume enables us 
to access the nucleon form factors at smaller momentum transfers. In this context, 
the LHPC calculation, where the simulations were performed with the largest spatial size 
of 5.6 fm with almost physical quark masses, is a favorite among the previous calculations.
Indeed, the LHPC results show better agreement with the experiments, though the electric RMS radius 
is slightly smaller than the PDG value.  

The simulated pion mass in the LHPC calculation is comparable to that of ours. 
Our lattice setup is, however, superior to the LHPC calculation with respect to 
our spatial size of 8.1 fm. The larger spatial volume provides rich information
about the momentum squared dependence of the nucleon form factors in the low $q^2$ region.
Raw data for $F_1$ and $F_2$ are available in the tables of Ref.~\cite{Green:2014xba}. 
Therefore, it is worth comparing our results for the $F_1$ and $F_2$ form factors with 
their results in the same plots. 

\begin{figure*}[hb]
\begin{tabular}{cc}
\begin{minipage}[t]{0.5\textwidth}
\includegraphics[height=6.0cm,keepaspectratio,clip]{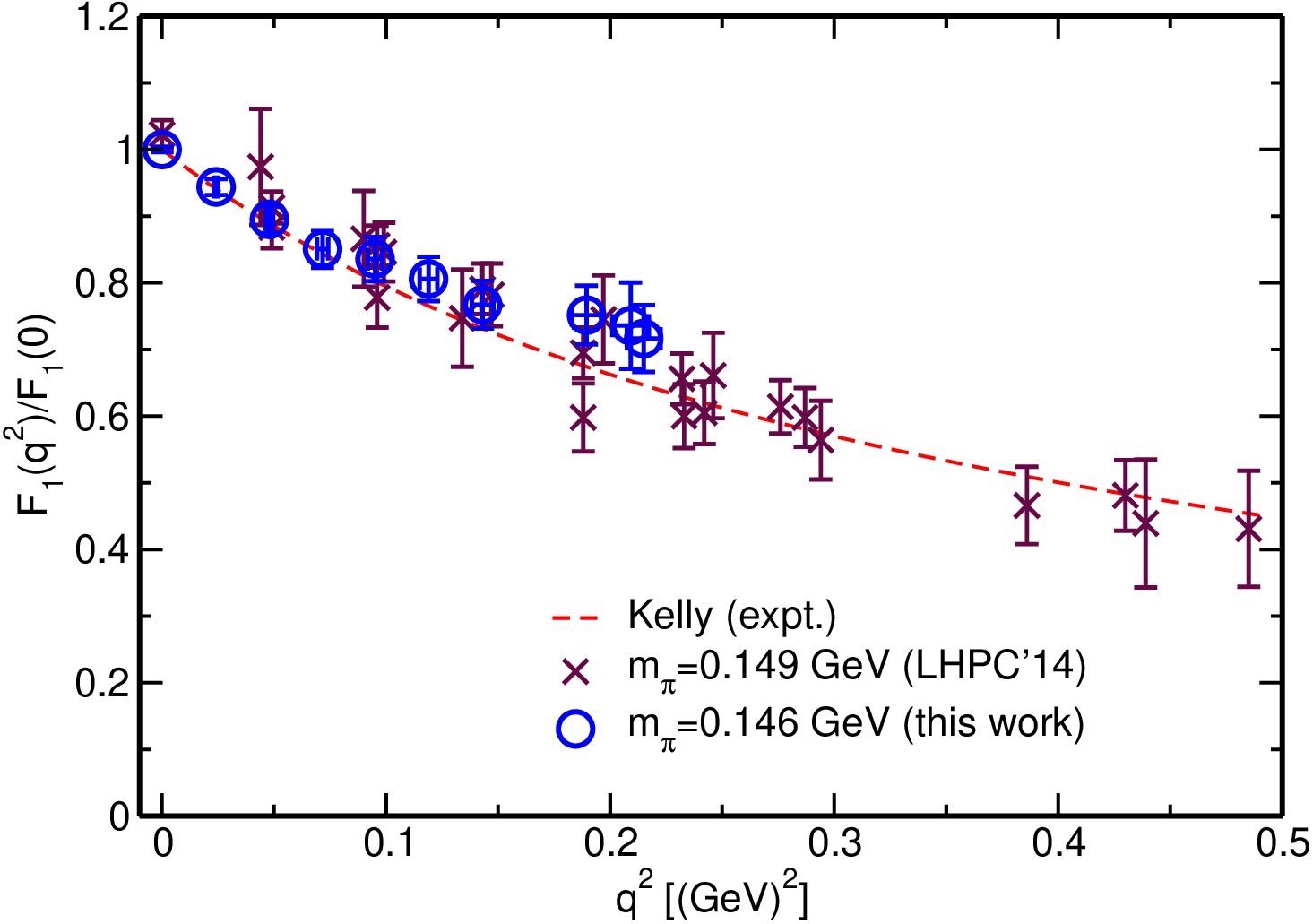}
\end{minipage}
&
\begin{minipage}[t]{0.5\textwidth}
\includegraphics[height=6.0cm,keepaspectratio,clip]{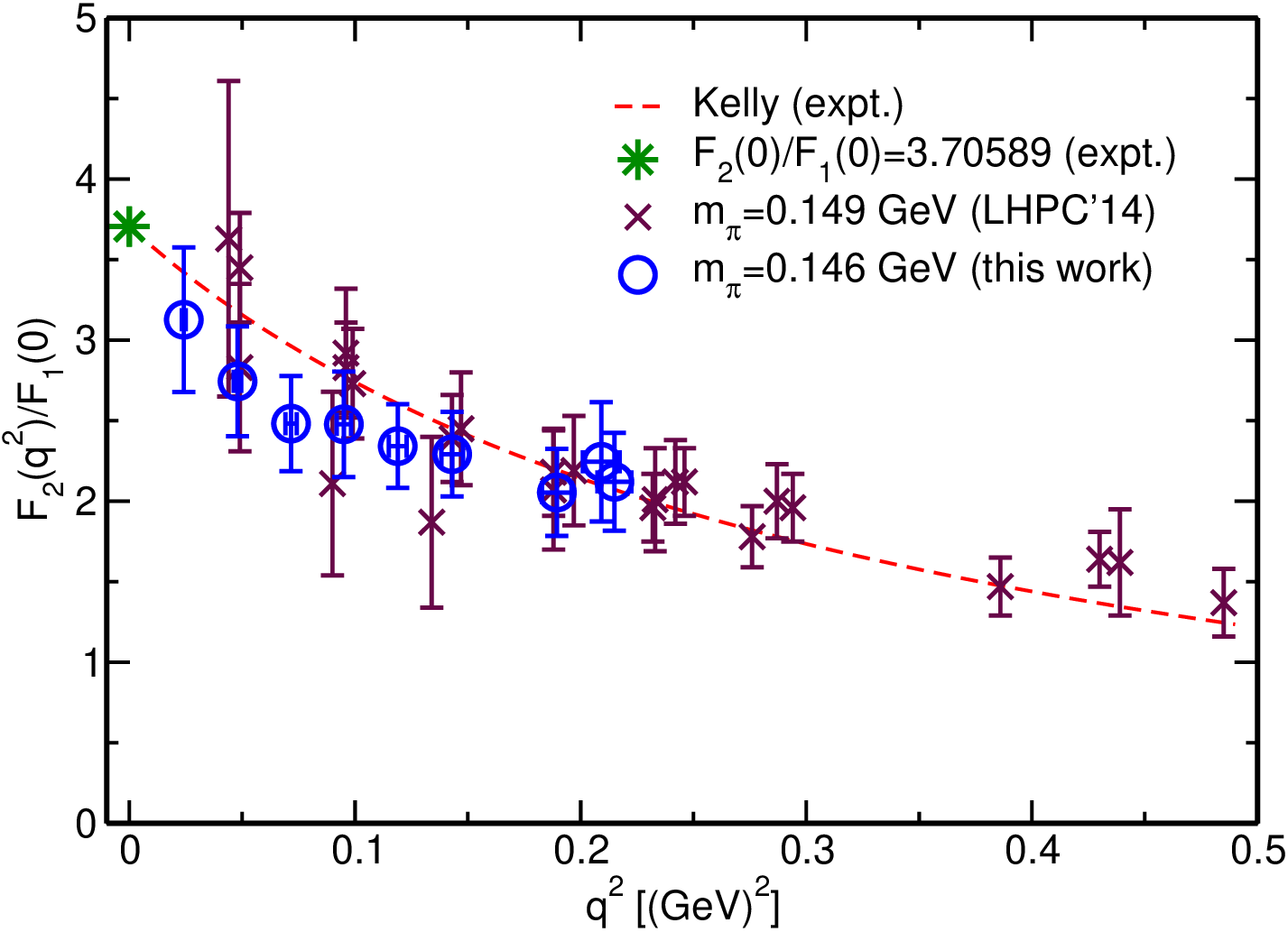}
\end{minipage}
\end{tabular}
\caption{Comparison of the {\it isovector} Dirac form factor $F_1$ 
(left) and Pauli form factor $F_2$ (right) results between 
from this work (circles) and LHPC'14 (crosses) taken from Ref.~\cite{Green:2014xba}.
The dashed curve in each panel shows Kelly's parametrization of the 
experimental data, while the asterisk in the right panel represents the experimental 
result.
}
\label{fig:F1F2_comp}
\end{figure*}

In Fig.~\ref{fig:F1F2_comp}, we show both the LHPC results (cross symbols) 
and our results (open circles) for the Dirac $F_1$ (left panel) and 
Pauli $F_2$ (right panel) form factors as a function of momentum squared $q^2$.
In each panel, the dashed curves correspond to Kelly's fit~\cite{Kelly:2004hm} 
as their experimental data. In the LHPC results, they use two types of kinematics 
regarding the momentum ${\bm p}^\prime$ for the final state of the nucleon state 
when constructing the three-point correlation functions~(\ref{Eq:RatioQ}). 
The smallest momentum transfer $q_{\rm min}^2=0.044$ [$({\rm GeV})^2$] 
in the LHPC calculation is given by the choice of ${\bm p}^\prime=2\pi/L\times (-1,0,0)$, 
while we consider only the rest frame of the final state with ${\bm p}^\prime={\bm 0}$.

Our results and the LHPC results are consistent with each other in both the $F_1$ and $F_2$ 
form factors within the statistical errors. 
We have found that the size of our statistical errors is similar to that of the LHPC data 
points which are calculated in the rest frame of the final state 
with ${\bm p}^\prime={\bm 0}$. It is clear that the $q^2$ dependence of the form 
factors at low $q^2$ are much constrained by our results that contain the smallest 
nonzero $q^2$ data point. We primarily focus on the
{\it isovector} Dirac RMS radius ($\sqrt{\la r^2_{1}\ra}$) and 
anomalous magnetic moment ($\kappa_v=F^v_2(0)$)
since the data for the $F_2$ form factor exhibits large statistical fluctuations 
in the low $q^2$ region.

We then extract the values of  $\sqrt{\la r^2_{1}\ra}$ and $\kappa_v=F^v_2(0)$
together with the Pauli RMS radius ($\sqrt{\la r^2_{2}\ra}$) from the $F_1$ and $F_2$ form factors 
from various uncorrelated fits
as summarized in Table~\ref{tab:RMS_DP}. 
Our results for $\sqrt{\la r^2_{1}\ra}$ and $\kappa_v$ obtained from the z-Exp fits 
are more consistent with both experiments albeit with large errors. 
The dipole fits tend to yield smaller errors in comparison with the other fits. 
For comparison, the LHPC results for $\sqrt{\la r^2_{1}\ra}$, $F^v_2(0)$ 
and $\sqrt{\la r^2_{2}\ra}$ are also listed in Table.~\ref{tab:RMS_DP}. 
Their quoted errors on both $\sqrt{\la r^2_{1}\ra}$ and $\kappa_v$ are, however, smaller by 
a factor of 2 or 3 in comparison with 
our results obtained using the z-Exp fit.
This is partly because the LHPC results are given by the dipole fits 
with large number of $q^2$ data points.
Indeed, if we adopt the dipole fit rather than the z-Exp fit, the statistical uncertainties on the obtained results become small as those of the LHPC results.
Although the LHPC results are also roughly consistent with experiments, 
the shorter $q^2$ interpolation (or extrapolation) that was achieved in our study 
by using the larger volume reduces the systematic uncertainties on the determination 
of $\sqrt{\la r^2_{1}\ra}$ (or $\kappa_v$).

%
%
\begin{table*}[ht] 
\caption{Summary of previous lattice calculations for the nucleon electric-magnetic form factors 
in comparison with this study. Here, $N_f$ denotes the number of dynamical quark flavors. 
In the fourth column, TM-Wilson (TM-Clover) stands for the twisted mass Wilson (clover) Dirac operator, DWF for the domain-wall fermions.
In the last column,``R'', ``S", ``TSF" and ``GPoF" stand for the standard ratio method, the summation method, the two-state fit method and
the generalized pencil-of-function method.
\label{tab:comp_previous_studies}
}
\begin{ruledtabular}
\begin{tabular}{lcccccccccc} \hline
\multicolumn{1}{l}{Publication}
& \multicolumn{1}{c}{$N_f$} 
& \multicolumn{1}{c}{Type} 
& \multicolumn{1}{c}{Fermion} 
& \multicolumn{1}{c}{$m_{\pi}$ [MeV]} 
& \multicolumn{1}{c}{$a$ [fm]} 
& \multicolumn{1}{c}{$La$ [fm]} 
& \multicolumn{1}{c}{$Lm_{\pi}$} 
& \multicolumn{1}{c}{$t_{\rm sep}/a$}
& \multicolumn{1}{c}{$t_{\rm sep}$ [fm]}
& \multicolumn{1}{c}{Method}
\\ \hline
PNDME'13~\cite{Bhattacharya:2013ehc} &2+1+1 &Hybrid~\footnotemark[1]\footnotetext[1]{Clover fermions on highly improved staggered quark (HISQ) ensembles}  & Clover  &220 & 0.120 & 3.8 & 4.4 & \{8,9,10,11,12\} &$\le 1.44$  & R, TSF\\
             & & & Clover &310 & 0.120 & 2.9 & 4.6 & \{8,10,12\} & $\le 1.44$ & R, TSF\\
LHPC'14~\cite{Green:2014xba} & 2+1 & Full & Clover & $\ge149$~\footnotemark[2]\footnotetext[2]{In Fig.~\ref{fig:Comp_RMS_Gm0}, we only quote the
results at the lightest pion mass.} & 0.116 & 5.6 & 4.21 & \{8,10,12\}  &$\le 1.39$  & R, S, GPoF\\
Mainz'15~\cite{Capitani:2015sba} &2 & Full & Clover & $\ge193$~\footnotemark[2] &  0.063 &  4.0 & 4.0 & ---  & $\le 1.1$ & R, S, TSF\\
ETMC'13~\cite{Alexandrou:2013joa} &2+1+1 & Full  & TM-Wilson & 213 & 0.064 & 3.1 & 3.35 & 18 & 1.15 & R \\
             & & &  TM-Wilson & 373 & 0.082 & 2.6 & 4.97 & 12 &  0.98 & R\\            
ETMC'17~\cite{Alexandrou:2017ypw} &2 & Full & TM-Clover& 130 & 0.094 & 4.5 & 2.97 & 
\{10,12,14,16,18\}~\footnotemark[3]\footnotetext[3]{The electric form factor determined 
with the projection operator ${\cal P}_t$ is evaluated up to $t_{\rm sep}/a=18$ ($t_{\rm sep}=1.69$ [fm]), 
while the magnetic, axial-vector and pseudoscalar form factors determined with 
the projection operator ${\cal P}_{5z}$ are evaluated only up to $t_{\rm sep}/a=14$
($t_{\rm sep}=1.32$ [fm]).} & $\le 1.69$~\footnotemark[3] & R, S, TSF\\
RBC-UKQCD'09~\cite{Yamazaki:2009zq} &2+1 & Full & DWF & $\ge 330$~\footnotemark[2] & 0.114 & 2.7 & 4.58 & 12 & 1.37 & R\\
RBC-UKQCD'17~\cite{Ohta:2017gzg} &2+1 & Full & DWF & 172 & 0.143 &  4.6 & 4.00 & \{7,9\} & $\le 1.29$ & R\\
  & & & DWF & 249 & 0.143 &  4.6 & 5.79 & \{7,9\} &  $\le 1.29$ & R\\
This work &2+1&Full  & Clover & 146 & 0.085 & 8.1 & 6.0 & 15 & 1.27 & R\\
\end{tabular}
\end{ruledtabular} 
\end{table*} 
%

%
%
%

%
%
%


%
%
\begin{figure*}[t]
\begin{tabular}{ll}
\begin{minipage}[t]{0.50\textwidth}
\includegraphics[height=6.0cm,keepaspectratio,clip]{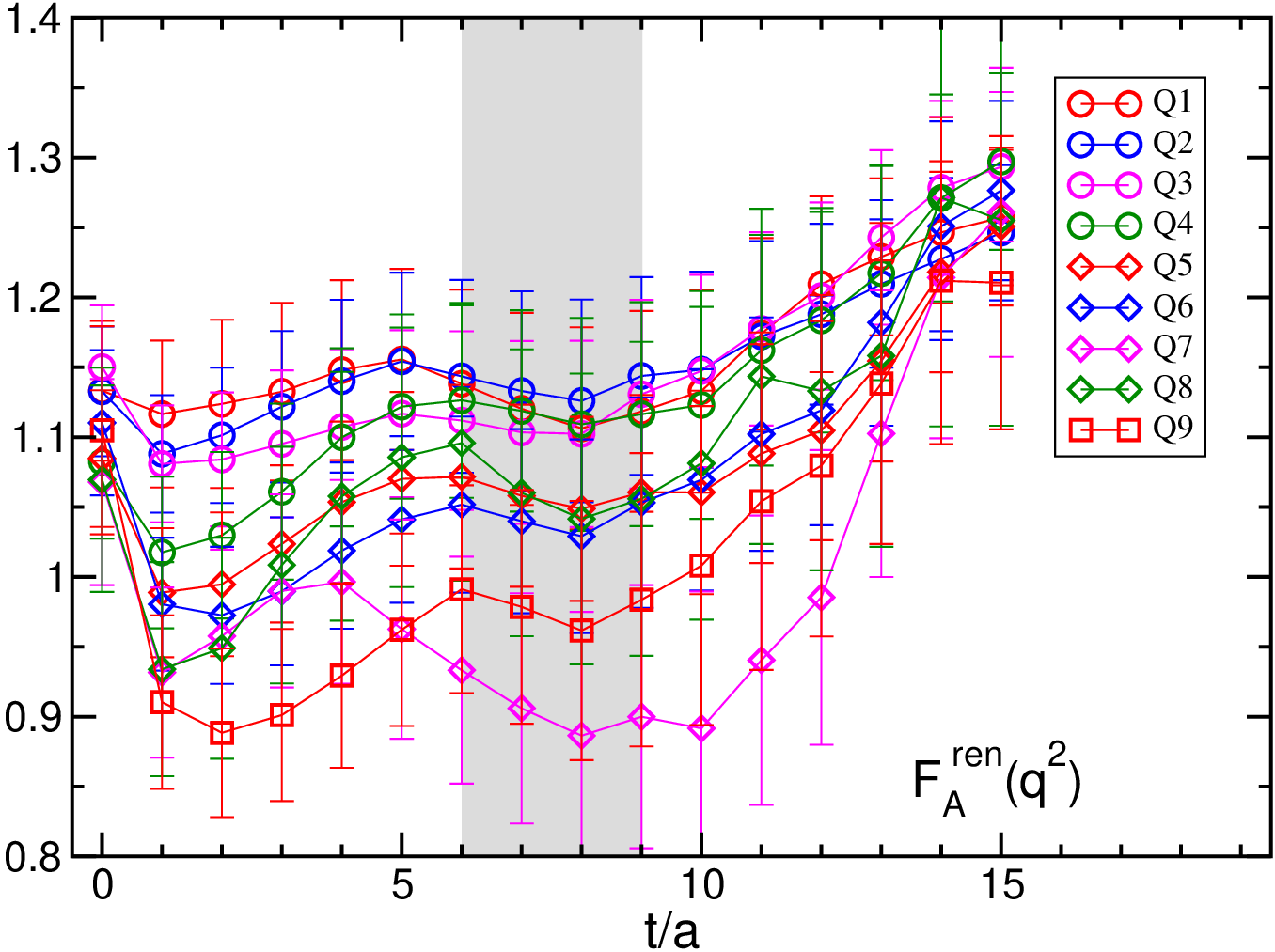}
\end{minipage}
&
\begin{minipage}[t]{0.50\textwidth}
\includegraphics[height=6.0cm,keepaspectratio,clip]{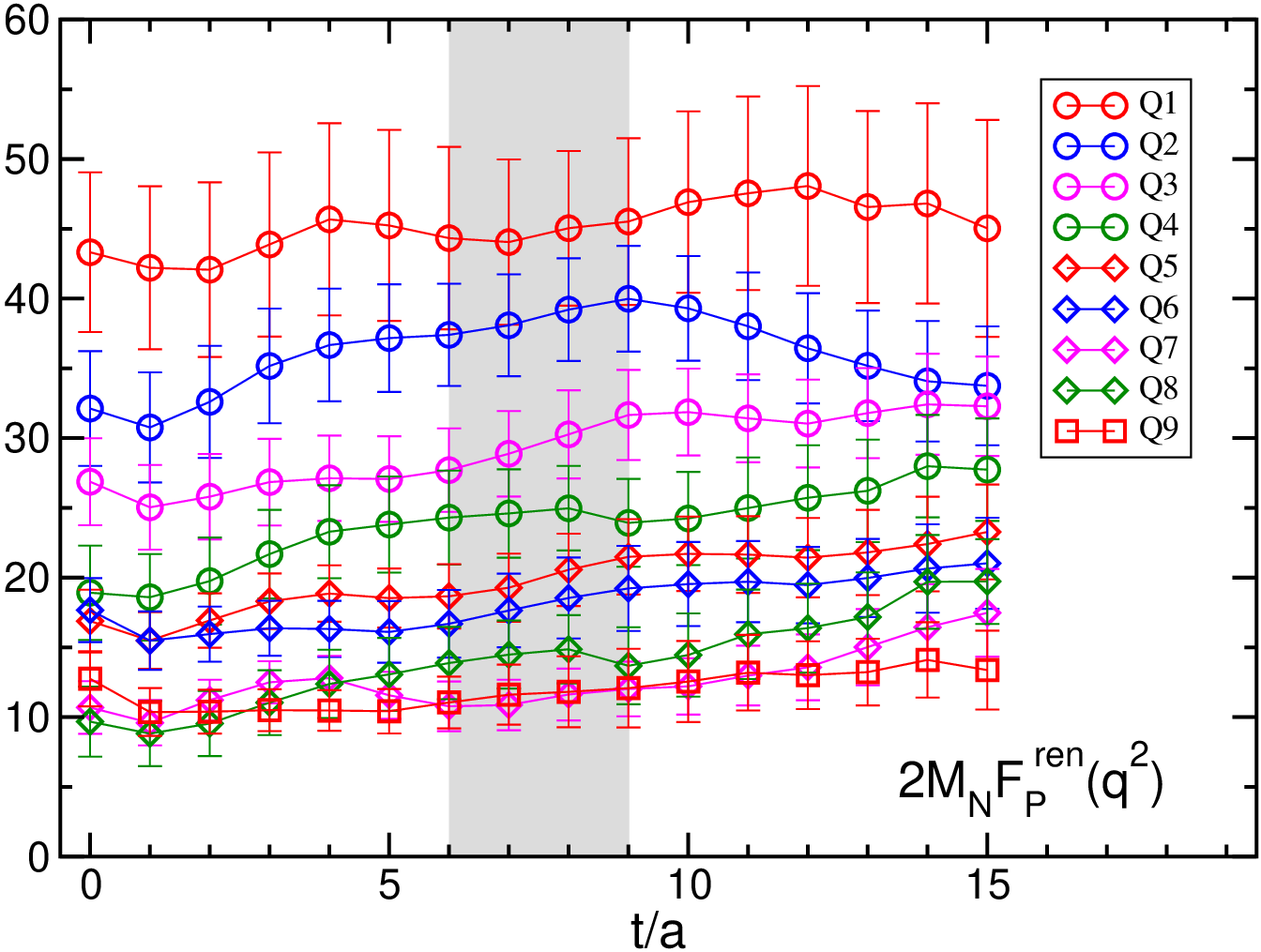}
\end{minipage}
\end{tabular}
\caption{Ratios of the three- and two-point functions (\ref{Eq:FA}) and (\ref{Eq:FP})
for the axial-vector form factor $F_A$ (left) and induced pseudoscalar form factor $F_P$ 
(multiplied by $2M_N$ to make it a dimensionless quantity) (right)
at the nine lowest nonzero momentum transfers.
Both ratios are renormalized with $Z_A=0.9650(68)(95)$~\cite{Ishikawa:2015fzw}. 
The gray shaded area ($6\le t/a \le 9$) in each panel shows the region where 
the values of the corresponding form factor are estimated.}
\label{fig:tdep_RFARFP}
\end{figure*}
%

%
%
\begin{figure*}[ht]
\begin{tabular}{cc}
\begin{minipage}[t]{0.5\textwidth}
\includegraphics[height=6.0cm,keepaspectratio,clip]{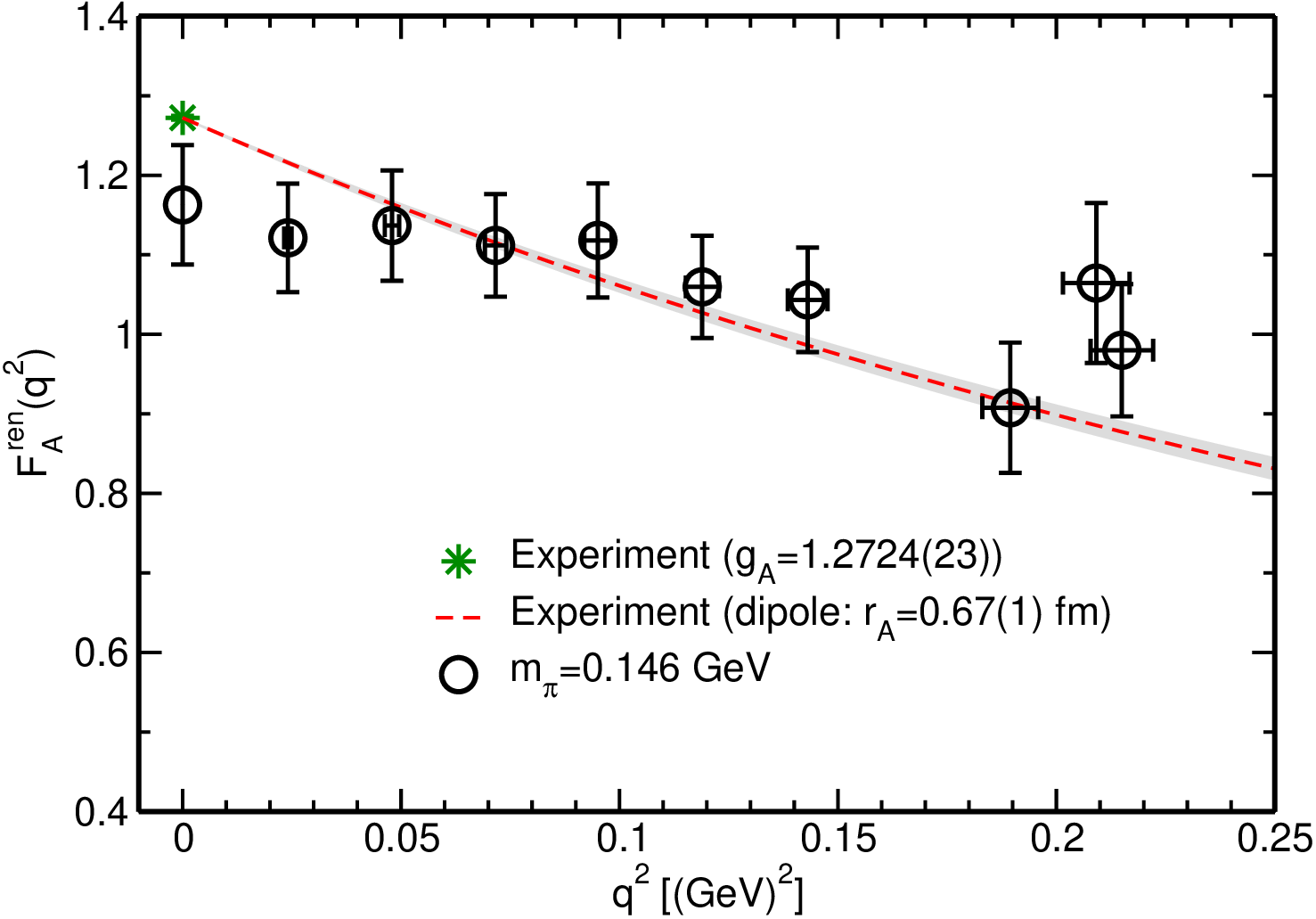}
\end{minipage}
&
\begin{minipage}[t]{0.5\textwidth}
\includegraphics[height=6.0cm,keepaspectratio,clip]{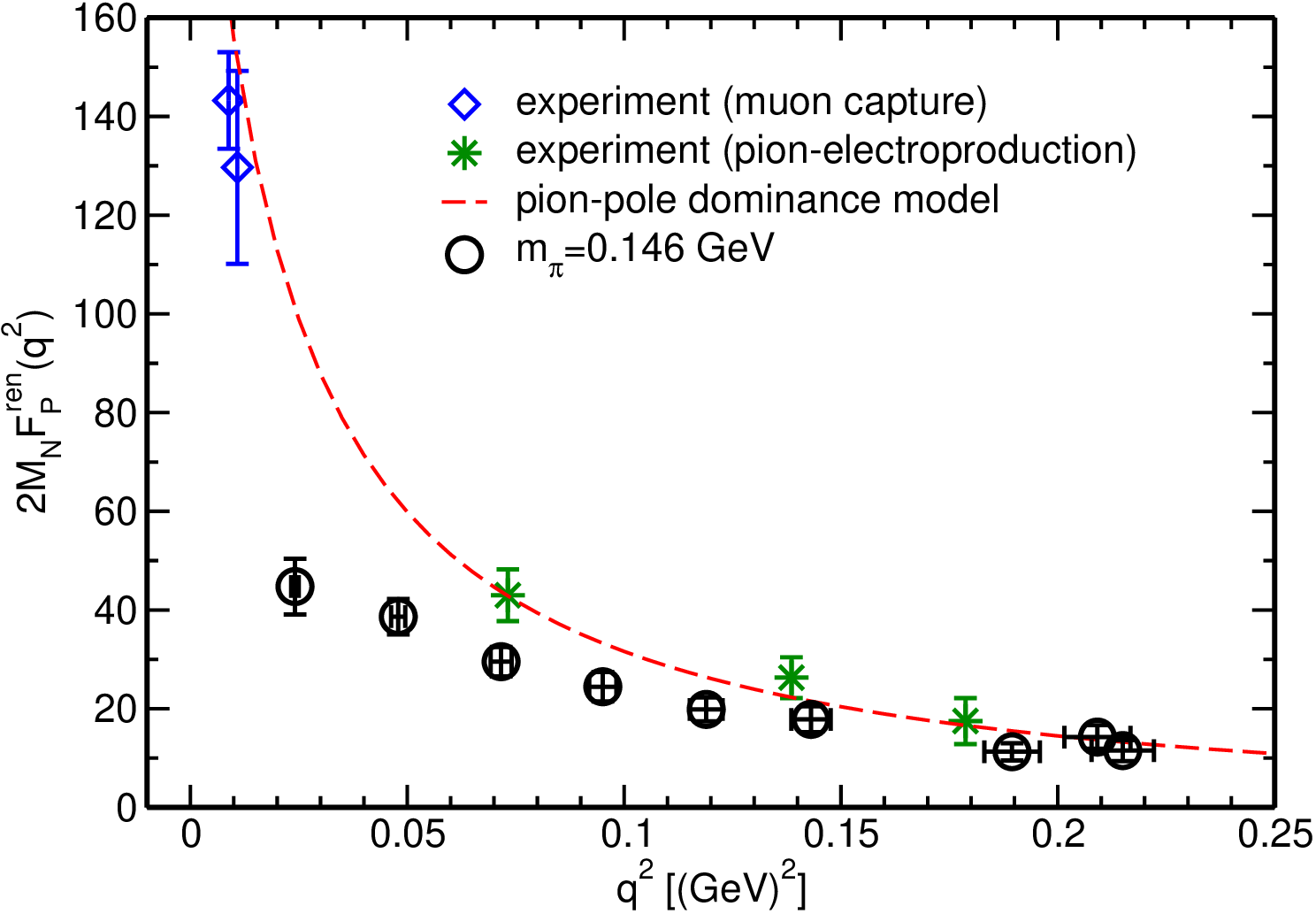}
\end{minipage}
\end{tabular}
\caption{Momentum-squared dependence of the 
    axial vector from factor $F_A$ (left) and induced pseudoscalar form
    factor $F_P$ (right). Both form factors are renormalized with
    $Z_A=0.9650(68)(95)$~\cite{Ishikawa:2015fzw}. 
    In the left panel, the experimental curve is given by a dipole form with the RMS
    radius of 
    0.67(1) fm~\cite{{Bernard:2001rs},{Bodek:2007ym}}, while the red dashed curve 
    in the right panel is given by the pion-pole dominance model. 
    The experimental data points in the right panel are given 
    by muon capture~\cite{{Andreev:2007wg},{Andreev:2012fj}} and 
    pion electroproduction~\cite{Choi:1993vt}.
     \label{fig:fafp_ren}}
\end{figure*}
%

\subsection{Results for nucleon form factors in the axial-vector and pseudoscalar channels}
\label{sec:4-C}

\subsubsection{Axial-vector $F_A$ and induced pseudoscalar $F_P$ form factors } 
\label{sec:4-C1}

In the axial-vector channel, two independent form factors $F_A$ and $F_P$
can be extracted from kinematically different types of three-point functions~(\ref{Eq:FA_FP}).
Recall that the $z$ direction is chosen as the polarized direction through
the definition of the projection operator ${\cal P}_{5z}$.
Therefore, the longitudinal momentum ($q_3$) dependence explicitly 
appears in Eq.~(\ref{Eq:FA_FP}). The three-point functions are 
classified with the transverse ($i=1$ or $2$) and longitudinal ($i=3$) 
components. Furthermore, for the case of $i=3$, there are
two types of kinematics: either $q_3\neq 0$ or $q_3=0$. 

We take advantage of the choice of either transverse or longitudinal
directions together with explicit $q_3$ dependence so as to separately 
obtain $F_A$ and $F_P$ from Eq.~(\ref{Eq:FA_FP}) in line with Ref.~\cite{Sasaki:2007gw}. 
For convenience, using the ratio of Eq.~(\ref{Eq:FA_FP}) we define 
%
%
\ben
\Lambda^{A}_L(t, {\bm q})=\sqrt{\frac{2E_N}{E_N+M_N}}
{\cal R}_{A, 3}^{5z}(t, {\bm q})
\label{Eq:Lamb_L}
\een
and
%
%
\begin{multline}
\Lambda^{A}_T(t, {\bm q})=-\frac{\sqrt{2M_N^2E_N(E_N+M_N)}}{2}\\
{}\times
\left(
\frac{{\cal R}_{A, 1}^{5z}(t, {\bm q})}{q_2q_3}
+\frac{{\cal R}_{A, 2}^{5z}(t, {\bm q})}{q_1q_3}
\right).
\label{Eq:Lamb_T}
\end{multline}
In the plateau region of $\Lambda^A_{L, T}(t,{\bm q})$
we determine the matrix elements of the axial-vector
current which has the following relation to the form factors:
%
%
\be
\Lambda^{A}_L({\bm q})=F_A(q^2)-\frac{q_3^2}{M_N+E_N}F_P(q^2),
\ee
%
%
\be
\Lambda^{A}_T({\bm q})=M_NF_P(q^2).
\ee
The simplest method is to obtain $F_A$ is obtained 
from $\Lambda^{A}_L({\bm q})$ with $q_3 = 0$, while $F_P$ is evaluated 
from $\Lambda^{A}_T({\bm q})/M_N$. However, due to the kinematics, $q_3 = 0$ is 
not available for Q3, Q6 and Q9 (labels defined in Table \ref{tab:q}), 
while $\Lambda_T^A({\bm q})$ is not calculable for Q1, Q4, Q8.
We then determine the two form factors through the following equations 
which explicitly depend on the longitudinal momentum $q_3$:
%
%
\begin{widetext}
\ben
F_A(q^2)&=&\left\{
\begin{array}{ll}
\Lambda_L^A(q_3\neq 0) + \frac{q_3^2}{M_N(M_N+E_N)}\Lambda_T^A({\bm q}) & \mbox{for Q3, Q6, Q9} \cr
\Lambda_L^A(q_3=0) & \mbox{for Q1, Q4, Q8} \cr
\end{array}
\right.\label{Eq:FA}\\
F_P(q^2)&=&\left\{
\begin{array}{ll}
\Lambda_T^A({\bm q})/M_N & \mbox{for Q3, Q6, Q9} \cr
\frac{M_N+E_N}{q_3^2}\left(
\Lambda_L^A(q_3= 0)-\Lambda_L^A(q_3\neq 0)
\right) & \mbox{for Q1, Q4, Q8} . \cr
\end{array}
\right.\label{Eq:FP}
\een
\end{widetext}
Although both ways are available for Q2, Q5 and Q7, we choose
the former in this study.

Figure~\ref{fig:tdep_RFARFP} shows the ratios of the 
three- and two-point functions (\ref{Eq:FA}) and (\ref{Eq:FP}) 
as a function of the current insertion time slice $t$ for
the axial-vector form factor ($F_A$) (left) and induced-pseudoscalar 
form factor ($F_P$) (right) at the nine lowest nonzero momentum transfers.
The latter is multiplied by $2M_N$ to make it a dimensionless quantity, while 
both ratios are renormalized with $Z_A=0.9650(68)(95)$, which is given in the
SF scheme~\cite{Ishikawa:2015fzw}.
We evaluate the values of both axial-vector and induced-pseudoscalar form factors 
by constant fits to the data points in the time region $6 \le t/a \le 9$ denoted 
by the gray shaded area.  

%
%
\begin{table}[ht] 
\caption{
Results for the nucleon form factors in the axial-vector channel
and pseudoscalar channel. The form factors $F_A^{\rm ren}$ and $F_P^{\rm ren}$ 
are renormalized ones, while $G_P$ is not yet renormalized.
\label{tab:formfactorAP}
}
\begin{ruledtabular}
\begin{tabular}{lllll} \hline
\multicolumn{1}{c}{$q^2$ [$({\rm GeV})^2$]}
& \multicolumn{1}{c}{$F^{\rm ren}_A(q^2)$} & \multicolumn{1}{c}{$2M_N F^{\rm ren}_P(q^2)$} 
& \multicolumn{1}{c}{$G_P(q^2)$}  \\ \hline 
0.000& 1.163(75) & N/A & N/A\cr
0.024(1)& 1.121(68)& 44.8(5.6) & 80.0(5.8)\cr
0.048(2)& 1.137(69)& 38.6(3.6) & 57.7(4.0)\cr
0.072(2)& 1.112(64)& 29.5(3.0) & 46.4(3.3)\cr
0.095(3)& 1.118(72)& 24.5(3.0) & 39.9(4.1)\cr
0.119(4)& 1.060(64)& 19.9(2.4) & 35.6(2.9)\cr
0.143(5)& 1.043(66)& 17.9(2.6) & 32.8(3.0)\cr
0.189(7)& 0.908(82)& 11.3(1.7) & 26.4(3.0)\cr
0.209(8)& 1.065(101)& 14.3(2.4)&18.6(4.0)\cr
0.215(8)& 0.980(83)& 11.5(2.1) & 26.8(3.7)\cr
\end{tabular}
\end{ruledtabular} 
\end{table} 

We next show the $q^2$ dependence of the renormalized form factors, $F^{\rm ren}_A$ (left panel) and $F^{\rm ren}_P$ (right panel), in Fig.~\ref{fig:fafp_ren}. 
The values of $F_A^{\rm ren}(q^2)$ and $2M_N F_P^{\rm ren}(q^2)$ 
are also summarized in Table~\ref{tab:formfactorAP}.
In the left panel, the experimental curve is given by a dipole form with a RMS radius 
of 0.67(1) fm~\cite{{Bernard:2001rs},{Bodek:2007ym}} for a comparison.
At a glance, the $F^{\rm ren}_A$ form factor is barely consistent with
the experimental curve in the whole region of measured momentum transfers 
within the current statistics, except for two points at the smallest and second 
smallest momentum transfers including the axial-vector coupling $g_A=F_A(0)$.

In the right panel, two experiments results from muon capture and pion-electroproduction are marked 
as blue diamonds and green asterisks. Our result for $F_P(q^2)$ is significantly underestimated 
in comparison with both experiments especially in the low-$q^2$ region. 
The $F_P$ form factor is expected to have a pion pole that dominates the behavior 
near zero momentum transfer.
The red dashed curve in the right panel is given by the pion-pole dominance (PPD) model, 
where the induced pseudoscalar form factor is given as 
%
%
\be
F_P^{\rm PPD}(q^2)=2M_N F_A(q^2)/(q^2+m_\pi^2),
\label{Eq:PPD_FP}
\ee
whose functional form becomes justified by the generalized GT relation~(\ref{Eq:GTrelation}) 
at least in the chiral limit ($\hat{m}=0$). Indeed, the two experimental results from muon capture and 
pion-photo production follow a curve given by the PPD model. According to the PPD model~(\ref{Eq:PPD_FP}), 
the observed quenching effect in the value of $F_P(q^2)$ might be {\it partly} 
due to the underestimation of the axial-vector coupling $g_A$ obtained in this study.

\begin{figure*}[ht]
\begin{tabular}{ccc}
\begin{minipage}[t]{0.33\textwidth}
\includegraphics[height=4.0cm,keepaspectratio,clip]{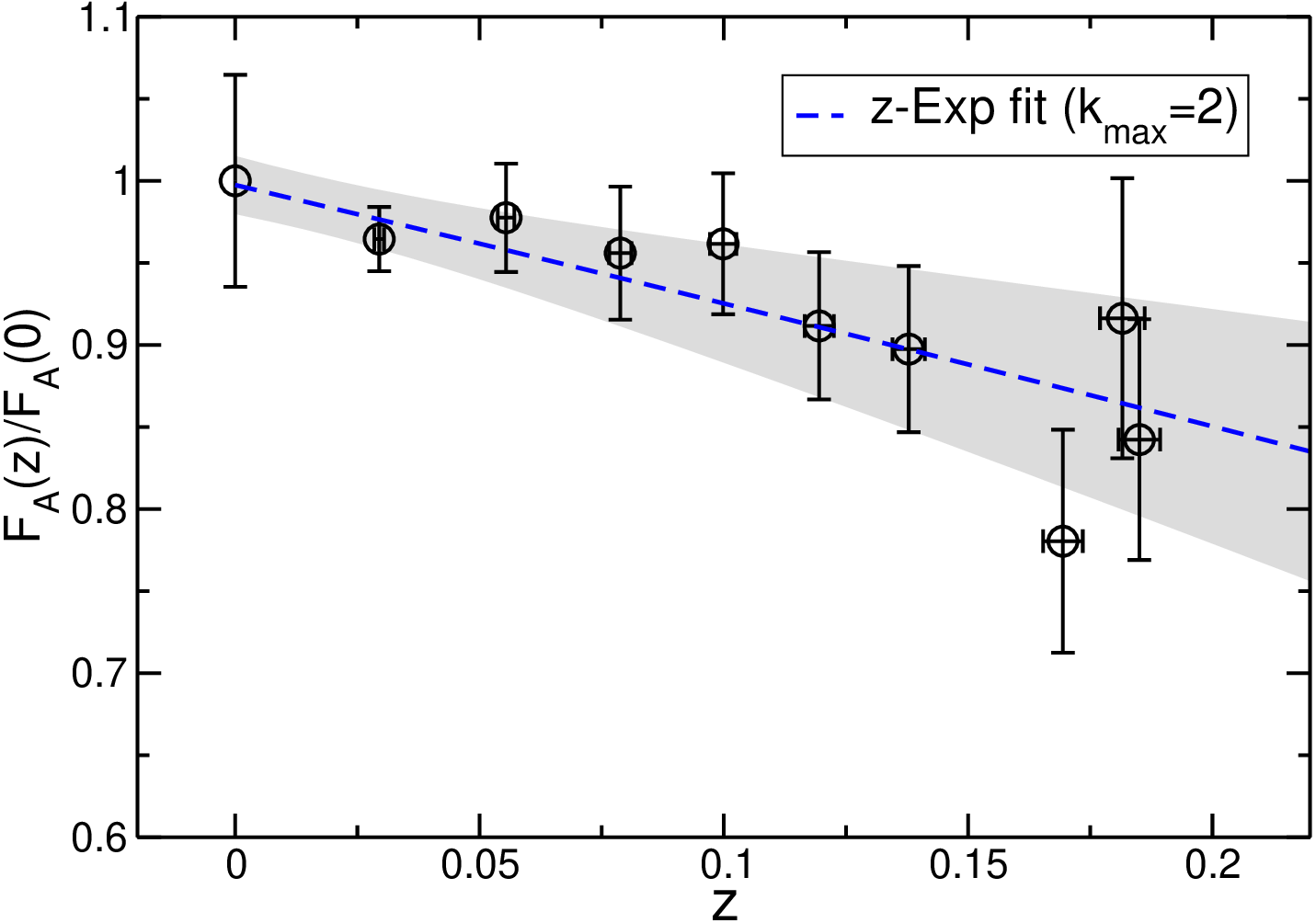}
\end{minipage}
&
\begin{minipage}[t]{0.33\textwidth}
\includegraphics[height=4.0cm,keepaspectratio,clip]{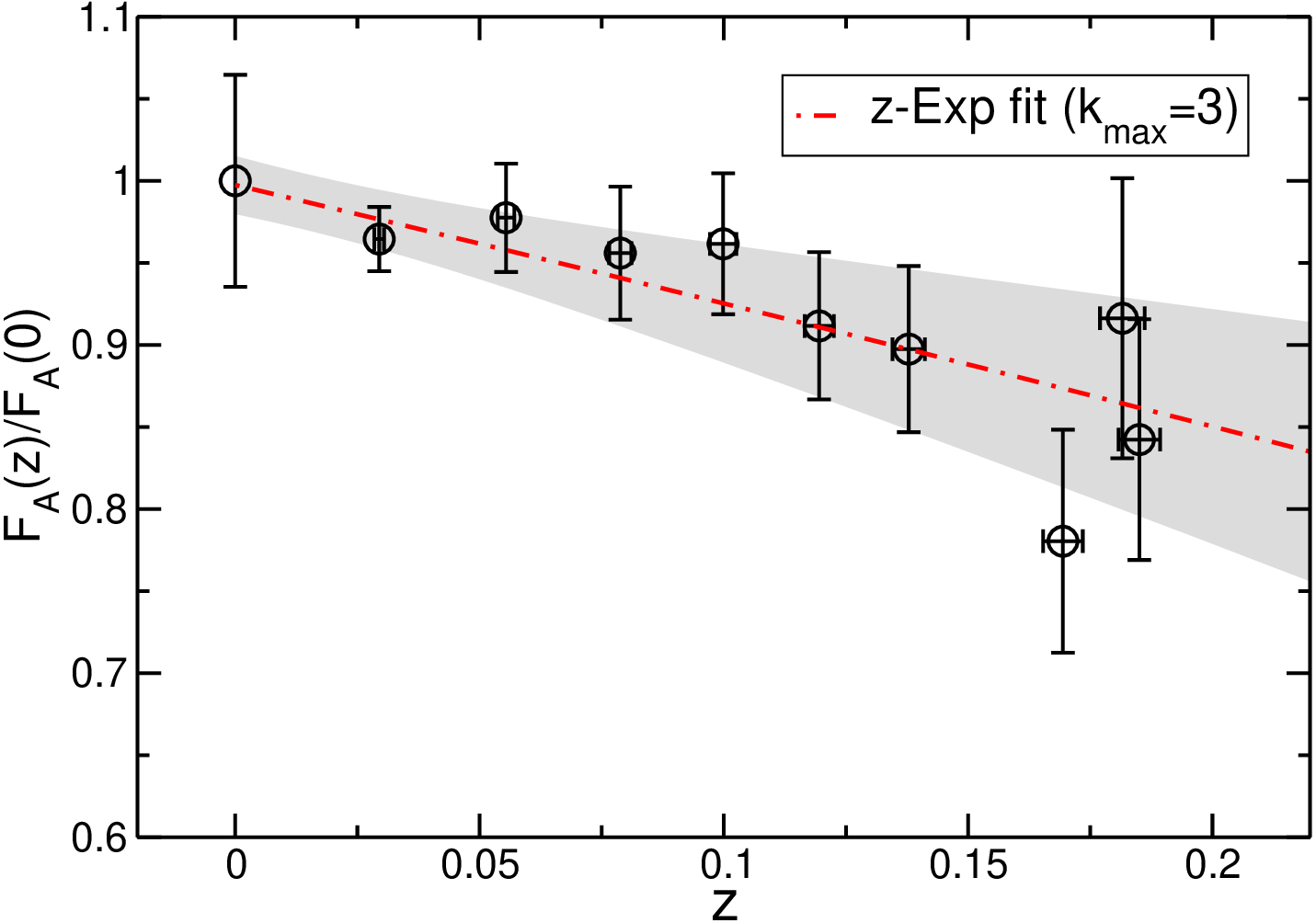}
\end{minipage}
&
\begin{minipage}[t]{0.33\textwidth}
\includegraphics[height=4.0cm,keepaspectratio,clip]{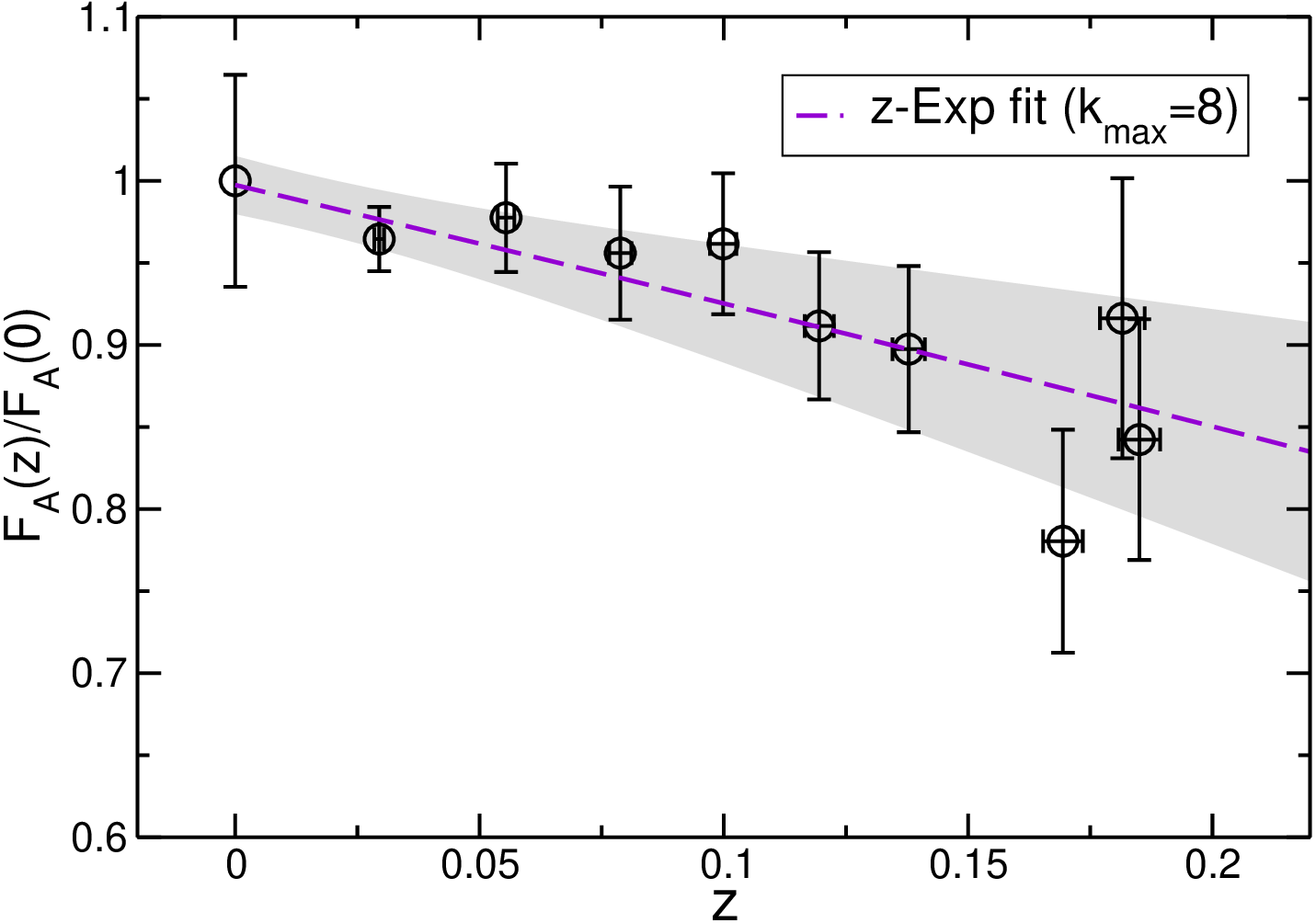}
\end{minipage}
\end{tabular}
\caption{Results for z-Exp fit of $F_A$ with $k_{\rm max}=2$ (left), 3 (middle) and 8 (right) using all ten data points.}
\label{fig:fit_A_zexp}
\end{figure*}

We next evaluate the axial RMS radius $\sqrt{\langle r_A^2\rangle}$ from the slope 
of $F_A(q^2)/F_A(0)$ at zero momentum transfer. Three types of fits are used 
as in the cases of $\sqrt{\la r^2_{E}\ra}$ and $\sqrt{\la r^2_{M}\ra}$.
First of all, Fig.~\ref{fig:fit_A_zexp} shows the z-Exp fit results
for all ten data points of $F_A(q^2)/F_A(0)$ with $k_{\rm max}=2, 3$ and 8 
in the $z$-expansion. In Fig.~\ref{fig:fit_A_zexp}, the ratio of $F_A(q^2)/F_A(0)$
is plotted as a function of the variable $z$, which is defined by Eq.~(\ref{Eq:z-value})
with $t_{\rm cut}=9m_{\pi}^2$. 

Before discussing the stability of the z-Exp fit, we remark that the statistical 
uncertainties on $F_A(q^2)/F_A(0)$ at the lower momentum transfers are 
extremely suppressed since statistical fluctuations in the numerator and 
denominator are highly correlated.
This observation indicates that the underestimation of $F_A(q^2)$ 
at the second smallest momentum transfer compared with the experimental value 
found in Fig.~\ref{fig:fafp_ren} can be attributed to the reduction of the axial-vector 
coupling $g_A$.

As shown in the three panels of Fig.~\ref{fig:fit_A_zexp}, 
we again confirm that the fit results are not sensitive to the choice 
of $k_{\rm max}$ as in the cases of $G_E$ and $G_M$. 
Therefore, the value of $k_{\rm max}=8$ is large enough
to guarantee that the z-Exp analysis makes a model-independent fit. 

%
%
%

%
%
\begin{table*}[ht]
\begin{ruledtabular}
\caption{
Results of axial radius $\sqrt{\langle r_A^2\rangle}$ obtained from various uncorrelated fits. 
\label{tab:RMSaxial}}
\begin{tabular}{ c  c c  c c c c c }
& & & \multicolumn{3}{ c }{z-Exp fit}  & Experimental value  \cr
\cline{4-6}
Observable & dipole fit & quadratic fit & $k_{\rm max}=2$ & $k_{\rm max}=3$ & $k_{\rm max}=8$  & \cr
\hline
$\sqrt{\langle r_A^2\rangle}$ [fm] 
& 0.40(12) & 0.22(49) & 0.46(11)& 0.46(11)& 0.46(11) 
& 0.67(1)
\cr
$\chi^2/{\rm dof}$ & 3.45/8 & 2.60/7 & 4.00/7 & 4.00/6 & 4.00/1 & --- \cr
\hline
\end{tabular}
\end{ruledtabular}
\end{table*}

\begin{figure}[ht]
\begin{minipage}[t]{0.50\textwidth}
\includegraphics[height=6.0cm,keepaspectratio,clip]{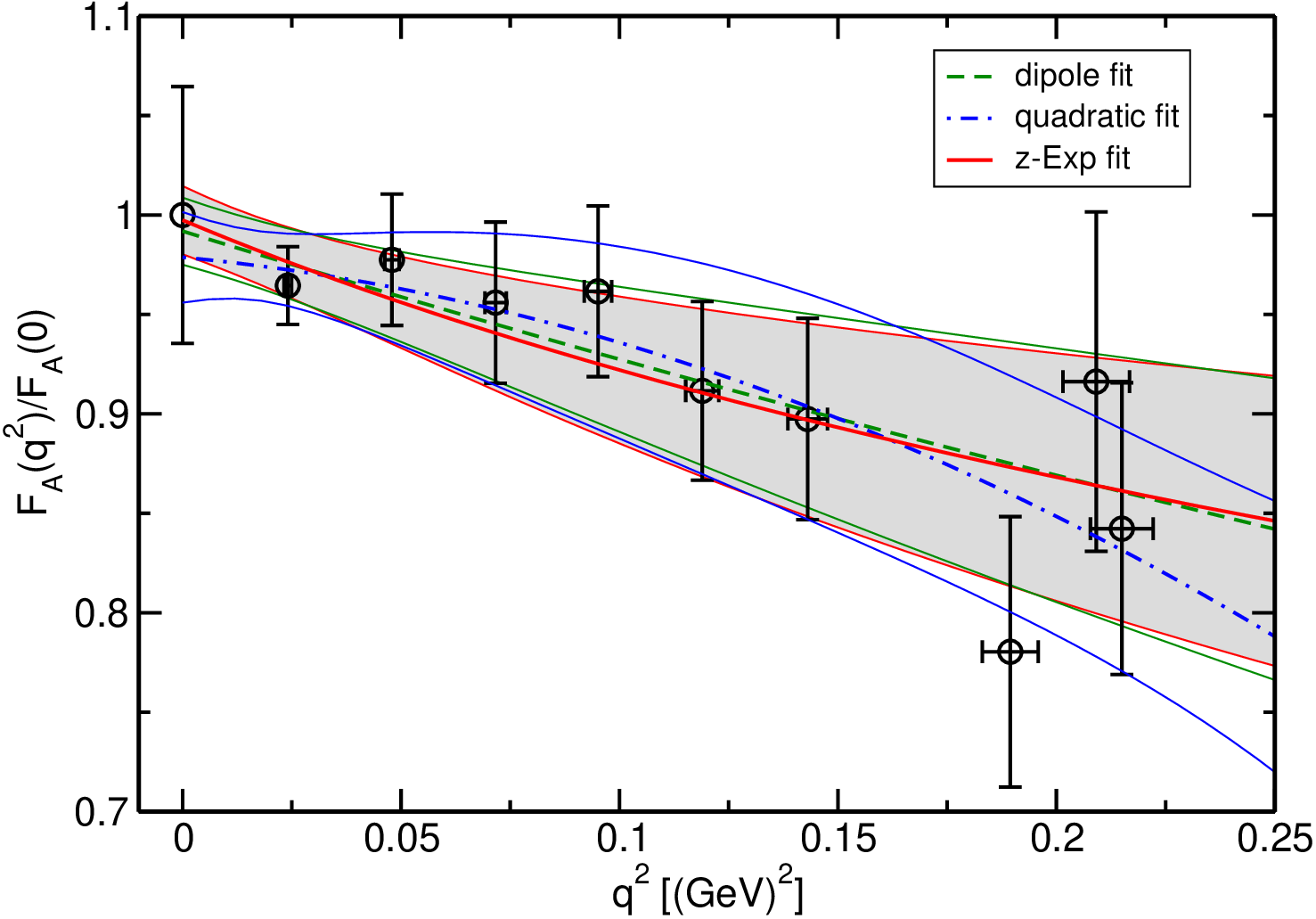}
\end{minipage}
\caption{
     Results for  $F_A$ with three types of fitting form ansatz: dipole (red),
     quadratic (blue) and z-Exp (green) fits.
     All fits are performed with all ten data points.\label{fig:gafit}}
\end{figure}

For comparison, in Fig.~\ref{fig:gafit}, 
we next show the fit results for $F_A(q^2)/F_A(q^2)$ with 
three types of fits: dipole (green dashed curve), quadratic (blue dot-dashed curve) 
and z-Exp (red solid curve) fits. 
All results of the axial RMS radius obtained
using the three fits are complied in Table~\ref{tab:RMSaxial}.
Although all three fits are mutually consistent with each other 
because of their large statistical errors, the z-Exp fit tends to give a
higher value, $\sqrt{\langle r_A^2\rangle}=0.46(11)$ fm, which is closer to 
the experimental value of 0.67(1) fm~\footnote{
Recently, Ref.~\cite{Hill:2017wgb} claims that the error on the experimental value of 0.67(1) fm~\cite{{Bernard:2001rs},{Bodek:2007ym}} is underestimated because of the model dependent analysis using the dipole ansatz. 
Instead, the value of $\sqrt{\langle r_A^2\rangle}=0.67(13)$ fm is quoted in Ref.~\cite{Hill:2017wgb}.}.

%
%
%

\subsubsection{Pseudoscalar $G_P$ form factor} 
\label{sec:4-C2}

As described in Sec.~\ref{sec:2}, it is theoretically known that the two form factors
$F_A$ and $F_P$ in the axial-vector channel are not fully independent.
Theoretically, the $q^2$ dependences of $F_A$ and $F_P$ should be constrained 
by the generalized GT relation~(\ref{Eq:GTrelation})
with the pseudoscalar form factor $G_P$ as a consequence of the axial Ward-Takahashi identity.
To test whether the correct behavior of the generalized GT relation is well satisfied in our simulations, 
we also calculate the pseudoscalar matrix element, which is described by the single form factor $G_P(q^2)$ defined in Eq.~(\ref{Eq:PSFF}). 

\begin{figure}[t]
	\begin{minipage}[t]{0.5\textwidth}	
        \includegraphics[height=6.0cm,keepaspectratio,clip]{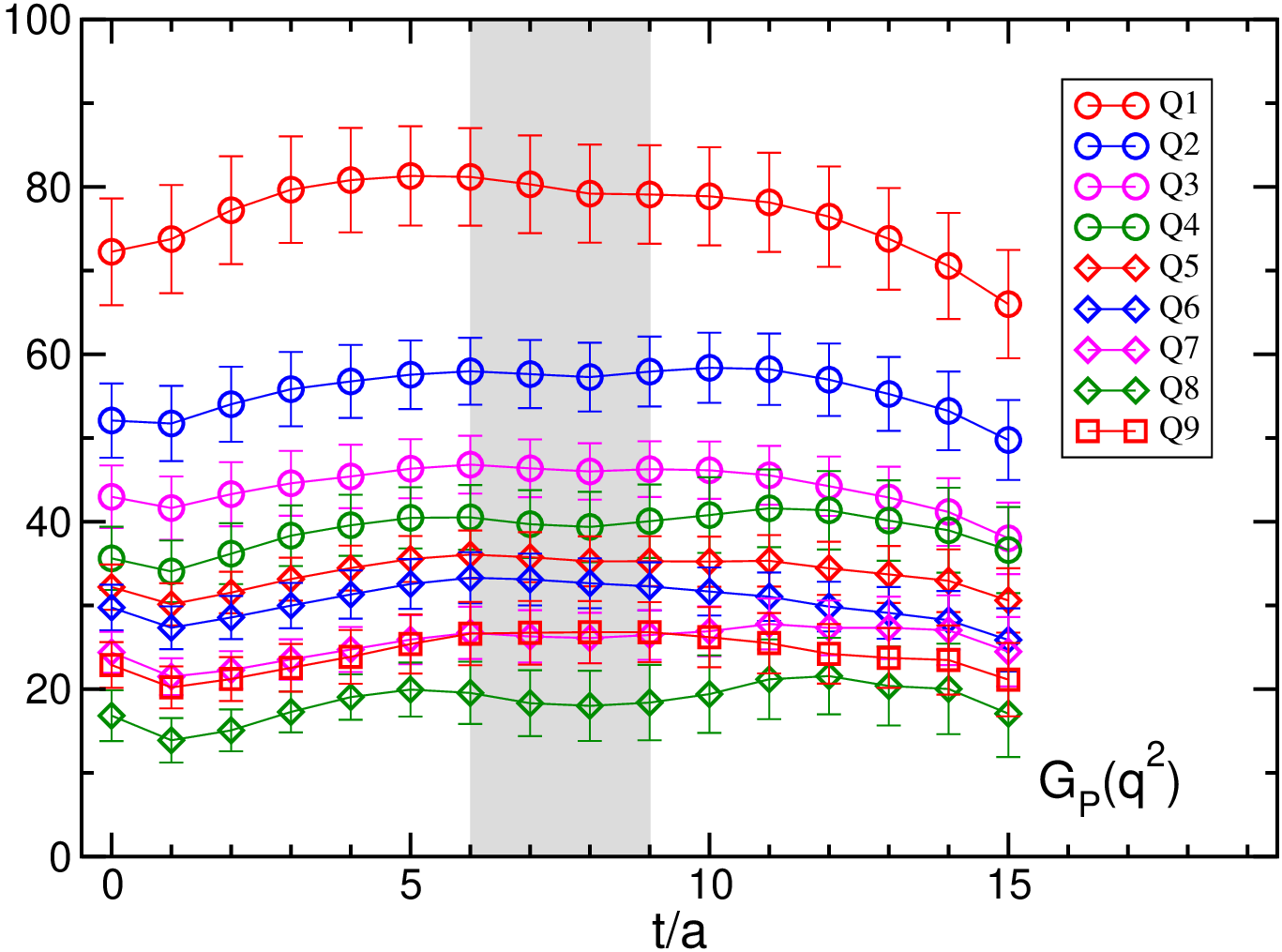}
	\end{minipage}
\caption{
Ratio of the three- and two-point functions (\ref{Eq:GP})
for the pseudoscalar form factor $G_P$ at the nine lowest nonzero momentum transfers
after extracting the relevant kinematical factors.
The gray shaded area ($6\le t/a \le 9$) in each panel shows the region where 
the values of the corresponding form factor are estimated.}
\label{fig:tdep_BGP}
\end{figure}

\begin{figure}[h]
	\begin{minipage}[t]{0.5\textwidth}	
        \includegraphics[height=6.0cm,keepaspectratio,clip]{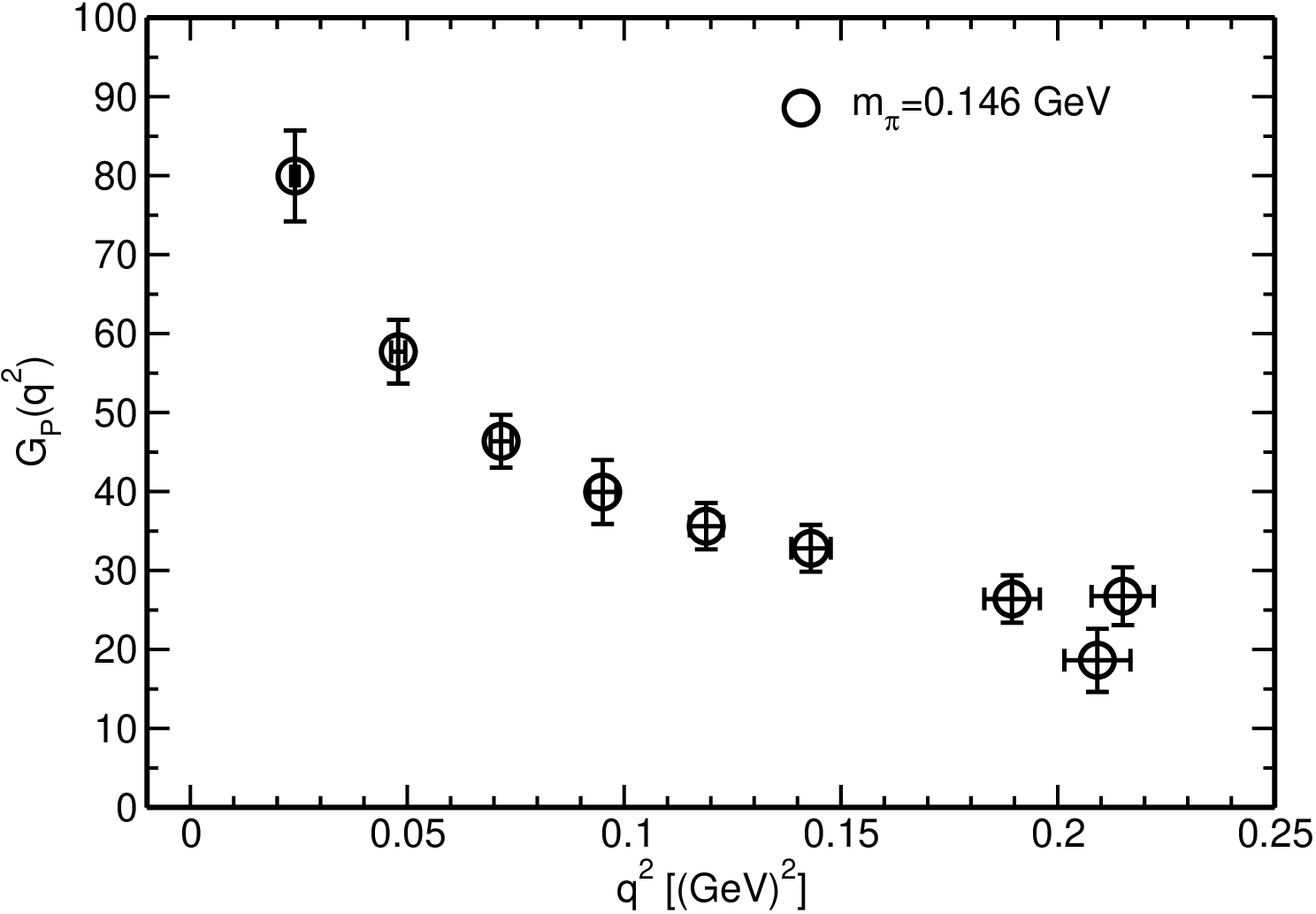}
	\end{minipage}
        \caption{Momentum-squared dependence of the pseudoscalar form factor, 
        which is the bare value, {\it i.e.} not renormalized.\label{fig:bgp}}
\end{figure}

Figure~\ref{fig:tdep_BGP} shows
the ratio of the three- and two-point functions (\ref{Eq:GP}) as a function of 
the current insertion time slice $t$ for the pseudoscalar form factor $G_P(q^2)$
at the nine lowest nonzero momentum transfers after 
extracting the relevant kinematical factors. 
The plateau region is clearly defined even at the higher momentum transfers
with our choice of source-sink separation. We thus evaluate the values of the 
pseudoscalar form factors by constant fits to the data points in the time region 
$6\le t/a \le 9$ denoted by the gray shaded area
as in the cases of the other form factors.
We next plot the bare pseudo-scalar form factor $G_P(q^2)$, which is not renormalized,
in Fig.~\ref{fig:bgp}.
The measured $q^2$ dependence of $G_P(q^2)$ resembles that of $F_P(q^2)$, where
the relatively strong $q^2$ dependence appears in the lower $q^2$ region 
as was expected from the pion-pole contribution.

In the PPD model, the pion-pole dominance holds even in $G_P(q^2)$. Combined with Eq.~(\ref{Eq:GTrelation})
and Eq.~(\ref{Eq:PPD_FP}), a naive pion-pole
dominance form $G_P^{\rm PPD}(q^2)$ is given as
%
%
\be
2\hat{m}G_P^{\rm PPD}(q^2)=2M_N F_A(q^2)\frac{m_{\pi}^2}{q^2+m_{\pi}^2}.
\ee
Thus one may realize that the ratio of the PPD forms, $G^{\rm PPD}_P$ and $F^{\rm PPD}_P$ 
does not depend on $q^2$ and gives the low-energy constant $B_0$ as
%
%
\be
\frac{G^{\rm PPD}_P(q^2)}{F^{\rm PPD}_P(q^2)}=B_0
\ee
with the help of the Gell-Mann$-$Oakes$-$Renner (GMOR) relation for the pion mass: $m_{\pi}^2=2B_0\hat{m}$.

\begin{figure}[thb]
     \begin{minipage}[t]{0.5\textwidth}
     \includegraphics[height=6.0cm,keepaspectratio,clip]{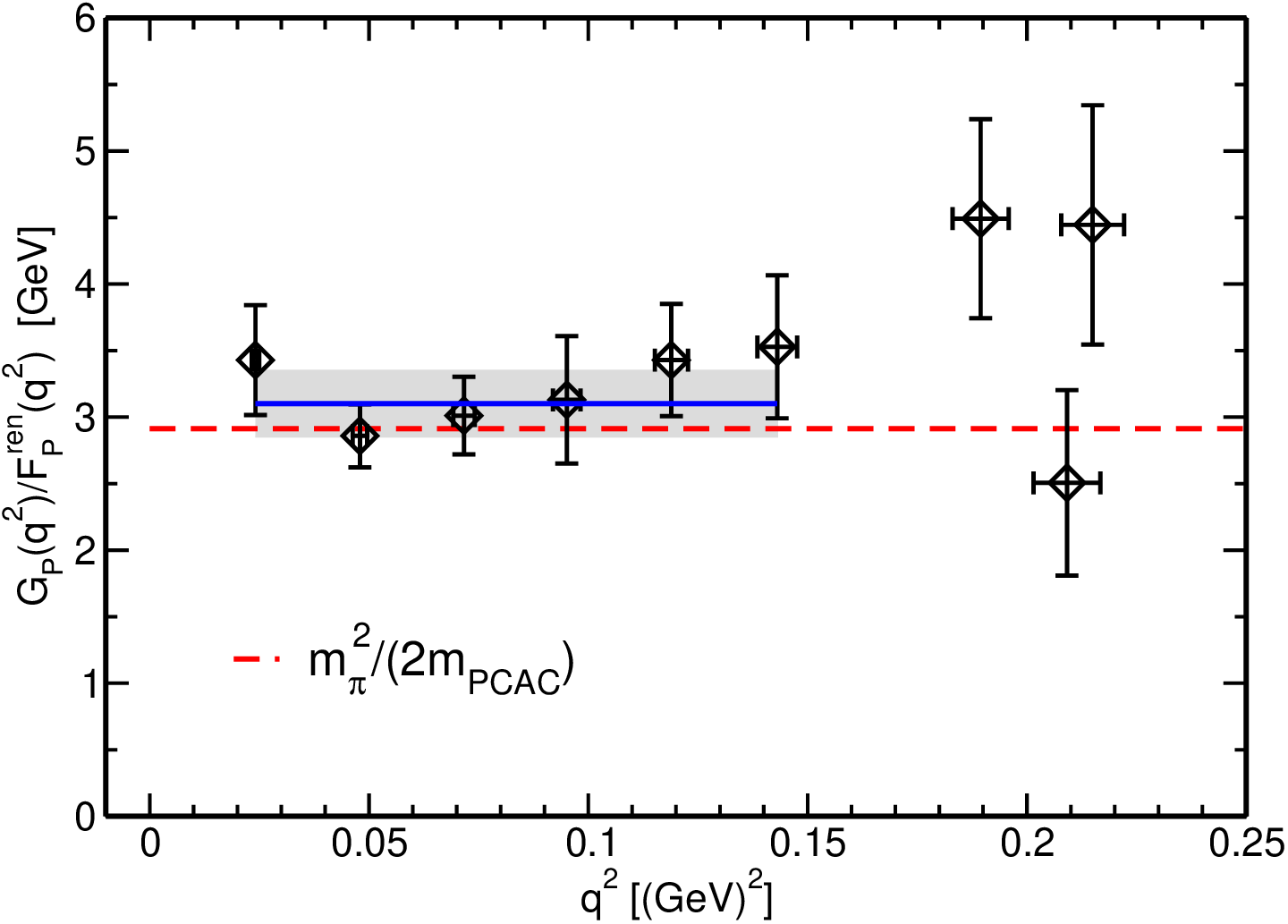}
     \end{minipage}
    \caption{The ratio of $G_P(q^2)/F^{\rm ren}_P(q^2)$  as a 
    function of momentum squared $q^2$. A flat $q^2$ dependence is observed due to the
    fact that the $G_P(q^2)$ form factor shares a similar pion-pole structure with the 
    $ F^{\rm ren}_P(q^2)$ form factor. The blue solid line represents the (constant) fit result 
    and the shaded band denotes the fit range and statistical error estimated using the jackknife 
    method, while a red dashed line represents the bare low-energy
    constant evaluated using the ratio of $m_\pi^2/(2m_{\rm PCAC})$.
    \label{fig:GPFPratio}}
\end{figure}

As shown in Fig.~\ref{fig:GPFPratio}, the ratio of $G_P(q^2)/F^{\rm ren}_P(q^2)$ indeed 
exhibits a flat $q^2$ dependence at lower $q^2$. We then estimate 
the low-energy constant $B_0$ by a constant fit to the plateau value
using six data points at the lower momentum transfer. We then get the bare value of the 
low-energy constant as $B_0=3.10(25)$ [GeV], which is represented by
blue solid line with a shaded band in Fig.~\ref{fig:GPFPratio} and tabulated in Table~\ref{table:summary_outputs}.
This value is fairly consistent 
with the one evaluated by the GMOR relation with the simulated pion mass
and a (bare) quark mass ($am_{\rm PCAC} =0.001577(10)$) obtained from 
the pion two-point correlation functions with the PCAC relation~\cite{Ishikawa:2015rho, Ishikawa:2015fzw}. 
This observation strongly indicates that the $G_P(q^2)$ form factor shares 
a similar pion-pole structure with the $ F^{\rm ren}_P(q^2)$ form factor
and the ratio of their residues is highly constrained by the GMOR relation.

%
%
%

%
%
%

%
%
%

\subsubsection{Test for the axial Ward-Takahashi identity} 
\label{sec:4-C3}

In order to verify the axial Ward-Takahashi identity in terms of the nucleon matrix elements,
we define the following ratio inspired by the generalized GT relation~(\ref{Eq:GTrelation})
%
%
\be
m_{\rm AWTI} = \frac{2M_NF^{\rm ren}_A(q^2) - q^2 F^{\rm ren}_P(q^2)}{2G_P(q^2)}
\label{Eq:MAWTI}
\ee
with the simulated nucleon mass $M_N$. If the ratio reveals a $q^2$-independent
plateau in the entire $q^2$ region, $m_{\rm AWTI}$
should be regarded as an alternative (bare) quark mass. 

\begin{figure}[ht]
	\begin{minipage}[t]{0.5\textwidth}	
     	\includegraphics[height=6.0cm,keepaspectratio,clip]{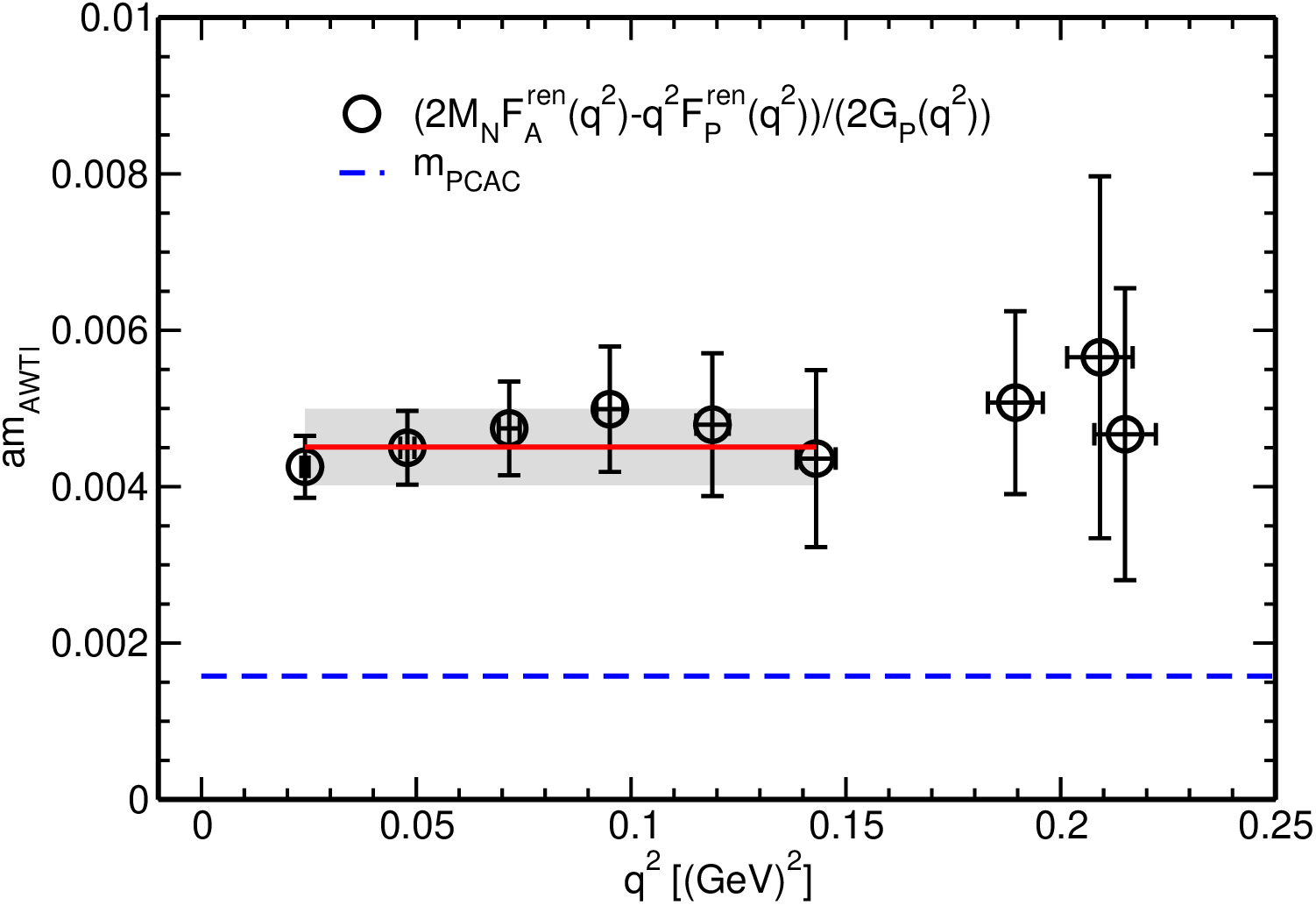}
	\end{minipage}
    	\caption{The ratio $m_{\rm AWTI}$ defined in Eq.~(\ref{Eq:MAWTI}) as a function of 
    	momentum squared $q^2$.  The blue dashed line denotes the PCAC quark 
	mass~\cite{Ishikawa:2015rho, Ishikawa:2015fzw} in lattice units.
    	\label{fig:awi}}
\end{figure}

As shown in Fig.~\ref{fig:awi},
the ratio $m_{\rm AWTI}$ has no appreciable $q^2$ dependence. 
It indicates that three form factors well satisfy the generalized GT relation. 
Using data points at the six lowest momentum transfers, 
we can read off $am_{\rm AWTI} =0.00451(48)$, 
which is roughly 3 times heavier than the PCAC quark mass~\cite{Ishikawa:2015rho, Ishikawa:2015fzw}. 
Since the PCAC relation is also a consequence of the axial Ward-Takahashi identity, 
the relation $m_{\rm PCAC}\approx m_{\rm AWTI}$ was expected. However, this is not the case.

Although the above consideration does not take into account ${\cal O}(a)$ improvement of
the axial-vector current, we next verify that this issue has nothing to do with finite lattice spacing artifacts.
The renormalized ${\cal O}(a)$-improved operator takes the form 
%
%
\be
{\cal A}^+_{\alpha}(x)= Z_A\left[A^+_{\alpha}(x)+ac_A \partial_{\alpha} P^+(x) \right],
\label{Eq:Aimp}
\ee
where $A^+_{\alpha}(x)=\bar{u}(x)\gamma_{\alpha}\gamma_5d(x)$
and $P^+(x)=\bar{u}(x)\gamma_5d(x)$. 
Strictly speaking, this improved axial-vector current
satisfies the axial Ward-Takahashi identity on the lattice:
$\partial_{\alpha}{\cal A}^+_{\alpha}(x)= 2\hat{m}P^{+}(x)$.
Therefore, the generalized GT relation~(\ref{Eq:GTrelation})
is slightly modified by the presence of the ${\cal O}(a)$ improvement term
in Eq.~(\ref{Eq:Aimp}). However, the term proportional to $c_A$ only contributes 
to the modification of the $F_P$ form factor as 
$F^{\rm imp}_P(q^2)=F_P(q^2)+ac_AG_P(q^2)$, 
and then the modified GT relation is expressed by 
%
%
\be
Z_A\left(2M_NF_A(q^2)-q^2F^{\rm imp}_P(q^2)\right)=2\hat{m}G_P(q^2).
\label{Eq:modGTrelation}
\ee

Starting from the modified GT relation~(\ref{Eq:modGTrelation}), the (bare) quark mass $\hat{m}$ becomes
%
%
%
\begin{multline}
 m^{\rm imp}_{\rm AWTI} = \frac{Z_A\left(2M_NF_A(q^2)-q^2F^{\rm imp}_P(q^2)\right)}{2G_P(q^2)} \cr
{}=  \frac{2M_NF^{\rm ren}_A(q^2) - q^2 F^{\rm ren}_P(q^2)}{2G_P(q^2)}-\frac{aZ_Ac_A}{2} q^2,
 \label{Eq:MAWTI-imp}
\end{multline}
where the first term is nothing but the ratio $m_{\rm AWTI}$ defined in Eq.~(\ref{Eq:MAWTI}) and the second term corresponds to
a correction term due to ${\cal O}(a)$ improvement of the axial-vector current. 
Although the ratio $m_{\rm AWTI}$ may indeed receive the ${\cal O}(a)$ correction, which yields 
a linear dependence on $q^2$,
the effect of ${\cal O}(a)$ improvement is insignificant in the low $q^2$ region.
Furthermore, the improvement coefficient $c_A$ is found to be very small
with our choice of $c_{SW}=1.11$~\cite{Taniguchi:2012kk}. 
The improvement coefficient $c_A$ is expected to be of the order of a few 0.01 
in lattice units~\cite{Taniguchi:2012kk}. 

\begin{figure}[ht]
	\begin{minipage}[t]{0.5\textwidth}	
     	\includegraphics[height=6.0cm,keepaspectratio,clip]{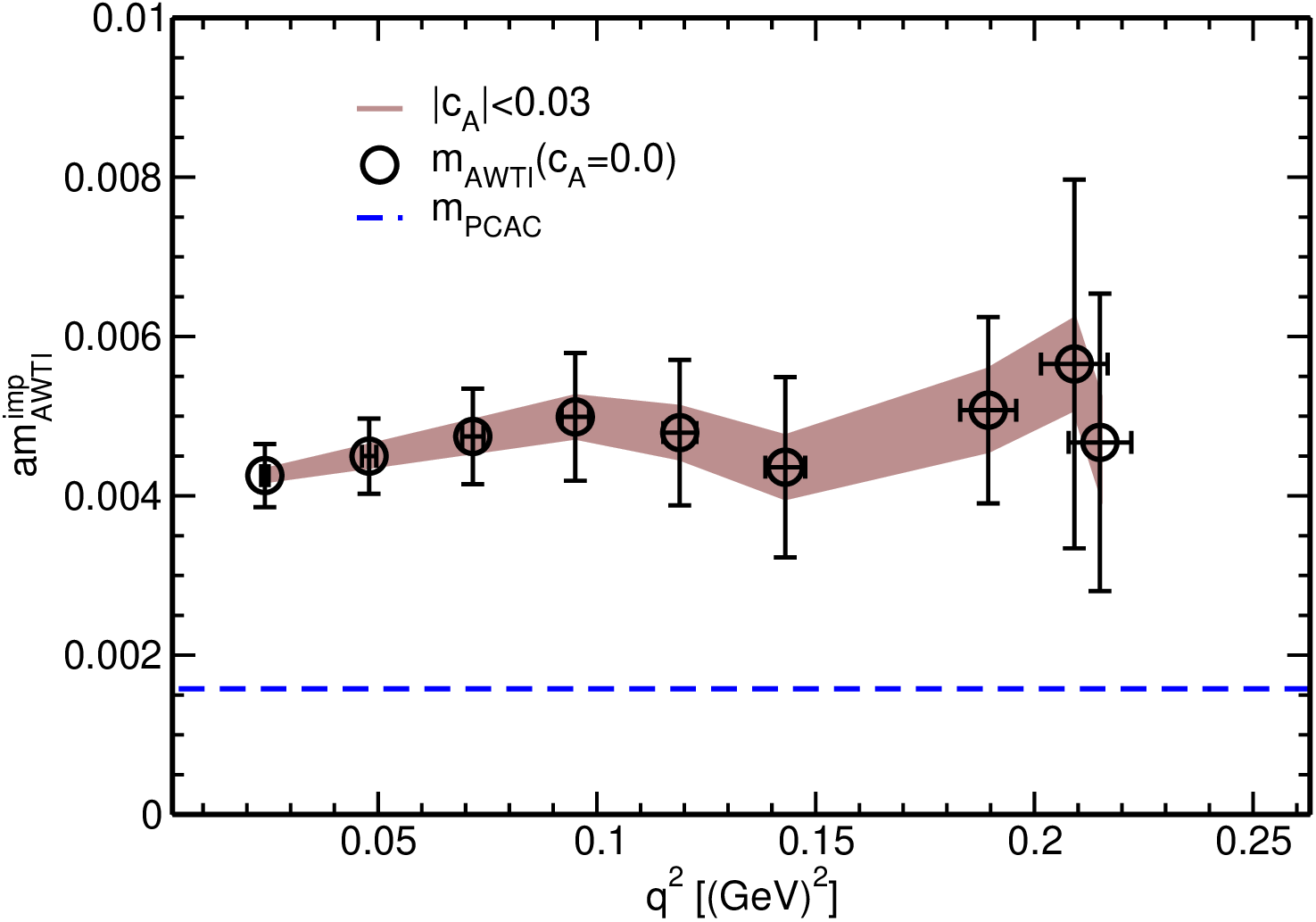}
	\end{minipage}
    	\caption{The ratio $m^{\rm imp}_{\rm AWTI}$ defined in Eq.~(\ref{Eq:MAWTI-imp}) 
	as a function of momentum squared $q^2$.
       The improvement coefficient $c_A$ is expected to be of the order of a few 0.01 in lattice units.
       The central value of the unimproved results (denoted by circles) at each $q^2$ point is
       likely to move in the brown band, which represents the region between the lower ($c_A=0.03$ in lattice units) and 
       upper ($c_A=-0.03$ in lattice units) limits.  
    	\label{fig:awi_imp}}
\end{figure}

In Fig.~\ref{fig:awi_imp}, we plot the ratio $m^{\rm imp}_{\rm AWTI}$ defined in Eq.~(\ref{Eq:MAWTI-imp}) 
as a function of momentum squared $q^2$. Circles represent the unimproved results obtained in 
Eq.~(\ref{Eq:MAWTI-imp}) with $c_A=0.0$ in lattice units. 
After adopting the renormalized ${\cal O}(a)$-improved operator~(\ref{Eq:Aimp}), 
the central value of the unimproved results at each $q^2$ point is likely to move 
in the brown band, which represents the region between the lower ($c_A=0.03$ in lattice units) and 
upper ($c_A=-0.03$ in lattice units) limits.  Figure~\ref{fig:awi_imp} indicates that 
the systematic uncertainties that arise from the ${\cal O}(a)$-improvement term in the
axial-vector current are much smaller than the current statistical errors in this study. The large deviation from 
the blue dashed line, which denotes the PCAC quark mass, mostly stays the same.
We thus conclude that the issue of $m_{\rm AWTI} \gg m_{\rm PCAC}$ 
is not directly related to finite lattice spacing artifacts.  We rather speculate that 
this issue is connected with the modification of the pion-pole structure in 
both the $F_P(q^2)$ and $G_P(q^2)$ form factors as will be described in the next subsection. 
Hereafter, the improvement coefficient is set to $c_A=0$ in our entire analysis.

%
%
%

%
%
%

%
%
\begin{table*}[ht] 
\caption{
Results for the bare low-energy constant ($B_0$), the uncorrected
quark mass ($m_{\rm AWTI}$), the pole mass of $F_P$ ($m_{\rm pole}$), the
modification factor $\alpha_{\rm pole}$ and the corrected quark mass ($m_{\rm quark})$.
\label{table:summary_outputs}
}
\begin{ruledtabular}
\begin{tabular}{cccccc} 
$B_0$ [GeV] &   $am_{\rm AWTI}$ & $am_{\rm pole}$ & $m_{\rm pole}$ [MeV]  & $\alpha_{\rm pole}$ & $am_{\rm quark}$ \cr
\hline
3.10(25)  &   0.00451(48)  &  0.1099(74) & 256(17) & 3.05(41) &    0.00145(12)\cr
\end{tabular}
\end{ruledtabular} 
\end{table*} 
%

\subsubsection{Distortion of pion-pole structure} 
\label{sec:4-C4}

In the previous subsections, we have observed that the $q^2$ 
dependences of $F_A$, $F_P$ and $G_P$ are well constrained by the generalized GT relation,
while the ``pion-pole" structures of $F_P$ and $G_P$ are likely the same. 
However, the bare quark mass ($m_{\rm AWTI}$) evaluated from the ratio (\ref{Eq:MAWTI})
is roughly 3 times heavier than the value of $m_{\rm PCAC}$.  This puzzle could be related 
to the discrepancy between our result for $F_P(q^2)$ and experiments.

We first speculate that the measured $F_P(q^2)$ can be described by the 
PPD inspired form:
%
%
\be
F_P(q^2)\approx \frac{2M_N F_A(q^2)}{q^2+m_{\rm pole}^2}
\ee
with the pole mass $m_{\rm pole}$, which is not necessarily constrained by our simulated pion mass $m_{\pi}$. 
If this functional form well describes the measured $F_P(q^2)$, we may define the effective ``pion-pole" mass from $F_P(q^2)$
as
%
%
\be
m_{\rm pole}=\sqrt{\frac{2M_NF_A(q^2)}{F_P(q^2)}-q^2},
\ee
which would exhibit a flat $q^2$ dependence. 

\begin{figure}[thb]
	\begin{minipage}[t]{0.5\textwidth}	
     	\includegraphics[height=6.0cm,keepaspectratio,clip]{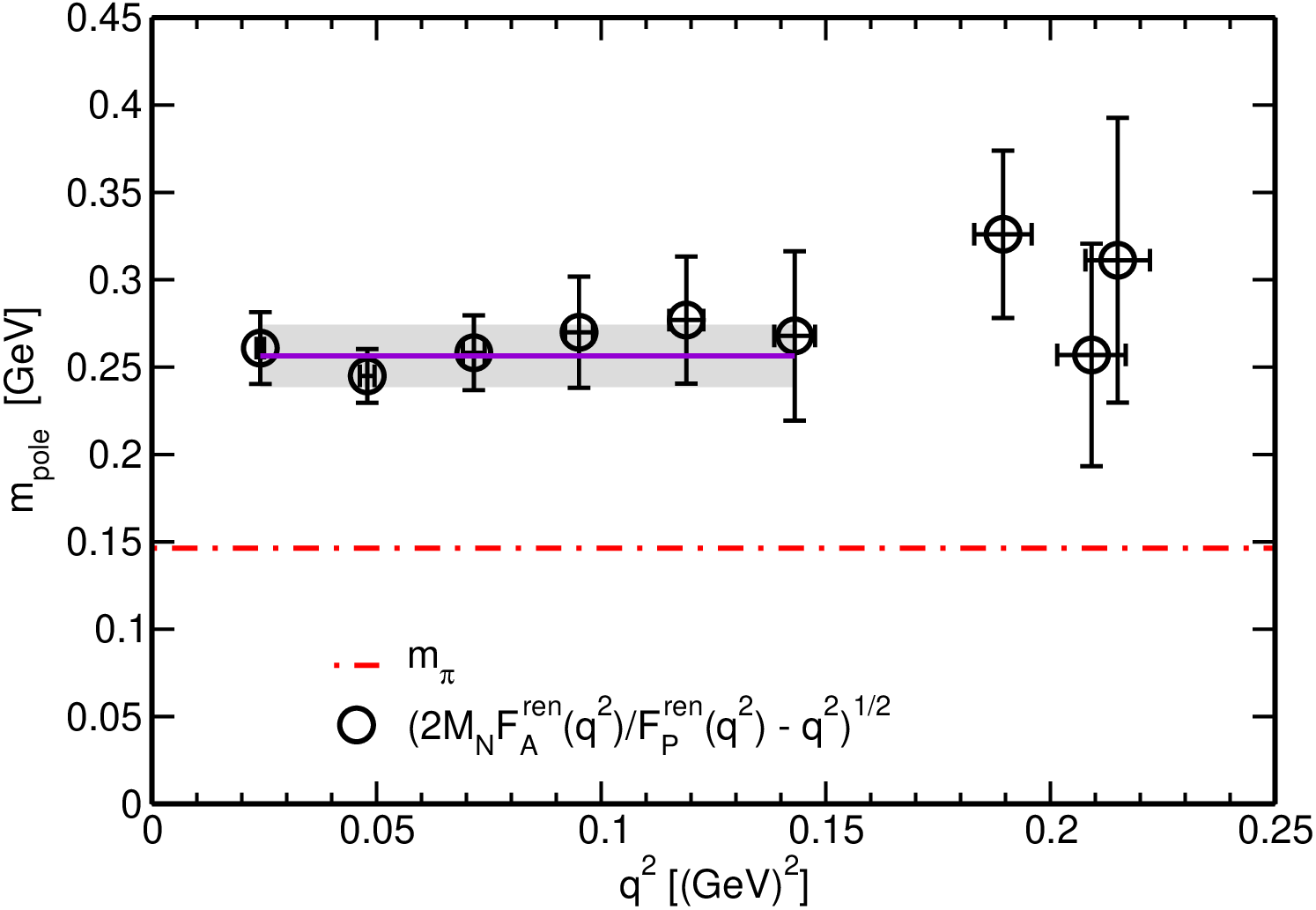}
	\end{minipage}
    \caption{Effective pole-mass plot as a function of
    momentum squared $q^2$. The horizontal dash-dotted line represents the value of 
    the simulated pion mass in physical units [GeV].
    \label{fig:pole}}
\end{figure}

In Fig.~\ref{fig:pole}, we plot the effective pole mass as a function of $q^2$.
The horizontal dash-dotted line represents the value of the simulated pion mass 
$m_{\pi}$ in lattice units.
Clearly, there is no appreciable $q^2$ dependence of the effective pole mass $m_{\rm pole}$.
In particular, the six data points at lower $q^2$ are fairly consistent within statistical errors. 
We then get the pole mass value as $am_{\rm pole}=0.1099(74)$,
which is given by a constant fit to data in the shaded region. 
However, the value $m_{\rm pole}=256(17)$ MeV is roughly twice as heavy as 
the simulated pion mass ($m_{\pi}\approx146$ MeV).
This indicates that the ``pion-pole" structure in $F_P(q^2)$ is modified by the larger pole mass. 
We then define the modification factor $\alpha_{\rm pole}$ as follows
%
%
\be
\alpha_{\rm pole}\equiv \frac{m_{\rm pole}^2}{m_{\pi}^2}
\ee 
and obtain its value as $\alpha_{\rm pole}=3.05(41)$ which 
can account for the discrepancy between $m_{\rm AWTI}$ and $m_{\rm PCAC}$ through 
the GMOR relation.
We thus estimate an improved value of the bare quark mass as
%
%
\be
m_{\rm quark} = m_{\rm AWTI}/\alpha_{\rm pole},
\ee
which should be very consistent with the value of $m_{\rm PCAC}$.  

\begin{figure}[thb]
	\begin{minipage}[t]{0.5\textwidth}	
     	\includegraphics[height=6.0cm,keepaspectratio,clip]{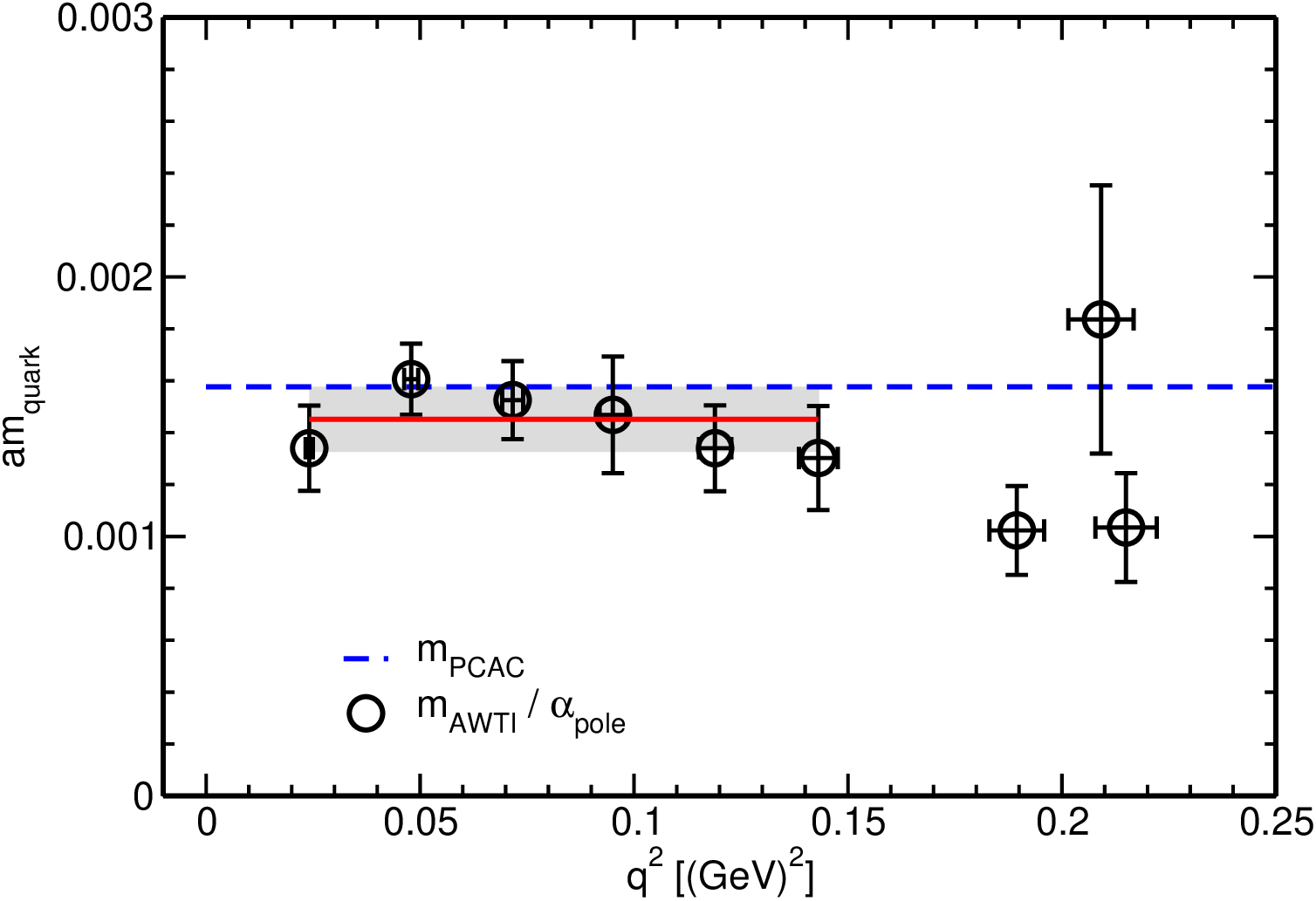}
	\end{minipage}
	\caption{The modified quark mass $m_{\rm quark}=m_{\rm AWTI}/\alpha_{\rm pole}$ 
	as a function of momentum squared $q^2$. 
	The blue dashed line denotes the PCAC quark mass~\cite{Ishikawa:2015rho, Ishikawa:2015fzw} 
	in lattice units.
    \label{fig:mpcac}}
\end{figure}

Next, we plot the quantity of $m_{\rm quark}$ in lattice units as a function of 
$q^2$ in Fig.~\ref{fig:mpcac}. Again, there is no appreciable $q^2$ dependence especially 
at low $q^2$. The horizontal dashed line represents the value of $m_{\rm PCAC}$, while 
the solid line indicates the fit result of $m_{\rm quark}$ and shaded band displays the fit range
and 1 standard deviation. The value of $m_{\rm quark}$ is obtained as
%
%
\be
am_{\rm quark}=0.00145(12),
\ee
which is in good agreement with the value $am_{\rm PCAC} =0.001577(10)$.

The importance of our findings is twofold: 1) our results for all three form factors, 
$F_A(q^2)$,  $F_P(q^2)$ and $G_P(q^2)$, are subjected to the generalized GT relation~(\ref{Eq:GTrelation}) 
as a consequence of the axial Ward-Takahashi identity, and
2) the discrepancy between our result for $F_P(q^2)$ and experiments is mainly caused 
by the distortion of the pion-pole structure in both $F_P(q^2)$ and $G_P(q^2)$, though 
the reason is not yet known. Therefore, in this work, we prefer not to estimate the pseudoscalar coupling $g_P$ 
and pion-nucleon coupling $g_{\pi NN}$, since both quantities are highly sensitive to the pole structure of $F_P(q^2)$.

%
%
%

%
%
%

\section{Summary and concluding remarks} 
\label{sec:5}

We have studied the nucleon form factors calculated in 2+1 flavor
QCD near the physical point ($m_\pi \approx 146$ MeV) 
in a large spatial volume $(8.1~\mathrm{fm})^3$
at a single lattice spacing of 0.085 fm.
We computed the relevant three-point correlation functions 
using the sequential source method with a fixed source-sink separation of 1.27 fm.

We first analyzed the vector nucleon matrix element, which is described by
the electric ($G_E$) and magnetic ($G_M$) form factors. 
We carefully examined the shapes of both the $G_E$ and $G_M$ form factor 
with a model-independent analysis based on the $z$-expansion method. 
As a result, we obtained the isovector electric 
RMS radius $\sqrt{\la r^2_{E}\ra}$ and magnetic moment $\mu_v$, which are consistent 
with experimental values. We would like to emphasize that the former quantity is, 
for the first time, consistent with both the experimental values from $ep$ 
scattering and $\mu$-$H$ atom spectroscopy, though 
further reduction of statistical and systematic errors (including
electromagnetic and isospin-breaking effects) down to less than 1\%
is required to resolve the experimental puzzle on the proton size.

We have also analyzed the axial-vector nucleon matrix element, which is described by
the axial-vector ($F_A$) and induced pseudoscalar ($F_P$) form factors. We have 
found that although the axial charge $g_A=F_A(0)$ is slightly underestimated in comparison 
with the experimental value, the axial vector form factor $F_A$ is barely consistent 
with experiments in the low-$q^2$ region ($0 \le q^2 \le  0.2$ [$({\rm GeV})^2$]).
However, the pseudoscalar form factor $F_P$ is considerably underestimated at very low $q^2$. 

To understand this issue, we have, in addition, calculated the pseudoscalar matrix element, which
is described by a single form factor of $G_P$ called the pseudoscalar form factor.  
We examined $q^2$ dependences of three form factors, $F_A$, $F_P$ and $G_P$,
which should be constrained by the generalized GT relation as a consequence of the axial 
Ward-Takahashi identity.
We have observed that the $G_P$ form factor shares a similar ``pion-pole" structure
with the $F_P$ form factor. The ``pion-pole" structure was, however, found to be distorted 
due to the pole mass being larger than
the simulated pion mass. If this effect is taken into account 
for an estimation of the bare quark mass by using three form factors through the 
generalized GT relation, we can fully verify the axial Ward-Takahashi identity in terms 
of the nucleon matrix elements. The bare quark mass obtained in this study as shown 
in Table~\ref{table:summary_outputs} is consistent 
with an alternative quark mass obtained from the pion two-point correlation functions 
with the PCAC relation. We, however, have not yet fully understood the origin of 
the ``pion-pole" distortion found in the $F_P$ and $G_P$ form factors. The similar issue 
in the axial-vector channel was reported in Refs.~\cite{{Alexandrou:2017hac},{Rajan:2017lxk}}.
After we completed this work, B\"ar has advocated that the distortion of the pion-pole structure 
that was observed in this work can be qualitatively explained by the $N\pi$ excited state contamination~\cite{Bar:2018akl}.

The PACS Collaboration is generating new ensembles in a significantly large spatial volume of 
$(10.8~{\rm fm})^3$ at the physical point ($m_\pi \approx 135$ MeV)~\cite{{128conf},{Ishikawa:2018jee}}.
Thus, we plan to develop the present calculation for a more precise determination of 
the nucleon form factors and also address all of the unsolved issues described in this paper. 
Such planning is now underway.

\begin{acknowledgments}
We thank Toshimi Suda, Eigo Shintani and Oliver B\"ar for useful discussions, 
and Yusuke Namekawa for his careful reading of the manuscript.
Numerical calculations for the present work have been carried out on 
the FX10 supercomputer system at Information Technology Center of the University of Tokyo, 
on the HA-PACS and COMA cluster systems under the ``Interdisciplinary Computational 
Science Program'' of Center for Computational Science at University of Tsukuba, on 
HOKUSAI GreatWave at Advanced Center for Computing and Communication of RIKEN, and on the 
computer facilities of the Research Institute for Information Technology of Kyushu University. 
This research used computational resources of the HPCI system provided by Information 
Technology Center of the University of Tokyo, Institute for Information Management and 
Communication of Kyoto University, the Information Technology Center of Nagoya University, 
and RIKEN Advanced Institute for Computational Science through the HPCI System Research 
Project (Project ID: hp120281, hp130023, hp140209, hp140155, hp150135, hp160125,
hp170022, hp180072).
We thank the colleagues in the PACS Collaboration for helpful 
discussions and providing us the code used in this work. 
This work is supported in part by MEXT SPIRE Field 5, and also by Grants-in-Aid for Scientific 
Research from the Ministry of Education, Culture, Sports, Science and Technology (No. 16H06002), 
and Grant-in-Aid for Scientific Research (C) (No. 18K03605).
\end{acknowledgments}


\end{document}